\documentclass[a4paper,11pt]{article}
\pdfoutput=1
\usepackage{jheppub} 
\usepackage{amsmath,amstext,amssymb}
\usepackage{hyperref}
\usepackage{comment,url}
\usepackage{rotating}
\usepackage{graphicx,subfigure}
\usepackage{placeins}
\usepackage{rotating}

\newcommand{\beq}{\begin{equation}}
\newcommand{\eeq}{\end{equation}}

\newcommand{\nn}{\nonumber}
\newcommand{\dd}{\mathrm{d}}

\makeatletter
\def\l@subsection#1#2{}
\def\l@subsubsection#1#2{}
\makeatother

\title{Quasinormal modes of charged magnetic black branes \& chiral magnetic transport}

\author[a]{Martin Ammon,}
\author[b]{Matthias Kaminski,}
\author[b]{Roshan Koirala,}
\author[a]{Julian Leiber}
\author[b]{Jackson Wu}

\affiliation[a]{Theoretisch-Physikalisches Institut, Friedrich-Schiller University of Jena, Max-Wien-Platz 1, 07743 Jena, Germany}
\affiliation[b]{Department of Physics and Astronomy, University of Alabama, Tuscaloosa, AL 35487, USA}

\emailAdd{martin.ammon@uni-jena.de}
\emailAdd{mski@ua.edu}
\emailAdd{rkoirala1@ua.edu}
\emailAdd{julian.leiber@uni-jena.de}
\emailAdd{jmwu@ua.edu}

\abstract{We compute quasinormal modes (QNMs) of the metric and gauge field perturbations about black branes electrically and magnetically charged in the Einstein-Maxwell-Chern-Simons theory. By the gauge/gravity correspondence, this theory is dual to a particular class of field theories with a chiral anomaly, in a thermal charged plasma state subjected to a constant external magnetic field, $B$. 
The QNMs are dual to the poles of the two-point functions of the energy-momentum and axial current operators, and they encode information about the dissipation and transport of charges in the plasma. 
Complementary to the gravity calculation, we work out the hydrodynamic description of the dual field theory in the presence of a chiral anomaly, and a constant external $B$. 
We find good agreement with the weak field hydrodynamics, which can extend beyond the weak $B$ regime into intermediate regimes. Furthermore, we provide results that can be tested against thermodynamics and hydrodynamics in the strong $B$ regime. 
We find QNMs exhibiting Landau level behavior, which become long-lived at large $B$ if the anomaly coefficient exceeds a critical magnitude. 
Chiral transport is analyzed beyond the hydrodynamic approximation 
for the five (formerly) hydrodynamic modes, including a chiral magnetic wave.}

\begin{document}
\maketitle
\flushbottom

\section{Introduction} \label{sec:introduction}

Recently, effects arising due to quantum anomalies have garnered much attention. The discoveries of the chiral magnetic effect (CME)~\cite{Kharzeev:2004ey,Fukushima:2008xe,Son:2009tf} and chiral vortical effect (CVE)~\cite{Erdmenger:2008rm,Banerjee:2008th,Son:2009tf}, both within quantum field theory and also from gauge/gravity duality or ``holography'', have raised a lot of interest in their experimental confirmation.\footnote{Note also the early work by Vilenkin~\cite{Vilenkin:1978is,PhysRevD.22.3080}.} 
Various other chiral transport effects, e.g.~the chiral separation effect (CSE), are conceptually analogous to the two prototypes CME and CVE, to which we restrict our attention henceforth (for an overview see for example~\cite{Jensen:2013vta,Kharzeev:2015znc}).
So far, the two most promising systems where the CME and CVE could be realized physically include the quark gluon plasma (QGP) generated in heavy ion collisions (HICs), and Weyl/Dirac semimetals. 
Experiments with Weyl semimetals report the observation of chiral transport effects in presence of magnetic fields~\cite{2014arXiv1412.6543L,2015arXiv150606577S,2015arXiv150407698Z,2015arXiv150600924W,2015arXiv150603190Y}. 
For HICs experimental signatures were proposed under some assumptions~\cite{Kharzeev:2010gr,Kharzeev:2015znc}. One such signature is that within a thermal plasma an interplay between the QCD vector current and the QCD axial current leads to a charge separation along the magnetic field created between the two colliding nuclei. 
However, a recent analysis of charge-dependent azimuthal correlations in a proton--lead collision from the CMS Collaboration~\cite{Khachatryan:2016got} challenged either this particular prediction or the assumptions leading to it. In proton--lead collisions the magnetic field is believed to be smaller than in nucleus-nucleus collisions, hence the charge separation effect should be suppressed, but this was not observed. Previous results from the STAR Collaboration~\cite{Abelev:2009ac,Abelev:2009ad,Adamczyk:2013kcb,Adamczyk:2014mzf,Adamczyk:2013hsi} at the Relativistic Heavy Ion Collider (RHIC) and the ALICE Collaboration~\cite{Abelev:2012pa} at the Large Hadron Collider remained inconclusive due to various possible background sources for the observed correlations~\cite{Wang:2009kd,Bzdak:2010fd,Schlichting:2010qia}. A better theoretical understanding of the physics governing the QGP in these HICs is likely to lead to prediction of more distinct and identifiable experimental signatures. This motivates our study 
of strongly coupled charged thermal states in presence of magnetic fields and anomalies in this paper.

On the field theory side, some results were previously obtained for plasmas which are anisotropic due to a magnetic field. Anomaly-driven effects such as CME and CVE reveal themselves in the transport properties of a system outside of equilibrium. 
The hydrodynamic description of chiral transport in weak magnetic fields\footnote{That is, magnetic fields which are of first order in derivatives. See Appendix~\ref{sec:hydrodynamics} for details on the definition of strong and weak magnetic fields and about the derivative counting.} in the case of a single axial current has been considered recently in~\cite{Abbasi:2015saa,Kalaydzhyan:2016dyr}. In these works, part of the hydrodynamic modes were given explicitely. In this paper, we perform an independent systematic and complete calculation of the relevant hydrodynamic modes at nonzero chemical potential, charge density, and magnetic field. Whenever the relevant quantities are provided, we find complete agreement with~\cite{Kalaydzhyan:2016dyr}, and we agree at vanishing chemical potential and charge density with the results shown in~\cite{Abbasi:2015saa} for $\omega_1,\, \omega_2,\, \omega_3$ (see also Appendix A for a discussion of the different hydrodynamic frames). A hydrodynamic analysis similar to~\cite{Abbasi:2015saa,Kalaydzhyan:2016dyr} for a system containing both the axial and a conserved vector current has also appeared very recently~\cite{Abbasi:2016rds}. 
There (as well as in~\cite{Abbasi:2015saa}), excitations propagating perpendicular to the magnetic field were also considered, while in~\cite{Kalaydzhyan:2016dyr} and the present work, the focus is on propagation parallel to the magnetic field. In the weak field case the chiral magnetic wave~\cite{Huang:2011dc} appears in a certain form, as is evident from~\cite{Kalaydzhyan:2016dyr,Abbasi:2016rds} and from our result in Appendix~\ref{sec:hydrodynamics}. 
Fewer reliable hydrodynamic results are available for plasmas in strong external magnetic fields\footnote{Strong magnetic fields are of zeroth order in a derivative expansion, i.e. they are $\mathcal{O}(1)$. See Appendix~\ref{sec:hydrodynamics}.}. One exception is a work considering polarization effects in equilibria with strong external (electro)magnetic fields~\cite{Kovtun:2016lfw}, generalizing~\cite{Israel:1978up}. An attempt at hydrodynamic constitutive relations in presence of a strong external magnetic field has been performed in~\cite{Huang:2011dc}. The striking finding of that latter paper are the five ``shear viscosities'' and the two ``bulk viscosities'' which appear because of the broken rotational symmetry due to the magnetic field. A strong claim was conjectured in~\cite{Kharzeev:2010gd}, where the authors argue for the chiral magnetic wave velocity to be of a particular form for any strength of the magnetic field. The authors also find that this velocity approaches the speed of light in the limit of large magnetic field at weak (and at strong coupling in the probe limit), and relate this limit behavior to Landau level physics using a weak coupling argument. In our work, we find evidence for the relation between Landau level physics and the chiral magnetic wave velocity at strong coupling, not assuming any probe limit. 

Despite previous field theory achievements, the study of non-equilibrium dynamics, especially for strongly interacting systems, has been difficult using traditional methods. However, the problem becomes particularly amenable in gauge/gravity duality~\cite{Maldacena:1997re,Ammon:2015wua,Nastase,Schalm}, because not only does it provide new tools to study strong interactions, but also the complicated quantum evolution problem can be  turned into a conceptually simple one of solving differential equations with given initial conditions. 
By now, it is evident that QGP is a strongly interacting system from the many experimental and theoretical studies supporting this~\cite{Heinz:2008tv}, and is thus naturally a candidate for the application of holography. For Weyl/Dirac semimetals, this is not yet clear. But then an exploration from the strongly-interacting perspective is even more interesting. 
In that vein, a host of papers explored the interplay between external magnetic fields and a chiral anomaly within holographic models of Weyl semimetals~\cite{Landsteiner:2015lsa,Landsteiner:2015pdh,Copetti:2016ewq,Ammon:2016mwa,Grignani:2016wyz,Kharzeev:2016mvi}.

Within holography transport is encoded in quasinormal modes (QNMs), 
see for example the reviews~\cite{Berti:2009kk,Konoplya:2011qq}.\footnote{Anomalous transport from holography was also recently discussed in a different approach~\cite{Bu:2016vum,Bu:2016oba}.} Numerous studies have shown that QNMs play an important role in the evolution dynamics of strongly interacting non-equilibrium systems, particularly in the relaxation process~\cite{Heller:2013oxa,Buchel:2015saa,Fuini:2015hba,Janiszewski:2015ura,Janik:2016btb,Attems:2016ugt}. On the gravity side, QNM frequencies are (quasi)eigenvalues of the linearized Einstein equations describing fluctuations around a black hole or black brane gravity background. 
Since the operator corresponding to the eigenvalue problem is not self-adjoint, the QNM frequencies are complex in general. 
These QNM frequencies correspond to the locations of poles of the retarded correlators in the dual field theory~\cite{Son:2002sd,Kovtun:2005ev}. They characterize the near-equilibrium behavior, encoding information about transport and dissipation. 
For correlators of conserved quantities such as the energy-momentum tensor, the QNM spectrum typically contains an infinite tower of gapped modes that are strongly damped, as well as a set of hydrodynamic modes $\omega = \omega(k)$ such that $\omega(k) \rightarrow 0$ as $k \rightarrow 0$~\cite{Kovtun:2005ev,Nunez:2003eq}. 
At vanishing magnetic field, quasinormal modes of charged black branes (also called Reissner-Nordstr\"om black branes) have been calculated, see~\cite{Janiszewski:2015ura}, references and the discussion therein. Those references worked either at vanishing momentum for the fluctuations or did not consider a Chern-Simons term, i.e. in those cases effects of the anomaly remained unknown. Fluctuations about a charged black brane in presence of a Chern-Simons term but at vanishing magnetic field were considered holographically in~\cite{Sahoo:2009yq,Matsuo:2009xn} in the hydrodynamic limit. Previously, that system was studied holographically in the hydrodynamic limit without the Chern-Simons term~\cite{Hur:2008tq,Matsuo:2009yu}. 
In the present work, we are closing the gaps in the gravitational calculations and present a systematic study of all metric and gauge field fluctuations with nonzero momentum along the external magnetic field in a background of charged magnetic black branes in presence of a Chern-Simons term within a particular theory which includes the aformentioned charged black brane setups as special cases.

In this paper, we consider (3+1)-dimensional systems at nonzero temperature, with a nonzero charge density, and in the background of a constant magnetic field. Most importantly, we consider such systems in the presence of an anomaly, and analyze perturbations with nonzero momentum and frequency in order to derive transport properties. A holographic model describing such a class of systems are charged magnetic black brane solutions with Einstein-Maxwell-Chern-Simons~(EMCS) theory~\cite{DHoker:2009ixq}. This model is ``top-down'' at least for a particular value of the Chern-Simons coupling, as it can then be derived from a consistent truncation of supergravity~\cite{Buchel:2006gb,Gauntlett:2006ai,Gauntlett:2007ma,Colgain:2014pha}, and it includes full backreaction from all the matter fields in its gravitational dynamics.
Previously, the QNM spectrum of this model with only the magnetic field turned on has been studied~\cite{Janiszewski:2015ura}, while~\cite{Finazzo:2016mhm,Critelli:2014kra} computed two shear viscosities in this anisotropic system (see also~\cite{Huang:2011dc}). The thermodynamics and the phase structure of this system at nonzero charge density and magnetic field has also been studied~\cite{Ammon:2016szz}. Here, we shall analyze the QNM spectrum of the latter system, and how it is influenced by the strength of electromagnetic background, as well as the anomaly. We solve for the background numerically using spectral methods, which we then use as input to compute the QNMs. We investigate the effects of changing external parameters: 
the temperature, $T$, chemical potential, $\mu$, and the strength of the magnetic field, $B$. 
We find interesting phenomena such as the appearance of Landau levels in some cases and various other gapped modes. 
In the hydrodynamic regime, we compare to our field theoretic calculation provided in Appendix~\ref{sec:hydrodynamics}. In particular, we provide dispersion relations for five modes in~\eqref{eq:helicity0poles} and~\eqref{eq:helicity1poles}. 
Following the modes (also beyond the hydrodynamic regime), we describe the evolution of the shear modes, see Fig.~\ref{fig:QNM2highB}, of two formerly hydrodynamic modes developing a (complex) gap, see Fig.~\ref{fig:overviewspin1}, and of three hydrodynamic modes, see Fig.~\ref{spinkb}, all for increasing $B$. 
Our results are summarized in Sec.~\ref{sec:discussion}.

The paper is organized as follows. In Sec.~\ref{sec:setup}, we discuss the setup of the holographic model dual to the (3+1)-dimensional system including a chiral $U(1)$-anomaly. The procedure of setting up the numerics by linearizing the field equation is presented in Sec.~\ref{sec:fluctuations}. In Sec.~\ref{sec:qnms} we compute the QNMs. We present how they behave as momentum and the strength of the magnetic field as well as the anomaly is varied, and we discuss their implications for transport and hydrodynamics with anomalies, and report on the Landau level behavior found in some cases. We conclude in Sec.~\ref{sec:discussion}. The construction of the hydrodynamic description is to be found in Appendix~\ref{sec:hydrodynamics}. Details of the numerical method employed are recorded in Appendix~\ref{sec:appendixConv}. 


\section{Holographic setup}\label{sec:setup}
Our goal is to study charged fluids in a particular class of quantum field theories with a chiral anomaly at strong coupling, and in the presence of a (strong) magnetic background field. Our treatment below is generally applicable, but let us first illustrate it using the $\mathcal{N}=4$ Super-Yang-Mills~(SYM) theory as a concrete example. The $\mathcal{N}=4$ SYM is coupled to a $U(1)$ symmetry, which is a subgroup of the global $SU(4)_R$ R-charge symmetry. Its matter content consists of one vector, four left-handed Weyl fermions, and six real scalars, all of which transform in the adjoint representation of the $SU(N_c)$ color group. The fermions and scalars are charged under the $U(1)$, while the vector is uncharged.

Consider the case in which an axial current, $J^\alpha$, of the $U(1)$ is coupled to a constant external magnetic field $B$, see e.g.~\cite{Fuini:2015hba}.
The action is then
\begin{equation}
S = S_{\text{SYM}} +\int d^4 x J^\alpha A_\alpha ^{\text{ext}} \,, \qquad F = dA^{ext} = B\, dx_1\wedge dx_2
\end{equation}
In this theory, the energy-momentum tensor, $T^{\mu\nu}$, and the axial current satisfy the following conservation equations:
\begin{eqnarray} \label{eq:hydroConservation}
\nabla_\mu T^{\mu\nu} &=& F^{\nu\mu} J_\mu\,,\\
\nabla_\mu J^{\mu} &=& \frac{3}{4} C \, \epsilon^{\alpha\beta\gamma\delta}F_{\alpha\beta} F_{\gamma\delta}\,,
\end{eqnarray}
where $\epsilon^{\alpha\beta\gamma\delta}$ is the totally antisymmetric Levi-Civita tensor in four-dimensional (flat) spacetime with $\epsilon^{0123} = 1$, and $C$ is the chiral anomaly coefficient~\cite{Adler:1969gk,Bell:1969ts,Son:2009tf}. The Greek indices here run from 0 to 3, and they indicate field theory coordinates $x^\mu = \{t,\, x_1,\, x_2,\, x_3\}$. For $\mathcal{N}=4$ SYM, $C = 1/(3\sqrt{3})$. However, for the more general theories which we consider below, $C$ is arbitrary. 

In order to obtain results for a charged fluid within this strongly coupled theory, we use the gauge/gravity correspondence, and we perform the relevant calculations in the dual gravitational description, which is given by an Einstein-Maxwell-Chern-Simons (EMCS) theory with an external magnetic field. The relevant fully backreacted gravitational solutions dual to the desired plasma states are the charged magnetic black branes~\cite{D'Hoker:2009mm,Ammon:2016szz}. Below, we first review the EMCS theory and its charged magnetic black brane solutions. We then discuss the thermodynamics in the gravity description, and match it to the dual field theory. 

\subsection{Einstein-Maxwell-Chern-Simons theory}\label{sec:EMCS}
The EMCS theory in five dimensions is defined by the action 
\begin{equation}\label{eq:actionS}
S_{grav}=\frac{1}{2\kappa^{2}}
\left[\int_{\mathcal{M}}\!\dd^{5}x\, 
\sqrt{-g}\left(R + \frac{12}{L^2} - \frac{1}{4}F_{mn}F^{mn}\right)
-\frac{\gamma}{6}\int_{\mathcal{M}} A\wedge F\wedge F\right]\,,
\end{equation}
where $L$ is the $AdS_5$ radius, $2\kappa^2 \equiv 16\pi G_5$, and $g \equiv \det g_{mn}$, with $G_5$ and $g_{mn}$ the five-dimensional Newton's constant and metric respectively. In the gauge sector, $F = dA$ is the five-dimensional Maxwell field strength, and $\gamma$ is the Chern-Simons coupling. We denote the five-dimensional gravitational bulk by $\mathcal{M}$, and 
$\partial \mathcal{M}$ its boundary. The coordinates of $\mathcal{M}$ are indicated by lower case Roman indices, which run from 0 to 4. The action given in~\eqref{eq:actionS} has to be amended by boundary terms~\cite{Henningson:1998gx,Balasubramanian:1999re,Taylor:2000xw} of the form
\begin{equation}\label{eq:actionSbdy}
S_{bdy}= \frac{1}{\kappa^2} \int_{\partial\mathcal{M}}\! \dd^4x \, \sqrt{-\hat{g}} \left( K - \frac{3}{L} + \frac{L}{4} R(\hat{g}) + \frac{L}{8} \ln\left( \frac{z}{L} \right) F_{\mu\nu} F^{\mu\nu} \right) \,.
\end{equation}
Here, $\hat{g}_{\mu\nu}$ is the metric induced by $g_{mn}$ on the conformal boundary of $AdS_5$, and $K$ is the trace of the extrinsic curvature with respect to $\hat{g}_{\mu\nu}$. The extrinsic curvature $K_{mn}$ is given by 
\begin{equation}
K_{mn} = \mathcal{P}_m^{\ \, o} \,  \mathcal{P}_n^{\ \, p} \, \nabla_o n_p \, , \qquad 
\mathcal{P}_m^{\ \, o} = \delta_m^{\ o} - n_m n^o \,,
\end{equation}
where $\nabla$ is the covariant derivative, and $n_m$ is the outward pointing normal vector of $\partial\mathcal{M}$. 

The equations of motion following from the action~\eqref{eq:actionS} read
\begin{align}
\label{eq:EOM1}
R_{mn} + 4g_{mn} - \frac{1}{2}\left(F_{mo}F_{n}{}^{o}-\frac{1}{6}g_{mn}F_{op}F^{op}\right) &= 0 \,,\\
\label{eq:EOM2}
\nabla_m F^{m n} + \frac{\gamma}{8 \sqrt{-g}} \, \tilde{\epsilon}^{nmopq} F_{mo} F_{pq} &= 0 \, ,
\end{align}
where $\tilde{\epsilon}^{mnopq}$ is the totally antisymmetric Levi-Civita symbol in five spacetime dimensions with $\tilde{\epsilon}^{01234}=1$. The Chern-Simons coupling $\gamma$ on the gravity side is related to the chiral anomaly coefficient $C$ on the field theory side via~\cite{Witten:1998qj,Bilal:1999ph}
\begin{equation}
C = \frac{\gamma}{6} \, .
\end{equation}
This identification will be crucial for the analysis of the thermodynamics below, and for the subsequent analysis of the chiral transport effects.

For simplicity, we will work in units where $L=1$ and $2\, \kappa^2 =1$ from now on.

\subsection{Charged magnetic black brane solutions} \label{sec:blackbranesolutions}
The charged magnetic black brane solutions of the EMCS equations, \eqref{eq:EOM1} and~\eqref{eq:EOM2}, are dual to the charged plasma of interest in a particular class of quantum field theories with a chiral anomaly at strong coupling, and in the presence of a magnetic background field of any strength. Due to the presence of various scales including the temperature, the chemical potential, and the magnetic field, conformal symmetry is broken by the state of such field theories.

As we shall see below when studying the fluctuations, it is convenient to use the Eddington-Finkelstein coordinates. Working in units where the horizon of the black brane is located at $z = 1$ and the conformal boundary at $z = 0$, the metric for the charged magnetic black brane can be expressed as
\begin{eqnarray}\label{eq:ansatzmetric_ef}
\dd s^{2}  &=&  \frac{1}{z^{2}}\left[\left(-u(z)+c(z)^2\,w(z)^2\right)\, \dd v^{2}-2\,\dd z\, \dd v + 2\, c(z)\,w(z)^2\dd x_3\, \dd v \right. \nonumber \\
 &&\left. + v(z)^{2}\, \left( \dd x_{1}^{2} +\dd x_{2}^{2}\right)  + w(z)^{2}\,\dd x_{3}^{2}\right] \, , \\
\label{eq:ansatzF_ef} 
F  &=&A_v'(z)\,\dd z \wedge \dd v +  B\,\dd x_{1}\wedge\dd x_{2}+ P'(z)\,\dd z \wedge \dd x_3 \, .
\end{eqnarray}
The field strength tensor~\eqref{eq:ansatzF_ef} may be obtained from a gauge field $A$ of the form
\begin{equation}\label{eq:ansatzA_ef}
A= A_v(z)\, \dd v + \frac{B}{2} \left(- x_2 \, \dd x_1 + x_1 \, \dd x_2 \right) +P(z)\, \dd x_3\, ,
\end{equation}
which is symmetric in $x_1$ and $x_2$, and points to the $SO(2)$ rotational symmetry in the $(x_1, x_2)$-plane.

Near the conformal boundary at $z=0$, the functions appearing in~\eqref{eq:ansatzmetric_ef} and~\eqref{eq:ansatzF_ef} may be expanded as
\begin{eqnarray}\label{eq:boundaryExpansionBackground}
&& u(z) = 1 + z^4\left[  u_4 +{\cal O}(z^2) \right]+ z^4\ln(z)\left[ \frac{B^2}{6}+ {\cal O}(z^2) \right]  \nn \, ,\\
&& v(z) = 1 + z^4\left[- \frac{w_4}{2} +{\cal O}(z^2)\right] + z^4\ln(z)\left[- \frac{B^2}{24}+ {\cal O}(z^2) \right] \, , \nn\\
&& w(z) = 1 + z^4\left[  w_4 +{\cal O}(z^2)\right] + z^4\ln(z)\left[ \frac{B^2}{12}+ {\cal O}(z^2) \right] \, , \nn\\
&& c(z) =z^4\left(  c_4 + {\cal O}(z^2) + z^4\ln(z) \left[ - \frac{B^2}{12}  {c_4} + {\cal O}(z^2)\right] \right) \, ,\nn  \\	
&&A_v(z)  =  \mu -  \frac{\rho}{2} z^2  -  \frac{\gamma B  p_1}{8} z^4  +  {\cal O}(z^6)\, ,\nn\\
&&P(z) = z^2\left( \frac{p_1}{2}  + \frac{\gamma B  \rho}{8}  z^2 + {\cal O}(z^4 ) \right) \,,
\end{eqnarray}
where $u_4,\, w_4,\, c_4, \, \rho, \, p_1$ are undetermined coefficients, while $B$ and $\mu$ are parameters (corresponding to sources in the field theory) determining the background solution. We see from this that the charged magnetic black brane is 
asymptotically $AdS$. Note that the leading term in $P(z)$ has been fixed such that no explicit source for a persistent current in the $x_3$-direction appears.

Near the horizon at $z=1$, expansion of the same functions yields
\begin{eqnarray}\label{eq:horizonExpansionBackground}
&u(z) = (1-z)\left[  \bar{u}_1 + {\cal O}(1-z) \right], \quad  &c(z) = (1-z)\left[  \bar{c}_1  + {\cal O}(1-z)\right], \nn \\
&v(z) = \bar{v}_0 + {\cal O}(1-z),  \nn  &A_v(z) = (1-z)\left[  \bar{A_v}_{\,0} + {\cal O}(1-z) \right] ,\nn\\
&w(z) =  \bar{w}_0 + {\cal O}(1-z),  &P(z) =  \bar{P}_0 + {\cal O}(1-z) \,,
\end{eqnarray}
where $\bar{u}_1,\, \bar{c}_1,\, \bar{w}_0,\, \bar{v}_0,\, \bar{A_v}_0$ and $\bar{P}_0$ are undetermined coefficients analogous to those in the near boundary expansion~\eqref{eq:boundaryExpansionBackground}. Note that regularity at the horizon fixes the leading coefficients in $u(z)$, $c(z)$, and $A_v(z)$.

The full solution of the functions $u,\, c,\, w,\, v,\, A_v$, and $P$ has to be obtained numerically. The main method used in this paper is the spectral method employed in~\cite{Ammon:2016szz}. For a detailed discussion of the numerical method, see Appendix~\ref{sec:appendixConv}. We stress that these magnetic black brane solutions are valid solutions within EMCS theory for any strength of the external magnetic field $B$.

\subsection{Thermodynamics from gravity}\label{sec:thermodynamicsGravity}
After a Wick-rotation of the time direction in the metric~\eqref{eq:ansatzmetric_ef}, requiring regularity at the horizon leads to the field theory temperature, $T$, being given by
\begin{equation}\label{eq:T}
T = \frac{|\bar{u}_1|}{4\pi} \, .
\end{equation}
The field theory entropy density, $s$, can be found from the Bekenstein-Hawking entropy formula, which gives (in units where $2\kappa^2 = 1$ which we work in)
\begin{equation}\label{eq:s}
s = 4\pi {\bar{v}_0}^2 \bar{w}_0 \, .
\end{equation}
Making use of standard recipes of gauge/gavity correspondence, in particular the relation~\cite{Balasubramanian:1999re}
\begin{equation}
\left\langle T_{\mu\nu} \right\rangle=\lim\limits_{z\rightarrow 0}\frac{1}{z^{2}}\left(-2K_{\mu\nu}+2(K-3) \, h_{\mu\nu}+\ln(z)\left(F_{\mu}^{\ \alpha}F_{\nu\alpha}
-\frac{1}{4} \, h_{\mu\nu}F^{\alpha\beta}F_{\alpha\beta}\right)\right) \, ,
\end{equation}
we can extract the energy-momentum tensor of the dual conformal field theory. Similarly, from the relation~\cite{D'Hoker:2009bc}
\begin{equation}
\left\langle J^{\mu} \right\rangle= \lim\limits_{z\rightarrow 0}\frac{1}{z^{3}}\left(h^{\mu\alpha}\partial_{z}A_{\alpha}+\frac{\gamma}{6}\epsilon^{\alpha\beta\gamma\mu}A_{\alpha}F_{\beta\gamma}\right) \,,
\end{equation}
we can obtain the expectation value of the current in the dual field theory. Given our ansatz, the energy momentum tensor and the current are given by
\begin{equation}
\label{eq:EnergyMomTensor}
\langle T^{\mu\nu}\rangle =\left( \begin{array}{cccc}
-3 \, u_4 & 0 & 0 &  
-4 \, c_4 \\
0 & -\frac{B^{2}}{4}-u_4-4\,w_4  & 0 & 0\\
0 & 0 & -\frac{B^{2}}{4}-u_4-4\,w_4  & 0\\
-4 \, c_4  & 0 & 0 & 8 \, w_4 - u_4 \end{array} \right) \, ,
\end{equation} 
\begin{equation}\label{eq:CFTcurrents}
\langle J^\mu \rangle = \left(\rho,\,0,\,0,\,p_1\right) \,.
\end{equation} 
Note that the trace of the energy momentum tensor is $\langle T_\mu{}^\mu \rangle = -B^2/2$. 

\subsection{Thermodynamics from field theory and matching to gravity data}\label{sec:thermodynamicsFieldTheory}
Eqs.~\eqref{eq:EnergyMomTensor} and~\eqref{eq:CFTcurrents} relate expectation values (or one-point functions) of the dual field theory energy momentum tensor and axial current on the left-hand side (LHS), to combinations of coefficients 
($u_4,\, w_4,\, c_4,\, \rho,\, p_1$) in the boundary expansion~\eqref{eq:boundaryExpansionBackground} of the metric and gauge field in the gravitational EMCS theory on the righ-hand side (RHS). However, each of these coefficients can also be related via gauge/gravity correspondence to physical quantities in the dual field theory by matching to the purely field theory expression of the RHS.\footnote{We thank P.~Kovtun for discussions about contents of this subsection.}

From a completely field-theoretic analysis in thermodynamic equilibrium, taking into account a strong background magnetic field of $\mathcal{O}(\partial^0)$ in the derivative expansion~\cite{Kovtun:2016lfw}, and a chiral anomaly~\cite{Jensen:2013kka} (see Appendix~\ref{sec:hydrodynamics} for more details), the energy-momentum tensor and the axial current are given by
\begin{equation}\label{eq:Tmn0Hydro}
\langle T^{\mu\nu}_\text{EFT}\rangle=\left( \begin{array}{cccc}
\epsilon_0 & 0 & 0 &  
\xi_V^{(0)} B\\
0 & P_0 - \chi_{BB} B^2 & 0 & 0\\
0 & 0 & P_0 - \chi_{BB} B^2 & 0\\
\xi_V^{(0)} B & 0 & 0 & P_0 \end{array} \right) +\mathcal{O}(\partial) \, ,
\end{equation} 
\begin{equation}\label{eq:Jm0Hydro}
\langle J^\mu_\text{EFT}\rangle = \left(n_0,\,0,\,0,\,\xi_B^{(0)} B\right) + \mathcal{O}(\partial) \,,
\end{equation} 
where a subscript ``EFT'' (for effective field theory) indicates the quantity is obtained from the field theory directly and not from the dual gravitational EMCS theory. A subscript ``$0$'' on the energy density, $\epsilon$, the pressure, $P$, and the charge density, $n$, indicates the quantity is evaluated in the thermodynamic equilibrium state. Similarly, the superscript ``$(0)$'' on the chiral transport coefficients, $\xi_{V,B} = \xi_{V,B}(\mu,T)$, indicates all thermodynamic quantities involved are evaluated in the thermodynamic equilibrium. Note that, at strong magnetic fields, the equilibrium partition function can depend on $T$, $\mu$ and on $B^2$~\cite{Kovtun:2016lfw}. Therefore, also the equilibrium quantities $\epsilon_0$, $P_0$, and $n_0$ are in general functions of $T$, $\mu$, and $B^2$.

Note that in a particular hydrodynamic frame (the thermodynamic frame~\cite{Jensen:2011xb,Jensen:2012jh}), $\xi_{V,B}$ are analytically known to be related to the anomaly coefficients: 
\begin{equation}
\xi_V =  -3 C \mu^2 + \tilde{C} T^2 \,, \qquad \xi_B = -6 C \mu \,.
\end{equation}
Here, $\tilde{C}$ is an undetermined coefficient related to the mixed gauge-gravitational anomaly~\cite{Landsteiner:2011cp,Gynther:2010ed,Landsteiner:2011iq,Amado:2011zx}, which is suppressed in the large-$N$ limit when derived from string theory setups. 
Note also that the trace of the energy momentum tensor is in general given by $\langle {T_\text{EFT}}_\mu{}^\mu \rangle = \alpha_0 B^2$ for any conformal field theory (such as ours) in an external $B$ field in flat spacetime, and the contribution on the RHS is the standard trace anomaly arising from the external field, with a coefficient $\alpha_0$, see e.g.~\cite{Fuini:2015hba}. In principle, our expression for the energy momentum tensor~\eqref{eq:Tmn0Hydro} receives derivative corrections. However, in our case $T$, $\mu$, and $B$, as well as the field theory metric are all constant in space and also time-independent. Hence, derivative corrections vanish~\cite{Kovtun:2016lfw} in this particular equilibrium state.
The trace of the energy momentum tensor is given by $\langle {T_\text{EFT}}_\mu{}^\mu \rangle = -\epsilon_0 + 3 P_0 -2 \chi_{BB}B^2$, where $\epsilon_0$ and $P_0$ again depend on $T$, $\mu$, and $B^2$.

Comparing the holographic result~\eqref{eq:EnergyMomTensor} and \eqref{eq:CFTcurrents} with the field theoretic expectation~\eqref{eq:Tmn0Hydro} and~\eqref{eq:Jm0Hydro}, 
we conclude that 
\begin{equation}
\chi_{BB} B^2= \langle T^{33}\rangle - \langle T^{11}\rangle = \frac{1}{4} B^2 + 12 w_4 (T,\mu,B^2)\, ,
\end{equation}
which can be understood as a defining relation for the magnetic susceptibility coefficient $\chi_{BB}$. 
Further, we are lead to identify those components arising entirely from the anomaly:
\begin{eqnarray}\label{eq:exactThermoIdentifications}
-4 c_4 =  \xi_V^{(0)} B , \, \qquad p_1  = \xi_B^{(0)} B \, .
\end{eqnarray}
For a detailed study of the identifications~\eqref{eq:exactThermoIdentifications}, see Sec.~2 and Appendix~C, both in~\cite{Ammon:2016szz}. The remaining equilibrium quantities can also be related to the holographic near boundary data:
\begin{eqnarray}
&n_0   = \rho \, ,\qquad
&\epsilon_0  = - 3u_4 \, , \nonumber\\
&P_0 - \chi_{BB} B^2  = -u_4 - 4 w_4 - \frac{B^2}{4} \, ,\qquad
&P_0  = -u_4 + 8 w_4\, ,\qquad
\end{eqnarray} 
where the right hand side is exact, and the left hand side receives no derivative corrections because our equilibrium state has $T$, $\mu$, and $B$ which are constant in space and time, i.e. derivatives on these quantities vanish.

\section{Fluctuations, quasinormal modes, and numerics}\label{sec:fluctuations}
\subsection{Fluctuations}
We consider fluctuations of the metric, $g_{mn}(z, x^{\mu})$, and the gauge field, $A_m(z, x^{\mu})$, on a fixed background solution, which we denote by $\bar{g}_{mn}(z)$ and $ \bar{A}_{m}(z)$ respectively. To derive the fluctuation equations, we write the metric and the gauge field as sums of a background and a fluctuation part:
\begin{equation}\label{eq:ansatzqnm}
g_{mn} = \bar{g}_{mn} + \varepsilon \,  h_{mn} \,, \qquad A_{m} = \bar{A}_{m} + \varepsilon \, a_m \,,
\end{equation}
and we expand the equations of motion to first order in $\varepsilon$. 
It is more convenient to work in momentum space, and so we perform a Fourier transformation on the fluctuations, $h_{mn}$ and $a_m$, along the spacetime coordinates of the dual field theory: 
\begin{equation}\label{eq:fouriertrafo}
h_{mn}(z,x^{\mu}) = \int\!\dd^4k \, e^{i k_{\mu} x^{\mu}}\,\tilde{h}_{mn}(z,k^{\mu}) \,, \qquad
a_{m}(z,x^{\mu}) = \int\!\dd^4k \, e^{i k_{\mu} x^{\mu}}\,\tilde{a}_{m}(z,k^{\mu}) \,,
\end{equation}
where $k_{\mu} x^{\mu} = -\omega t + \vec{k} \cdot \vec{x}$. For notation simplicity, we will drop the tilde on the Fourier transformed fields from now on.

The presence of the magnetic field $B$ defines a preferred direction. For simplicity, we shall consider only the case of the momentum $k$ being aligned with $B$. We can then choose a coordinate system in which $k$ is orthogonal to the 
$(x_1,x_2)$-plane, and the $SO(2)$ symmetry of the background is preserved. If $k$ were not aligned with $B$, the $SO(2)$ would be broken by the fluctuations, and they would all be coupled together in the fluctuation equations. 

Given that the $SO(2)$ stays unbroken, fluctuations may be classfied in accordance to how they transform under it. The metric and gauge fluctuations transform as helicity-2, helicity-1, and helicity-0 modes under rotations about the 
$x_3$-axis:
\begin{center}
\begin{tabular}{c|c}
Helicity \hspace{1mm} & Fluctuation modes \\ \hline
2\, &$ \, h_{12},\,h_{11}-h_{22}  $\\
1\, &$ \, h_{t1},\,h_{13},\,a_{1},\,h_{z1}  $\\
    &$ \,h_{t2},\,h_{23},\,a_{2},\,h_{z2} $\\
0\, &$ \, h_{tt},\,h_{t3},\,h_{33},\,h_{11}+h_{22},\,h_{zt},\,h_{z3},\,h_{zz},\,a_t,\,a_3,\,a_z $
\end{tabular}
\end{center}
The equation of motions for modes of different helicities decouple, thus each helicity sector can be treated independently.

In order to consider only the physical modes of the system, we have to fix the gauge freedom. To do so, we choose a gauge  where $a_z = 0$ and $h_{mz} = 0$. The equations of motion for these fields then correspond to constraints, which are first order ordinary differential equations (ODEs). We then expect constraints to arise from the following modes:
\begin{center}
\begin{tabular}{c|c}
Helicity \hspace{1mm} & \hspace{1mm} Constraint modes\\
\hline
2\,& \,  none\\
1\,& \, $h_{z1},\,h_{z2}$ \\
0\,& \, $h_{zt},h_{z3},\,h_{zz},a_{z}$
\end{tabular}
\end{center}

Consider the helicity-2 sector. The two fluctuations $h_{12}$ and $h_{11} - h_{22}$ are decoupled from each other, and their equations of motion are identical. Next, the fluctuations within the helicity-1 sector can be further decoupled. In particular, the physical modes split into helicity-$1^\pm$ subsectors consisting of modes 
\begin{equation}
a_\pm =  a_1 \pm i a_2 \,, \quad h_{t\pm} = h_{t1} \pm i h_{t2} \,, \quad  h_{3\pm} = h_{31} \pm i h_{32} \,.
\end{equation}
The $\pm$ subsectors are decoupled from each other, and fluctuations of each subsector satisfy three second-order differential equations and one constraint equation. Lastly, there are six physical modes, 
$h_{tt},\,h_{t3},\,h_{33},\,h_{11}+h_{22},\,a_t,\,a_3$, in the helicity-0 sector, and they satisfy six second order differential equations and four constraint equations. 

In each of the three helicity sectors, the set of fluctuations is invariant under the following two transformations:
\begin{equation}\label{eq:Rtransfs}
\mathcal{R}_1:
\begin{cases}
B \mapsto -B \\
\,\,\gamma \mapsto -\gamma
\end{cases}
\,, \qquad\qquad
\mathcal{R}_2: 
\begin{cases}
B \mapsto -B \\
\,\,k \mapsto -k
\end{cases} 
\,.
\end{equation}
Under the transformation $\mathcal{R}_1$, all other parameters such as $\mu$, $T$, and $k$, as well as the background fields are remain unchanged. The equations of the helicity-2 and helicity-0 sectors are invariant under $\mathcal{R}_1$, while the equations of the helicity-$1^\pm$ subsectors are mapped into each other. Under $\mathcal{R}_2$, the sign flips in the background functions $c(z)$ and $P(z)$ as can be seen from their near boundary expansions in~\eqref{eq:boundaryExpansionBackground} given~\eqref{eq:exactThermoIdentifications}; all fluctuations with one leg in the $x_3$-direction also flip their signs. The equations of the helicity-2 and helicity-0 sectors then stay invariant under $\mathcal{R}_2$ (up to an overall sign flip), while the helicity-$1^\pm$ subsectors are again mapped into each other.

\subsection{Quasinormal modes and numerical details}\label{subsec:QNMandnumericaldetails}
QNMs are solutions to linearized fluctuation equations about the charged magnetic brane background here, subject to specific boundary conditions. Since QNMs correspond to poles of the retarded Greens functions in the dual field theory~\cite{Kovtun:2005ev}, incoming boundary conditions are imposed at the horizon of the brane. With the black brane solution written in Eddington-Finkelstein coordinates, incoming boundary conditions are imposed by requiring regularity at the horizon. Because QNMs do not source any dual operators, at the conformal boundary we have to set the non-normalizable modes to zero. 

In order to find the QNMs, we have to find the QNM frequency $\omega$ for which there is a non-trivial regular solution to the fluctuations equations, subject to the boundary conditions mentioned above. The problem can be recast into a generalized eigenvalue problem for $\omega$.{\footnote{For a method to compute the residues using spectral methods see~\cite{Ammon:2016fru}.}${}^,$\footnote{In the helicity-zero sector, there are terms quadratic in $\omega$ in the fluctuation equations. However, as explained in Appendix~\ref{sec:appendixConv}, by introducing auxiliary fields we can still put the problem into the linear form shown in~\eqref{eq:geneigenv}.}} In particular, in each helicity sector the resulting second order differential equations may be schematically written as
\begin{equation}\label{eq:geneigenv}
\left(A[\bar{g}_{mn}, \bar{A}_m, \partial_z, k] + \omega \,  B[\bar{g}_{mn}, \bar{A}_m, \partial_z, k] \right) \left( \begin{array}{cc} h_{mn}(z) \\ a_m(z) \end{array}\right) = 0 \,,
\end{equation}
where $A$ and $B$ are differential operators involving the background fields, $\bar{g}_{mn}$ and $\bar{a}_m$, derivatives with respect to $z$, and the momentum $k$. 

The generalized eigenvalue problem can be solved numerically by using spectral methods, in which the differential operators, $A$ and $B$, are represented by matrices, $\hat{A}$ and $\hat{B}$. The QNM frequencies are then the generalized eigenvalues associated with these differential matrices, and the QNM functions the corresponding eigenvectors. Below, we will 
use the QNM functions to check explicitly whether the constraint equations are satisfied in the helicity-1 and helicity-0 sectors. For more details, we refer to Appendix~\ref{sec:appendixConv}.

\section{Quasinormal mode results} 
\label{sec:qnms}

We determined the frequencies $\omega(k)$ for the QNMs in all three helicity sectors. Of particular interest are QNMs corresponding to hydrodynamic modes which satisfy $\omega(k)\rightarrow 0$ for $k \rightarrow 0$ and modes corresponding to long-lived quasi-particles with $\mbox{Im}(\omega)$ being small.

As discussed above, the sets of fluctuation equations are invariant under the transformations $\mathcal{R}_{1,2}$ given in Eq.~\eqref{eq:Rtransfs}. We can use these two transformations to restrict ourselves to only positive values of $\gamma$ and 
$B.$ The momentum $k$ of the QNM is taken to be real. 

Moreover, since we consider a conformal field theory, instead of considering $T$, $\mu$, $k$ and $B$ separately, we use dimensionless quantities by normalizing $\mu, k$ and $B$ to appropriate powers of temperature, i.e. $\tilde{\mu} = \mu / T$ and $\tilde{B} = B / T^2$ as well as $\tilde{k} = k/ T.$ Sometimes, it is also convenient to normalize the dimensionful quantities to the chemical potential $\mu$ by introducing $\bar{T} = T / \mu = 1 / \tilde{\mu}$, $\bar{B} = B / \mu^2$ and $\bar{k} = k / \mu$.

\subsection{Helicity-2 sector under $SO(2)$ rotations}

First, let us consider the helicity-2 sector. In order to find the QNM frequencies, we rewrite the ordinary differential equation for $h_{12}$ in terms of a generalized eigenvalue problem and represent the derivatives and background fields using spectral methods. It turned out that we had to use of order $N=100$ Chebyshev polynomials in order to get good convergence\footnote{The convergence is discussed in Appendix~\ref{sec:appendixConv}. In order to improve accuracy, we use a mapping $z \mapsto z^2$. This mapping improves the convergence of the QNM frequencies and avoids the observed oscillatory behavior of the QNM frequency as a function of $N$. See Appendix~\ref{sec:appendixConv} for more details.} of the QNM frequencies. How many of the QNMs we can trust depends highly on the values $\tilde{B}$ and $\tilde{\mu}$ characterizing the background as well as $\tilde{k}$ which specifies the momentum of the QNM.  

\begin{figure}[ht]
	\centering
  \includegraphics[width = 0.5\textwidth]{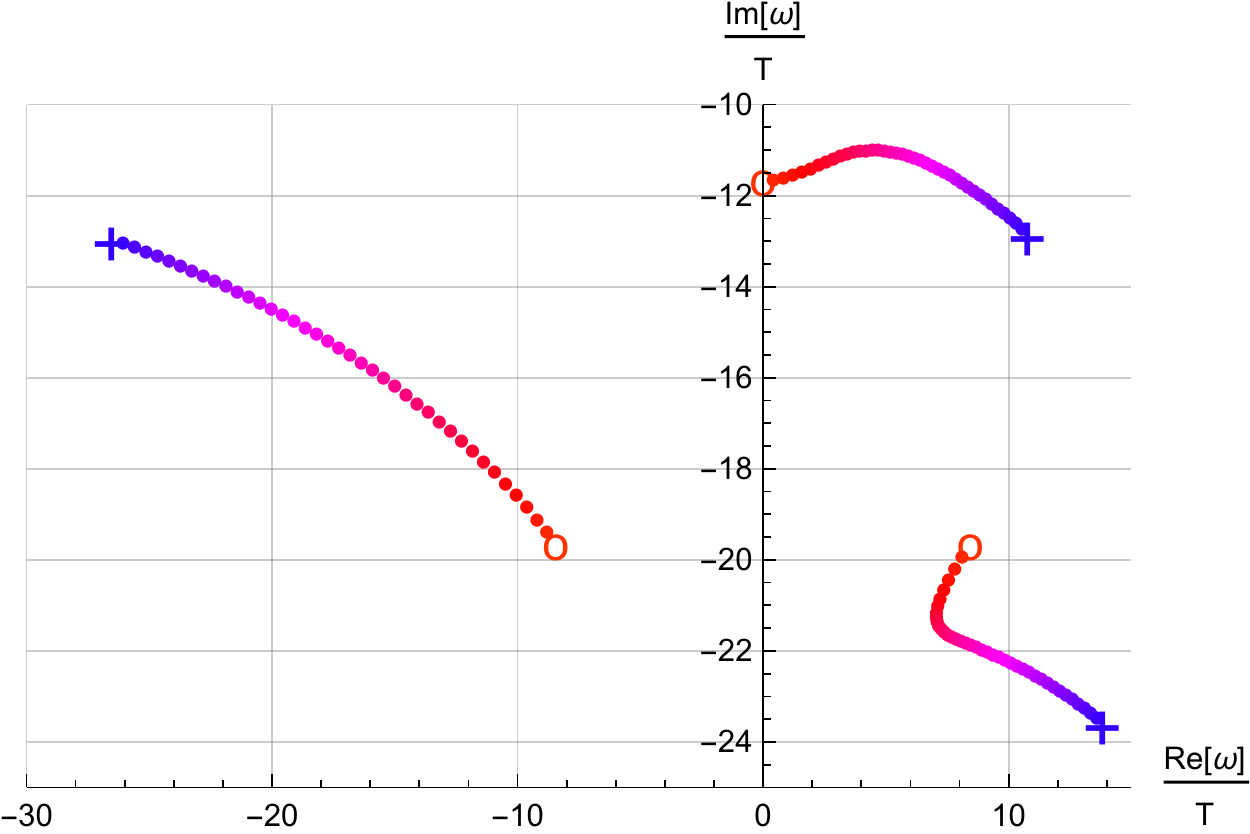}
	\caption{Three lowest helicity-2 QNMs for $\tilde{\mu}=10, \tilde{B}=65$ and $\gamma = 3/2.$ The momentum $\tilde{k}$ varies from $\tilde{k}=0$ (denoted by a circle) to $\tilde{k}=20$ (denoted by a cross).}
	\label{overviewspin2}
\end{figure}

For example, in Fig.~\ref{overviewspin2} we show the lowest lying QNMs for $\tilde{\mu} = 10$ and $\tilde{B}=65$ (corresponding to $\bar{T} = 0.1$ and $\bar{B}=0.65$). The momentum varies from $\tilde{k} \in [ 0, 20 ]$ and we have chosen equidistant values for $\tilde{k}$ in the figure. 
The convergence plots for these QNMs are shown in Appendix~\ref{sec:appendixConv}. In particular, we conclude that we can trust the values of the QNM frequencies up to $10^{-7}$. 

For large magnetic fields, the QNMs are more sensitive to $\tilde{k}$, and some of them approach the real frequency axis. Moreover, we see that in the extremal limit, $\tilde{\mu} \rightarrow \infty$ (i.e.~$T\rightarrow 0$), the QNMs coalesce along the imaginary axis, presumably forming a branch cut for the exact extremal case, as was discussed already in~\cite{Janiszewski:2015ura}. This would be analogous to the near-extremal $AdS_4$ case studied thoroughly in~\cite{Edalati:2010hk,Edalati:2010pn}.

\begin{figure}[ht]
	\centering
  \includegraphics[width = 0.45\textwidth]{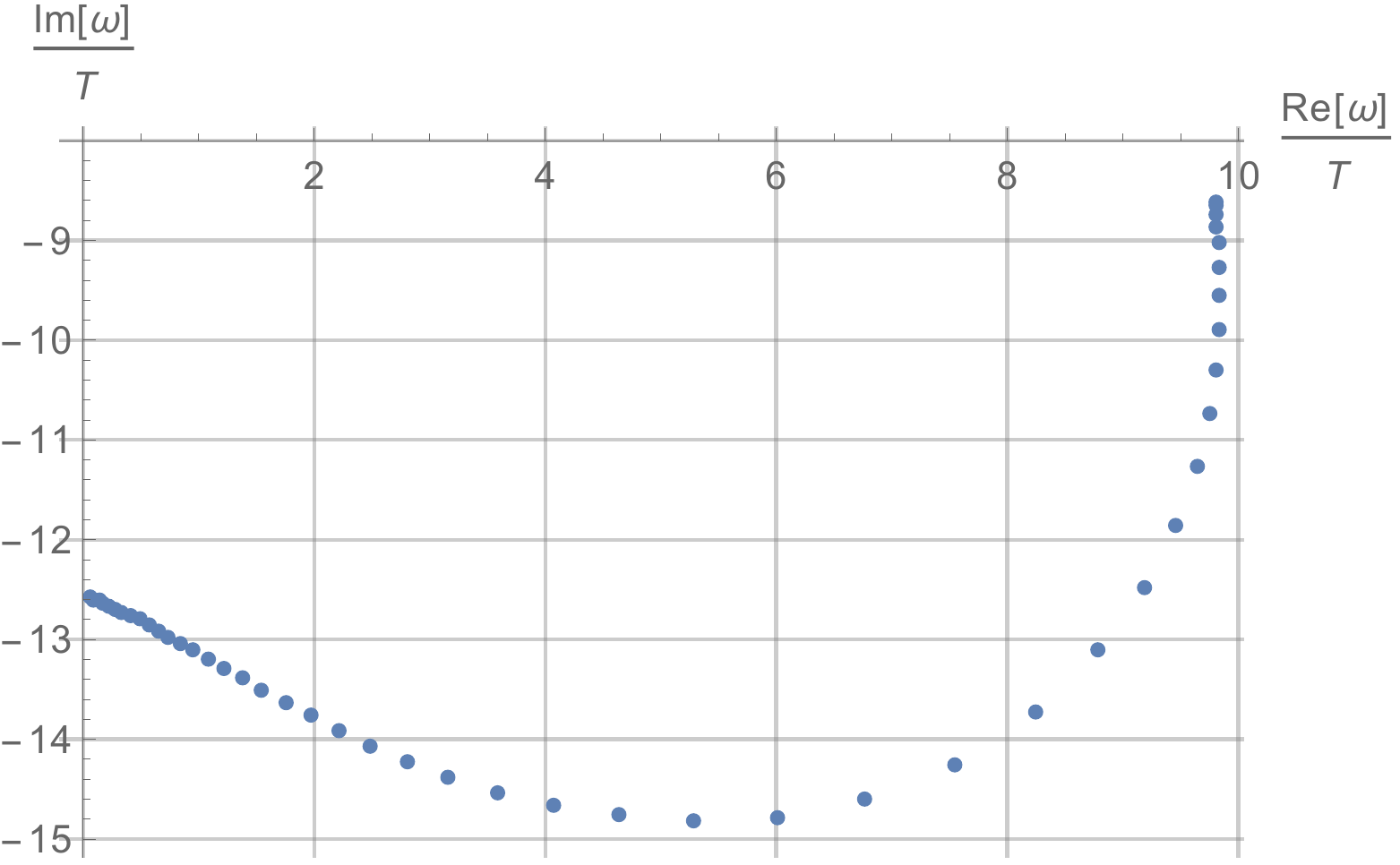}
  \includegraphics[width = 0.45\textwidth]{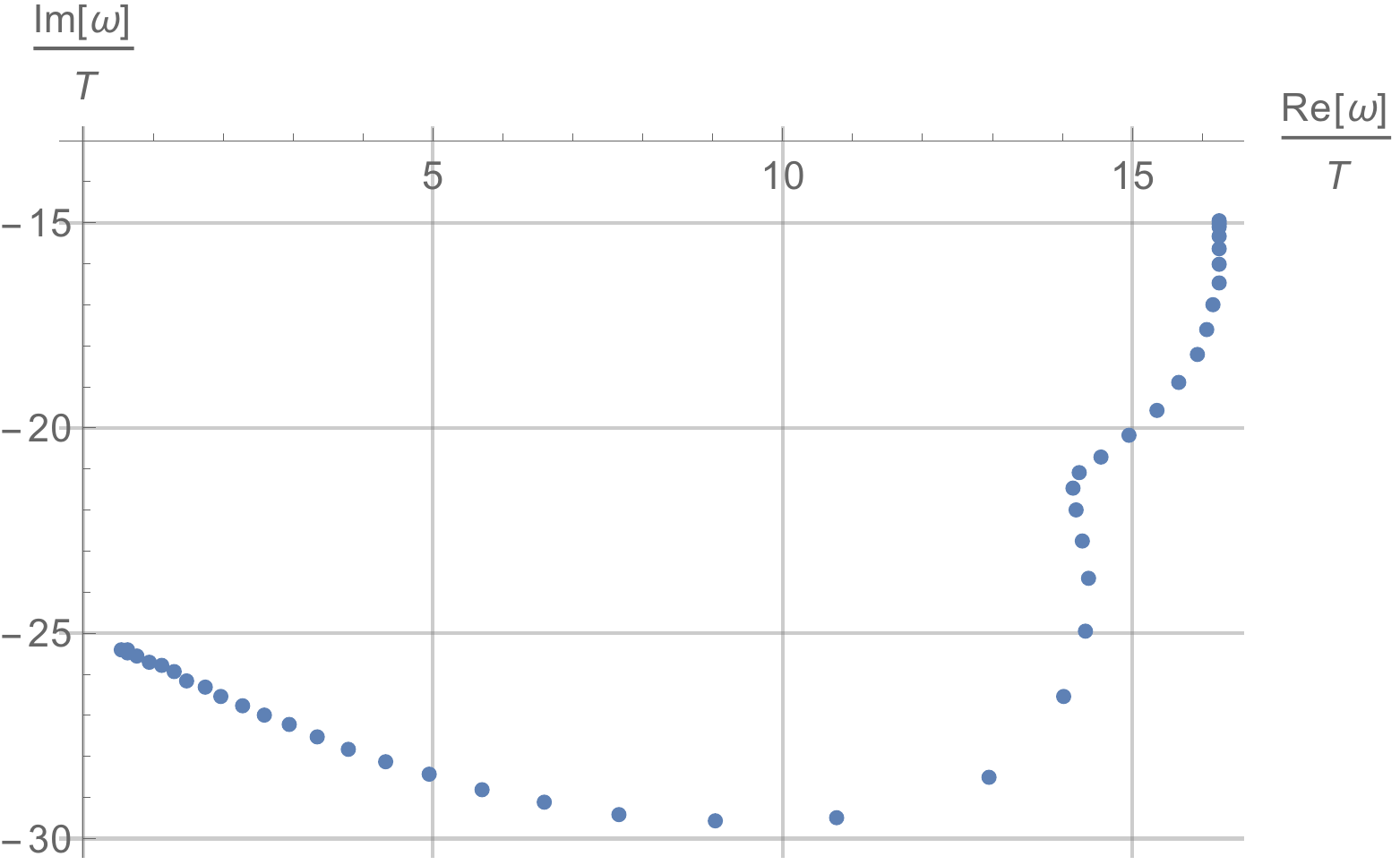}
	\caption{Two helicity-2 QNMs of the magnetic black brane, i.e.~with $\mu=0$, $\tilde{k}=0$, for $B$ from $0$ to $9$ in equidistant steps of $\Delta B = 0.2$. The value $B=9$ corresponds to $\tilde{B} \sim 3 \cdot 10^4$.}
	\label{fig:QNM2highB}
\end{figure}

Moreover, we observe that some of the QNMs approach the imaginary axis for very large magnetic fields $\tilde{B}.$ This holds at zero as well as at finite chemical potential $\mu$. In Fig.~\ref{fig:QNM2highB} we show two of these modes for magnetic branes,\footnote{This result is in agreement with \cite{Janiszewski:2015ura}, in particular Fig.~12, where their numerical results were inconclusive at those large magnetic field values.} i.e. with $\mu=0.$

\FloatBarrier

\subsection{Helicity-1 sector under $SO(2)$ rotations}
Next we consider the helicity-1 sector which is of particular interest since the Chern-Simons coupling constant enters there explicitly in the fluctuation equations. As explained above, the helicity-1 sector contains two decoupled subsectors, denoted by helicity-$1^+$ and helicity-$1^-$. Both sectors consists of three second order ordinary differential equations as well as one constraint. We reformulate the three second order equations in terms of a generalized eigenvalue problem.  

\begin{figure}[ht]
	\centering
  \includegraphics[width = 0.47\textwidth]{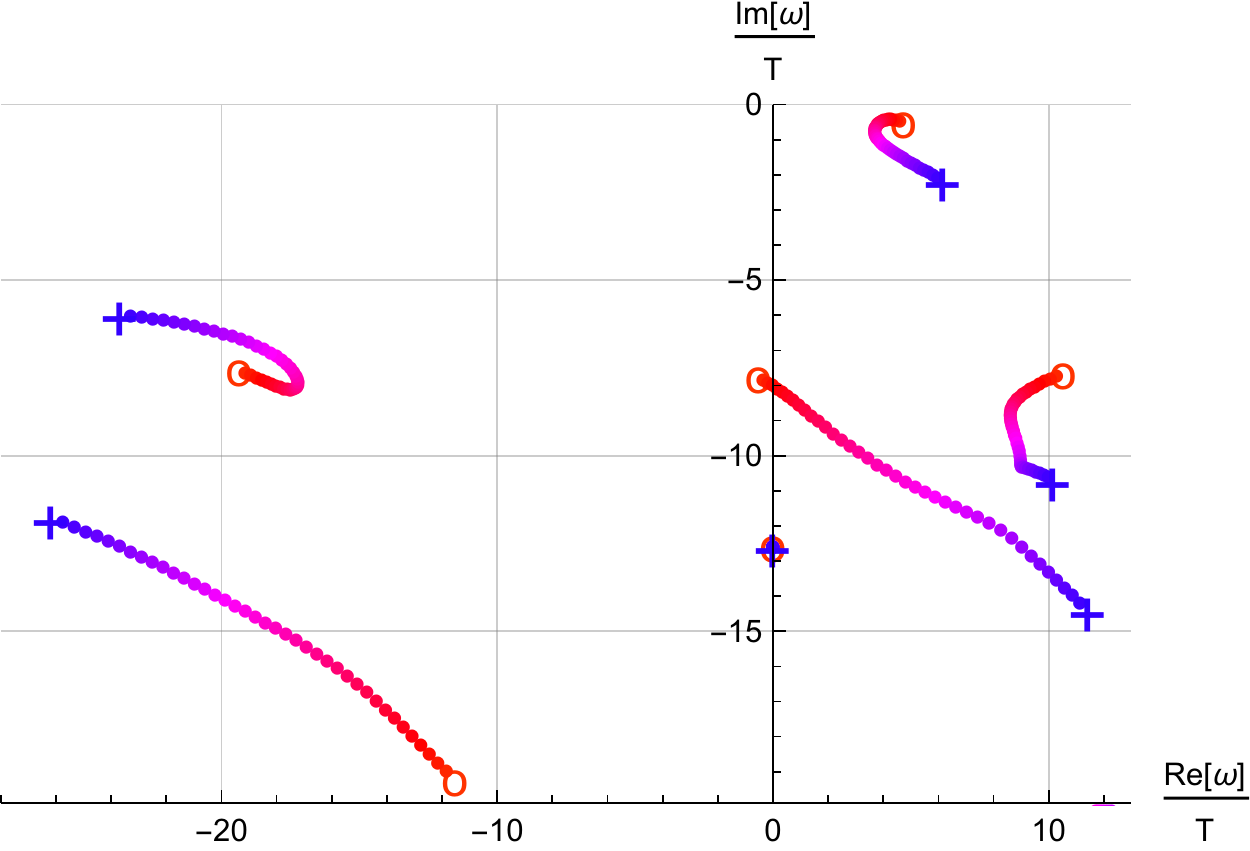}
  \hspace{4mm}\includegraphics[width = 0.47\textwidth]{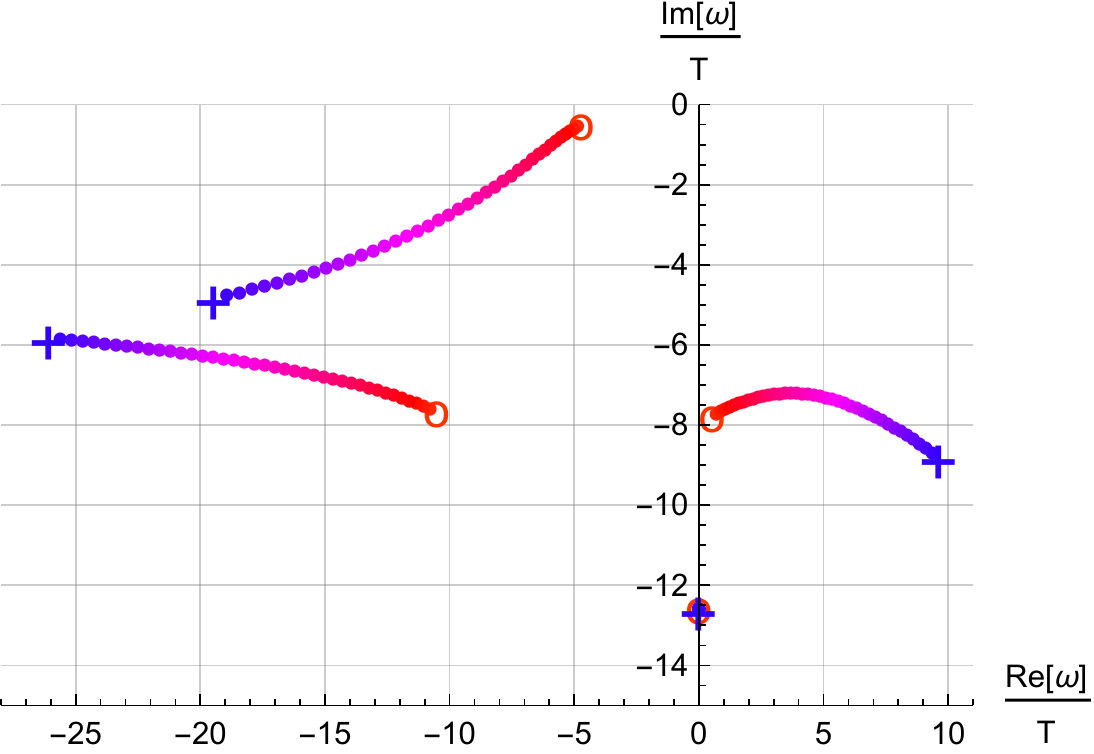}
	\caption{Lowest QNMs in the helicity-$1^+$ (left) and helicity-$1^-$ (right) sector. In particular, we find two fake QNMs which do not depend on $\tilde{k}$.
	The momentum varies from $\tilde{k}=0$ (denoted by a circle) to $\tilde{k}=20$ (denoted by a cross). The remaining parameters are chosen like in Fig.~\ref{overviewspin2}.}
	\label{fig:overviewspin1}
\end{figure}

We show in Fig.~\ref{fig:overviewspin1} the lowest lying QNMs for $\tilde{\mu} = 10$ and $\tilde{B}=65$ (corresponding to $\bar{T} = 0.1$ and $\bar{B}=0.65$). The momentum varies over $\tilde{k} \in [ 0, 20 ]$ and we have chosen equidistant values for $\tilde{k}.$ In Fig.~\ref{fig:overviewspin1} we find in each sector one QNM which does not move with $\tilde{k}$ and is located at $\omega = - 2 \pi i n T$ with $n$ being an integer. These are fake QNMs satisfying the three coupled second order differential equations, but not the constraint.  
See Appendix~\ref{sec:appendixConv} for a discussion of fake QNMs.
From now on, we will discard these modes. 

\subsubsection{Hydrodynamic modes}
Let us first discuss the hydrodynamic modes in the helicity-1 sector. For vanishing magnetic field, $B=0$, we find two ungapped hydrodynamic QNMs, 
which can be identified as the momentum diffusion modes. In presence of the anomaly a term of third order in momentum $k$ affects the momentum diffusion as was already discovered in holographic models~\cite{Sahoo:2009yq,Matsuo:2009xn}, whereas we clarify the field theoretic origin of this term in Appendix~\ref{sec:hydrodynamics}.
In the case of $B \neq 0$, the QNMs are gapped, i.e. their dispersion relations satisfy $\omega(k) \neq 0$ for $k \rightarrow 0$. However, it is remarkable that we are still able to find agreement with hydrodynamics, at least for small magnetic fields $B$ and counting $B$ as first order in the derivatives, i.e.~$\mathcal{O}(B)\sim \mathcal{O}(k) \sim \mathcal{O}(\partial)$, as we explain below. In particular, analyzing the numerical QNMs for small $k \ll 1$, we can fit our results to the hydrodynamic predictions.

For the numerical data, we use backgrounds with fixed temperature $\bar{T}= 0.2$  and magnetic field $B$ and scan the momentum $k$ from zero to one. We then perform a polynomial fit to the real- and imaginary part of the QNMs respectively. The upper limit for the fit range of $k$, denoted by $k_{max}$, is chosen such that the coefficients of the fit polynomials do not change when lowering $k_{max}$.

\begin{figure}[ht]
	\centering
  \includegraphics[width = 0.44\textwidth]{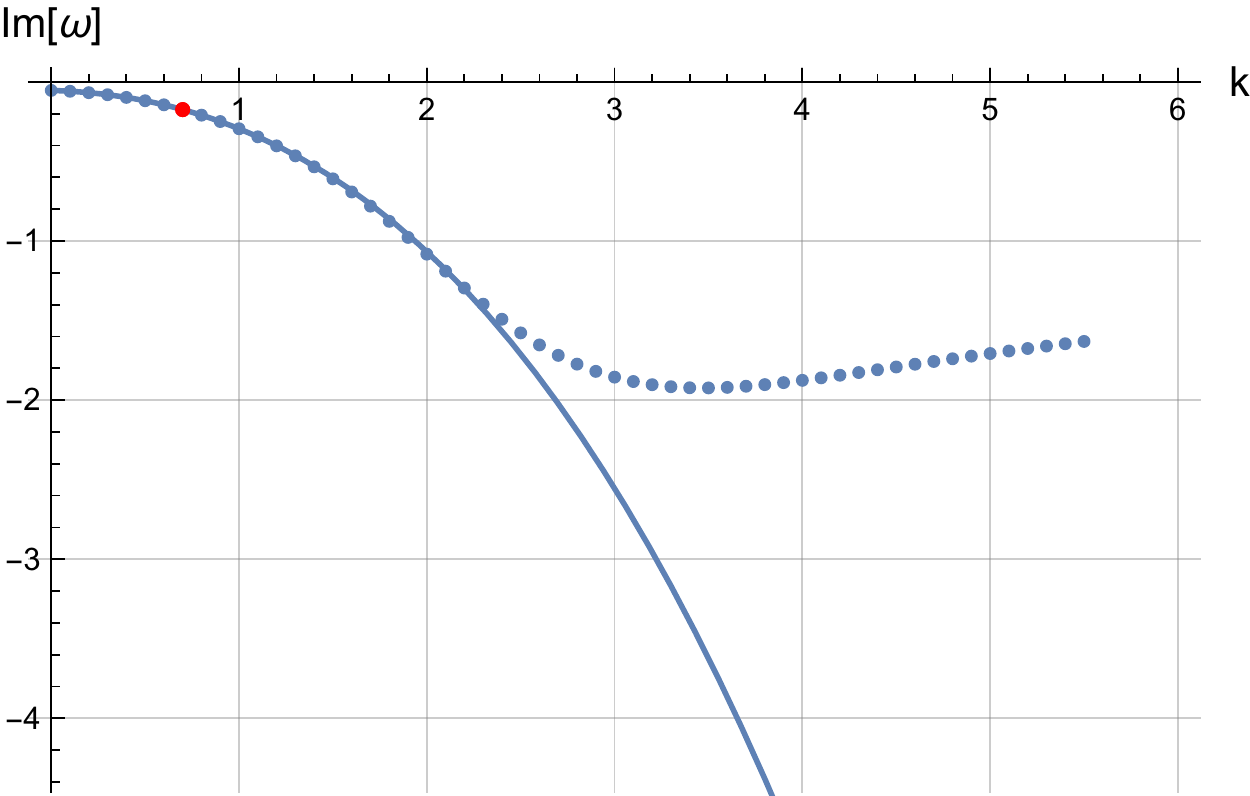}
	\caption{Comparison of a polynomial fit to the imaginary part of the dispersion relation of the lowest QNM in the helicty-$1^-$ sector for $\bar{T}=0.2$, $B=1$ and $\gamma = 3/2.$}
	\label{figeta}
\end{figure}

We take the helicity-$1^-$ sector to be exemplary for helicity-1. We split our analysis in real and imaginary part of the hydrodynamics and numerical results for QNMs, respectively. In Fig.~\ref{figeta} we show an example for a polynomial fit to the imaginary part of the dispersion relation of the lowest QNM in the helicity-$1^-$ sector for $\bar{T}=0.2$ and $B=1$. For small $k$, the imaginary part of the dispersion relation is fitted by the polynomial $\mbox{Im}(\omega) = -0.054 - 0.0034 k -0.18 k^2 - 0.031 k^3.$ The coefficients of the fit are also displayed in Table~\ref{comps1IM}. For larger $k$, the third-order polynomial fit highly deviates from the imaginary part of the QNM dispersion relation, see Fig.~\ref{figeta}. We compare the coefficients of the fit polynomial to the hydrodynamic prediction of the dispersion relation 
at weak $B$ (i.e.~$\mathcal{O}(B)\sim \mathcal{O}(k)\sim\mathcal{O}(\partial)$)
\begin{eqnarray}\label{eq:helicity1polesmain}
\omega &=& \mp \frac{B  n_0}{\epsilon_0 + P_0} - i k^2  \frac{\eta}{\epsilon_0+P_0} +k \frac{B n_0 \xi_3}{(\epsilon_0+P_0)^2} - \frac{i B^2\sigma}{\epsilon_0+P_0} \nonumber \\
&& \pm k^3 \frac{i \eta \xi_3}{(\epsilon_0+P_0)^2}
\mp k^2 B \frac{n_0 \xi_3^2}{(\epsilon_0+P_0)^3}
\pm k B^2 \frac{i \sigma \xi_3}{(\epsilon_0+P_0)^2}+ \mathcal{O}(\partial^3) \, ,
\end{eqnarray}
as discussed in Appendix~\ref{sec:hydrodynamics}. 
In~\eqref{eq:helicity1polesmain} we have included some terms of third order, but not all of them, as indicated by $\dots + \mathcal{O}(\partial^3)$.
Note that the energy density $\epsilon _0$, the pressure $P_0$, the charge density $n_0$ as well as transport coefficients such as the shear viscosity $\eta$ and the conductivity $\sigma$ enter the prediction from hydrodynamics. While the energy density and the pressure can be determined from the numerical background solutions, the transport coefficients are not known explicity for our setup. Hence, we approximate the transport coefficients for the charged magnetic brane by its values for zero magnetic field, i.e. for Reissner-Nordstr{\"o}m black brane. In particular, we use $\eta/s = 1/(4\pi)$ and $\sigma = (\mu^2-6)^2/[4(\mu ^2+3)^2]$\cite{Sahoo:2009yq,Matsuo:2009yu}. This additional approximation is justified for small enough magnetic field $B$.

With increasing magnetic field $B$ we can clearly track the disagreement between hydrodynamic prediction and numerics. The breakdown for increasing $B$ shows the limits of both of our assumptions: weak field hydrodynamics and the RN approximation for the transport coefficients.

First, we show the imaginary part of the dispersion relation in Table~\ref{comps1IM}. We display the fit coefficients and the corresponding hydrodynamic predictions for different B, comparing order by order in $k$. In particular, we find good agreement for the zeroth, second and third order in $k$ up to surprisingly large $B$. Also at nonzero $B$, the aforementioned $k^3$-term arises in presence of the anomaly and is approximating the numerical data surprisingly well. The first order in $k$ does not fit as expected, because we have not taken into account all contributions of $\mathcal{O}(\partial^3)$ in our hydrodynamic expansion.
\begin{table}[ht]
\begin{equation*}
\begin{array}{c||cc|cc|cc|cc}
 B & -\frac{B^2 \sigma }{P_0+\epsilon _0} & \text{QNM fit} & \frac{B^2 k \xi_3 \sigma }{\left(P_0+\epsilon _0\right)^2} & \text{QNM fit} & -\frac{\eta  k^2}{P_0+\epsilon_0} & \text{QNM fit} & \frac{\eta  k^3 \xi_3}{\left(P_0+\epsilon _0\right)^2} & \text{QNM fit} \\
\hline
 0 & 0 & 3.8 \cdot 10^{-6} & 0 & -0.000071 & -0.17 & -0.17 & -0.025 & -0.024 \\
 0.10 & -0.00044 & -0.00044 & -0.000065 & -0.00040 & -0.17 & -0.17 & -0.025 & -0.024 \\
 0.20 & -0.0018 & -0.0018 & -0.00026 & -0.0014 & -0.17 & -0.17 & -0.025 & -0.024 \\
 0.30 & -0.0040 & -0.0040 & -0.00058 & -0.0030 & -0.17 & -0.17 & -0.025 & -0.025 \\
 0.40 & -0.0070 & -0.0073 & -0.0010 & -0.0053 & -0.17 & -0.17 & -0.025 & -0.025 \\
 0.50 & -0.011 & -0.012 & -0.0016 & -0.0084 & -0.17 & -0.17 & -0.025 & -0.026 \\
 0.60 & -0.016 & -0.017 & -0.0024 & -0.012 & -0.17 & -0.17 & -0.025 & -0.027 \\
 0.70 & -0.022 & -0.024 & -0.0032 & -0.017 & -0.17 & -0.17 & -0.025 & -0.028 \\
 0.80 & -0.029 & -0.032 & -0.0042 & -0.022 & -0.17 & -0.17 & -0.025 & -0.029 \\
 0.90 & -0.036 & -0.042 & -0.0054 & -0.028 & -0.17 & -0.17 & -0.025 & -0.030 \\
 1.0 & -0.045 & -0.054 & -0.0067 & -0.034 & -0.17 & -0.18 & -0.025 & -0.031 \\
\end{array}
\end{equation*}
\caption{Imaginary part of the dispersion relation of the hydrodynamic mode in the helicity-$1^-$ sector for $\bar{T}=0.2$ and $\gamma=3/2.$ Here we display the coefficients expected from hydrodynamics and from the polynomial fit to the imaginary part of the dispersion relation of the lowest QNM. }
\label{comps1IM}
\end{table}

Second, we consider the real part of the dispersion relation of the lowest QNM in Table~\ref{comps1RE}. The zeroth and first order coefficients in $k$ are well described by our prediction from hydrodynamics. Note that these coefficients can be calculated exactly with the numerical background without approximating transport coefficients by its RN values, and hence tests only the weak field approximation within our hydrodynamics. Both give very good agreement with the data up to relatively large $B$.  
The second order in $k$ does not fit the data, again due to unknown $\mathcal{O}(\partial^3)$ contributions in the hydrodynamic approximation. The third order in $k$ is predicted to be zero in $\mathcal{O}(\partial^2)$ hydrodynamics, while our numerical data indicates a $k^3$-term increasing monotonically with the magnetic field $B$.
\begin{table}[ht]
\begin{equation*}
\begin{array}{c||cc|cc|cc|cc}
 B & \frac{-B n_0}{P_0+\epsilon _0} & \text{QNM fit} & \frac{B k n_0 \xi_3}{\left(P_0+\epsilon _0\right)^2} & \text{QNM fit} & \frac{-B k^2 n_0 \xi_3^2}{\left(P_0+\epsilon _0\right)^3} & \text{QNM fit} & k^3 & \text{QNM fit} \\
 \hline
 0 & 0 & 0 & 0 & 0 & 0 & 0 & 0 & 0 \\
 0.10 & -0.041 & -0.041 & -0.0060 & -0.0060 & -0.00089 & -0.0044 & 0 & -0.00039 \\
 0.20 & -0.081 & -0.081 & -0.012 & -0.012 & -0.0018 & -0.0089 & 0 & -0.00081 \\
 0.30 & -0.12 & -0.12 & -0.018 & -0.018 & -0.0027 & -0.014 & 0 & -0.0013 \\
 0.40 & -0.16 & -0.16 & -0.024 & -0.024 & -0.0036 & -0.018 & 0 & -0.0019 \\
 0.50 & -0.20 & -0.20 & -0.030 & -0.030 & -0.0045 & -0.023 & 0 & -0.0027 \\
 0.60 & -0.25 & -0.24 & -0.036 & -0.035 & -0.0054 & -0.029 & 0 & -0.0037 \\
 0.70 & -0.29 & -0.28 & -0.043 & -0.041 & -0.0063 & -0.035 & 0 & -0.0049 \\
 0.80 & -0.33 & -0.32 & -0.049 & -0.047 & -0.0073 & -0.041 & 0 & -0.0066 \\
 0.90 & -0.37 & -0.36 & -0.055 & -0.053 & -0.0082 & -0.048 & 0 & -0.0088 \\
 1.0 & -0.41 & -0.40 & -0.062 & -0.059 & -0.0092 & -0.055 & 0 & -0.012 \\
\end{array}
\end{equation*}
\caption{Real part of dispersion relation versus polynomial fit to the lowest QNM for $\bar{T}=0.2$ and $\gamma=3/2.$}
\label{comps1RE}
\end{table}

\subsubsection{Long-lived modes}
Besides the hydrodynamic modes, it is important to characterize other long-lived modes with a small imaginary part in the dispersion relation for large magnetic fields. 

\begin{figure}[ht]
	\centering
  \includegraphics[width = 0.5\textwidth]{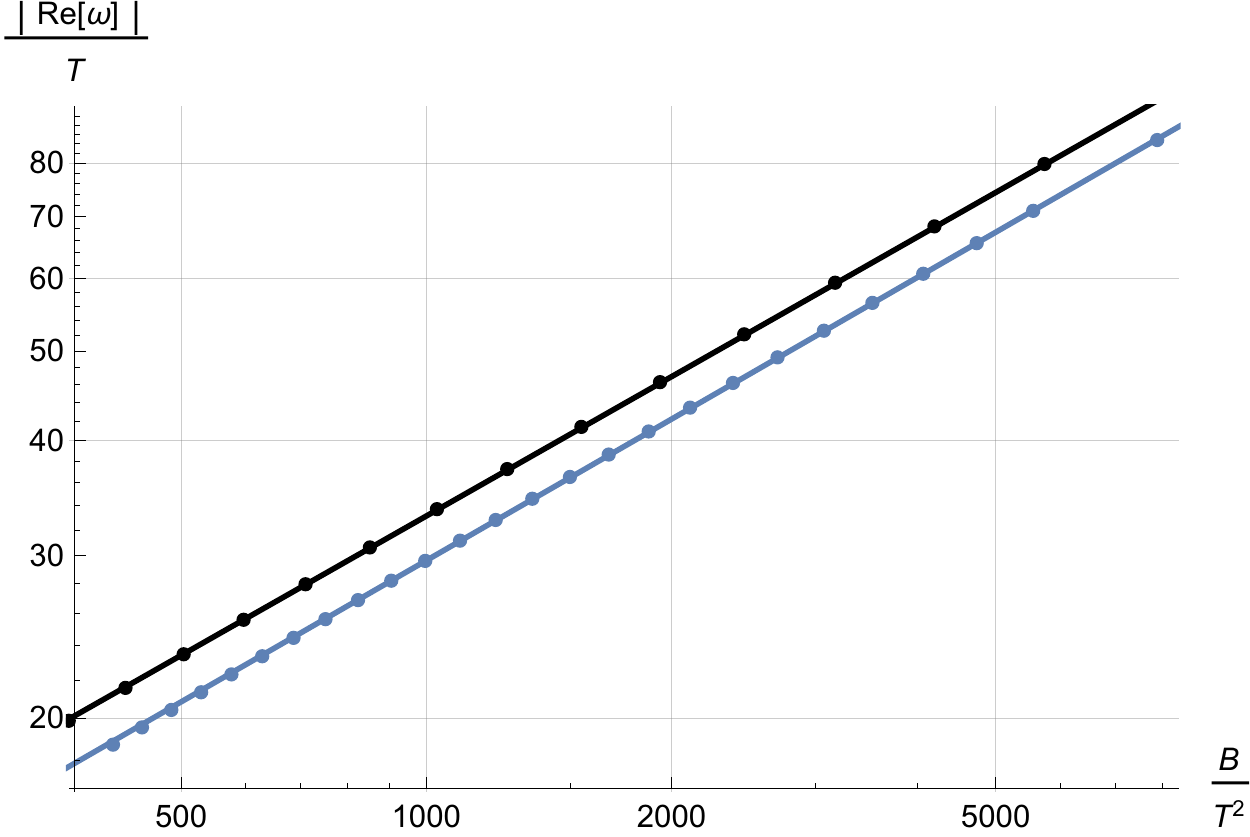}
	\caption{Real part of a QNM versus the magnetic field for $\mu = 1$, $\tilde{k}=0$, and $\gamma = 3/2$ (blue line) as well as $\gamma=5$ (black line), respectively. In particular, we find that $\mbox{Re}(\omega) \propto \sqrt{B}$ for both values of $\gamma.$
	}
	\label{landau1}
\end{figure}

For $\gamma > \gamma_c \approx 4$, we find long-lived modes in the limit of large magnetic fields.\footnote{In \cite{Ammon:2016fru}, with a different convention for the Chern-Simons coupling~$\kappa=\gamma/8$, the critical value of $\kappa$ was determined numerically to be $1/2$ for helicity-0 QNMs; consistently, we have $\gamma_c=4$ for $\mu=0$.} In particular, we find that the real part of the QNM dispersion relation is proportional to $\sqrt{B}$ for large magnetic fields, see Fig.~\ref{landau1}. Moreover, for $\gamma > \gamma_c$, the imaginary part of these QNMs stay small  for large $B$. Hence these QNMs are alongside the hydrodynamic modes the longest-lived modes. For example, for $\gamma = 5 > \gamma_c$, the imaginary part of this QNM at $\tilde{B} \approx 5 \cdot 10^3$ is smaller than $|\mbox{Im} \, (\tilde{\omega})| < 10,$ while the imaginary part of all other QNMs is much larger.

In contrast to that, for $\gamma < \gamma_c$, the imaginary part of the QNMs for these modes diverges for large $B$ and they are not the longest lived modes in our system. However, the real part of the dispersion relation is still proportional to $\sqrt{B}$ for large magnetic fields $B$.

\subsection{Helicity-0 sector under $SO(2)$ rotations}
Finally, we discuss the helicity-0 QNMs and their frequencies. Note that as in the helicity-1 sector, the Chern-Simons coupling $\gamma$ also appears explicitly in the equations of motion of the helicity-0 sector. 

\begin{figure}[ht]
	\centering
  \includegraphics[width = 0.5\textwidth]{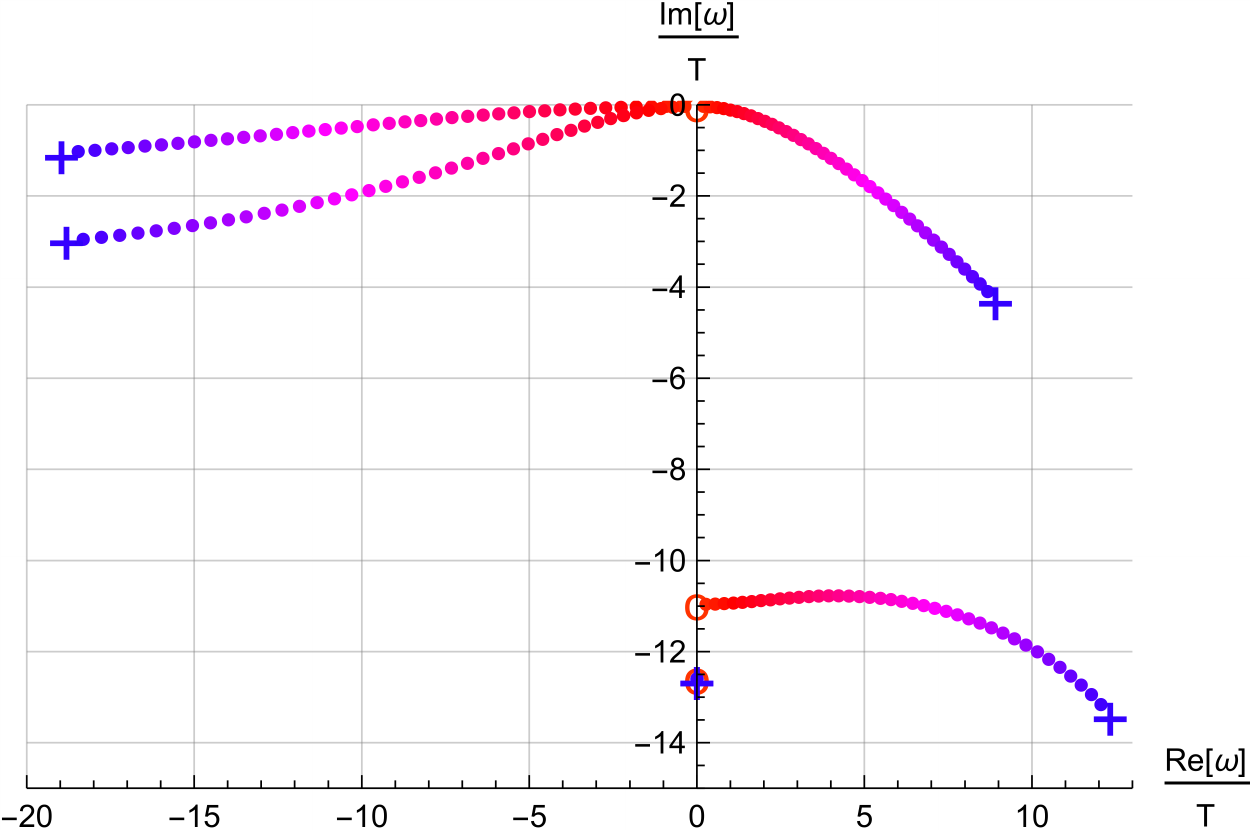}
	\caption{Lowest four QNMs in the helicity-0 sector as well as one fake mode. 
	The momentum $\tilde{k}$ varies from $\tilde{k}=0$ (denoted by a circle) to $\tilde{k}=20$ (denoted by a cross). The remaining parameters are chosen like in Fig.~\ref{overviewspin2}.
	}
	\label{overviewspin2a}
\end{figure}

In Fig.~\ref{overviewspin2a} we display the four lowest QNMs in the helicity-0 sector for $\tilde{\mu}=10$ and $\tilde{B}=65.$ The momentum varies from $\tilde{k}=0$ to $\tilde{k}=20$, while the Chern-Simons coupling is $\gamma=3/2$. In particular, we identify three hydrodynamic modes and one fake mode. 
Two of these modes originate from the well-known sound modes and the third one originates from the charge diffusion mode. Note that these three modes stay hydrodynamic modes at $B\neq 0$, i.e.~they remain gapless.

\subsubsection{Hydrodynamic modes}

For the three hydrodynamic QNMs identified, we perform again a fit in $k$ to the numerical results for the QNM dispersion relation of the form
\begin{equation}
\omega_0 = v_0 \, k - i D_0 \, k^2 \, , \qquad \omega_\pm = v_\pm \, k - i \Gamma_\pm \, k^2 \, .
\end{equation}
In particular, we find that one of the modes, e.g. $\omega_0$, is purely diffusive for $B=0$, in agreement with analytical results for the Reissner-Nordstr{\"o}m case and with the hydrodynamic predictions summarized in Appendix~\ref{sec:hydrodynamics}. This agreement with hydrodynamics holds even for small magnetic fields $B$ as we discuss now. Taking the hydrodynamic prediction for the diffusion constant $D_0$ and sound velocity $v_0$, see eq. \eqref{eq:v0}, together with the Reisser-Nordstr{\"o}m approximation for the thermodynamic quantities such as entropy density and the enthalpy, as well as for the conductivity we get 
\begin{equation}
v_0 = -\frac{B \gamma}{4}\frac{6+\mu^2}{3+\mu^2}  \,, \qquad   D_0 = \frac{6+\mu^2}{12+4\mu^2} \,.
\end{equation}
We compare this hydrodynamic prediction, approximated by RN thermodynamics, to the fit data of the dispersion relation of the QNM and obtain very good agreement up to $B=0.3$ as it is evident from the table
\begin{equation*}
\begin{array}{c|cccccccccc}
B& 0. & 0.1 & 0.2 & 0.3 & 0.4 & 0.5 & 0.6 & 0.7 & 0.8 & 0.9 \\ \hline \hline
D_0 &0.418 & 0.418 & 0.418 & 0.418 & 0.418 & 0.418 & 0.418 & 0.418 & 0.419 & 0.419 \\
\text{fit} &0.418 & 0.415 & 0.407 & 0.393 & 0.375 & 0.353 & 0.328 & 0.302 & 0.275 & 0.248 \\ \hline
v_0 &-0. &- 0.063 &- 0.126 &- 0.188 &- 0.251 &- 0.314 &- 0.377 &- 0.439 & -0.502 &- 0.565 \\
\text{fit} &-0. &- 0.063 &- 0.126 &- 0.187 &- 0.246 &- 0.303 &- 0.356 &- 0.406 &- 0.451 &- 0.491 
\end{array}
\end{equation*}
The same computation can be done for the hydrodynamic modes $\omega_\pm$. Using the hydrodynamic prediction \eqref{eq:vPM}, and again approximating by thermodynamic expressions of Reisser-Nordstr{\"o}m black branes, we get 
\begin{equation}
v_{\pm} = \pm \frac{1}{\sqrt{3}} + B \frac{\xi_V}{6+2 \mu^2} \,,\qquad \Gamma_{\pm} = \frac{1}{6+2 \mu^2} 
\end{equation}
which compares to real part of the first order in $k$ fit coefficient:
\begin{equation*}
\begin{array}{c|cccccccccc}
B & 0. & 0.1 & 0.2 & 0.3 & 0.4 & 0.5 & 0.6 & 0.7 & 0.8 & 0.9 \\ \hline\hline
v_{-}&  -0.577 & -0.59 &- 0.602 & -0.614 & -0.626 & -0.639 &- 0.651 & -0.663 &-0.675 & -0.687 \\
\text{fit} &- 0.577 &- 0.59 & -0.603 &- 0.616 &- 0.629 & -0.643 &- 0.657 &- 0.672 & -0.687 & -0.704 \\ \hline
v_{+}& 0.577 & 0.565 & 0.553 & 0.541 & 0.528 & 0.516 & 0.504 & 0.492 & 0.48 & 0.467 \\
\text{fit} & 0.577 & 0.565 & 0.554 & 0.543 & 0.533 & 0.523 & 0.515 & 0.507 & 0.501 & 0.497 \\ \hline
\Gamma_{\pm}& 0.112 & 0.112 & 0.112 & 0.112 & 0.112 & 0.112 & 0.112 & 0.112 & 0.112 & 0.112 \\
\text{$-$~fit} & 0.112 & 0.112 & 0.112 & 0.113 & 0.113 & 0.113 & 0.113 & 0.114 & 0.114 & 0.114 \\
\text{$+$~fit} & 0.112 & 0.112 & 0.112 & 0.113 & 0.113 & 0.113 & 0.114 & 0.114 & 0.113 & 0.112 \\
\end{array}
\end{equation*}
Again, we have good agreement up to $B=0.3$. 

\begin{figure}[ht]
	\centering
  \includegraphics[width = 0.47\textwidth]{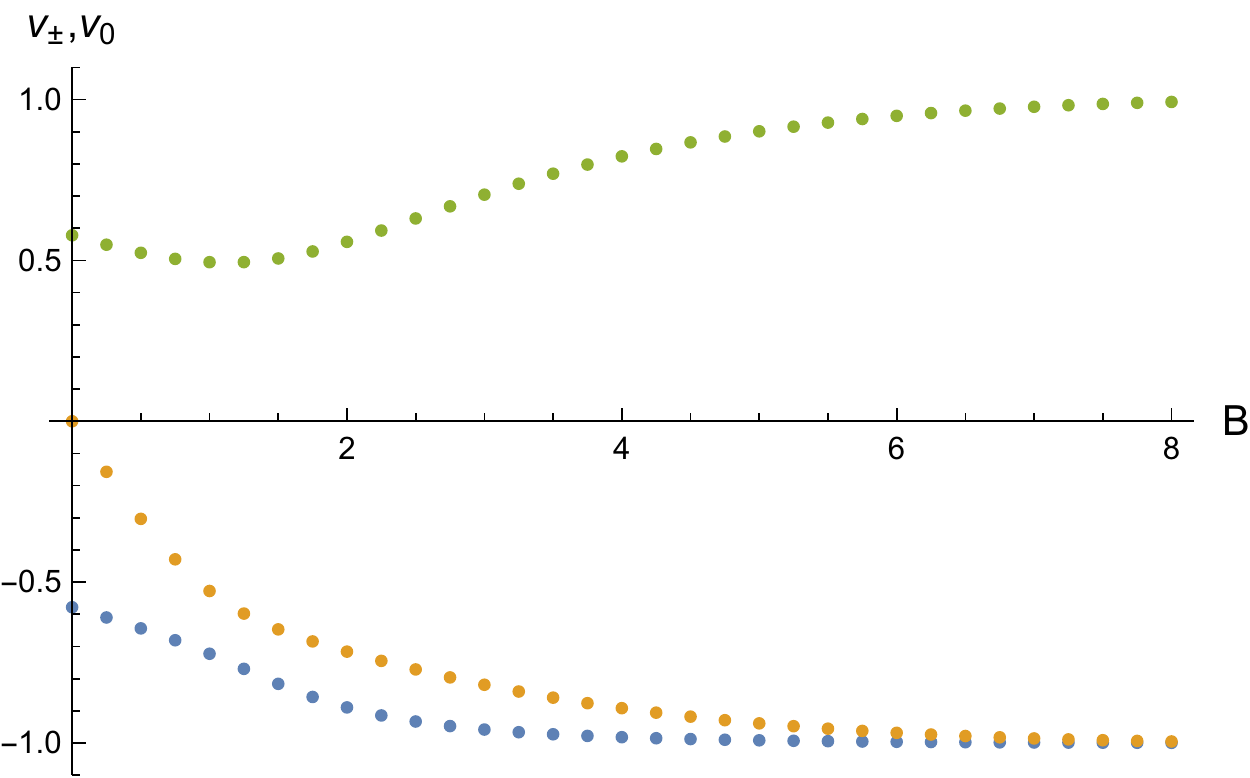}\hspace{4mm}
  \includegraphics[width = 0.47\textwidth]{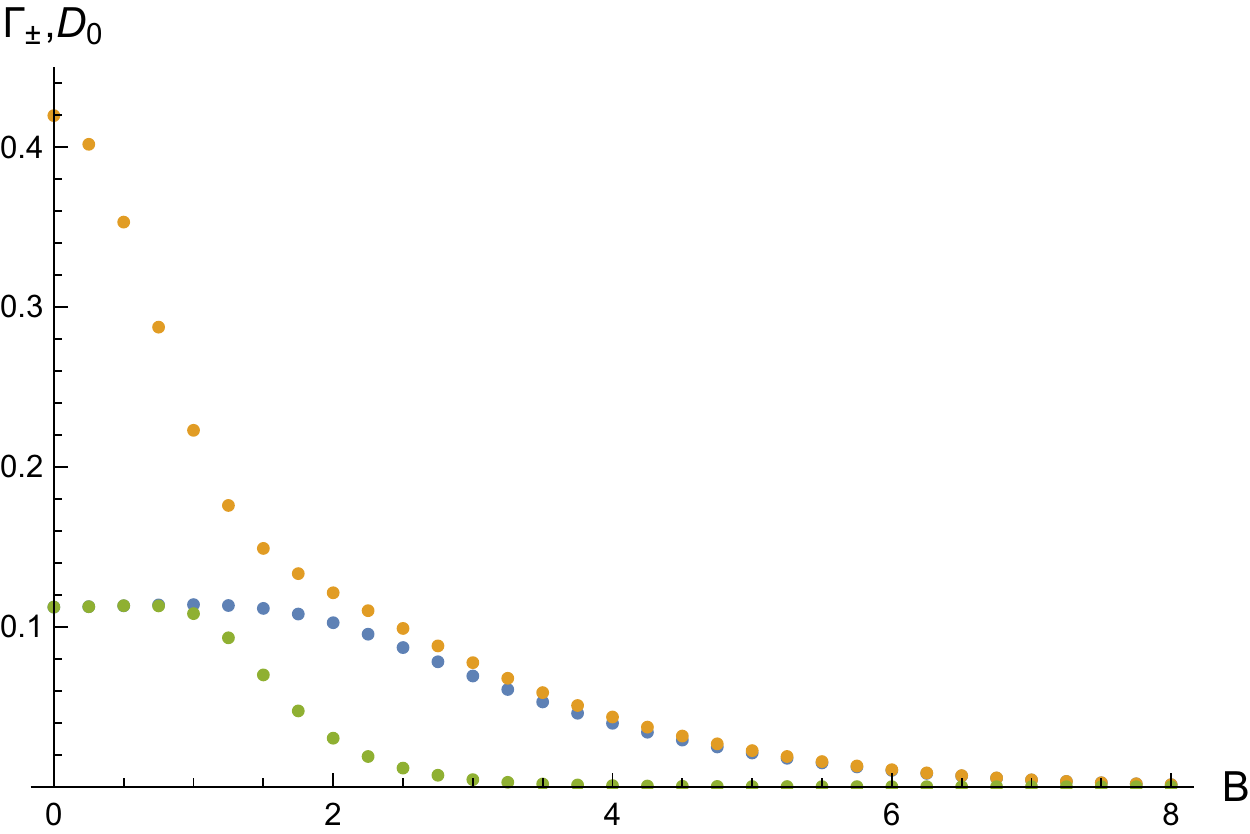}
	\caption{The velocities $v_0$~(orange), $v_+$~(green), $v_-$~(blue) and the attenuation coefficients $D_0$~(orange), $\Gamma_+$~(green), $\Gamma_-$~(blue) for the three hydrodynamic modes in the helicity-0 sector as a function of $B$ and with a Chern-Simons coupling $\gamma=3/2 <\gamma_c$.}
	\label{spinkb}
\end{figure}

The holographic calculation reveals also some interesting properties for hydrodynamics at strong magnetic fields. For any value of the magnetic field, we find three hydrodynamic modes $\omega_0$ and $\omega_\pm$. Let us first consider the imaginary part of each dispersion relation, i.e. $\Gamma_\pm$ and $D_0$. 
While counting magnetic fields of order derivative, i.e.~at weak $B$, we conclude in \eqref{eq:vPM} that the attenuations are the same for the sound modes $\omega_\pm$, i.e. $\Gamma_+ = \Gamma_-$. This, however, is not true for strong $B$, as can be seen in Fig.~\ref{spinkb}. For $B \gtrsim 1$, we see that $\Gamma_+ \neq \Gamma_-$. 
In particular, the figure suggests that $\Gamma_+$ approaches zero for large enough magnetic fields, while $\Gamma_-$ first approaches $D_0$, before both approach zero together. From the real part of the dispersion relation we can read off the sound velocities $v_\pm$ and $v_0$. If we treat the magnetic field of order derivative, the velocities $v_+$ and $v_-$ satisfy $v_+ - v_- 
= 2 c_s = 2/\sqrt{3}.$ This is indeed the case as we can see from Fig.~\ref{spinkb}. However, for magnetic fields $B \gtrsim 1$, we clearly see deviations from it. 
It appears that $v_-$ approaches $v_0$ in the limit of large magnetic fields. While performing the fit, we noticed that a term linear in $k$ is sufficient to fit the real part of the frequency, and a term quadratic in $k$ is sufficient to obtain a reliable fit for the imaginary part up to fairly large values of $B\approx 8$. While we show the evolution of velocities and attenuations for $\gamma=3/2<\gamma_c$ in Fig.~\ref{spinkb}, we have checked that their behavior does not change significantly when we perform the same calculation for $\gamma = 5 >\gamma_c$.

One of the helicity-0 modes (the former charge diffusion mode) turns into a chiral magnetic wave mode~\cite{Kharzeev:2010gd}\footnote{\cite{Kharzeev:2010gd} works with an axial and a vector current, defining the chiral magnetic wave as an excitation that involves a coupling between axial and electric charge. In this present work we have only one current and thus only one charge, namely the axial one. Nevertheless, the hydrodynamic dispersion relations of the relevant excitations are identical as can be seen from Appendix~\ref{sec:hydrodynamics}. \label{fn:chiralMagWaveCaveat}
} 
with velocity $v_0$ and attenuation $D_0$ displayed from zero to large $B$ in Fig.~\ref{spinkb}. 
At large magnetic field the chiral magnetic wave velocity $v_0$ asymptotes to -1, i.e.~to the magnitude of the speed of light. This is particularly intriguing as that same chiral magnetic wave QNM shows Landau level behavior at large magnetic field, i.e. the real part of the frequency is proportional to $\sqrt{B}$ and the imaginary part asymptotes to zero. This provides evidence at strong coupling for the relation between Landau level occupation and the large $B$ behavior of $v_0$. That relation was proposed by Kharzeev and Yee in~\cite{Kharzeev:2010gd} based on weak coupling reasoning. This can also be understood as supporting Kharzeev and Yee's conjecture that the form of the chiral magnetic velocity is valid at arbitrary $B$.

\subsubsection{Long-lived modes: Landau Levels in the helicity-0 sector}

\begin{figure}[ht]
	\centering
  \includegraphics[width = 0.4\textwidth]{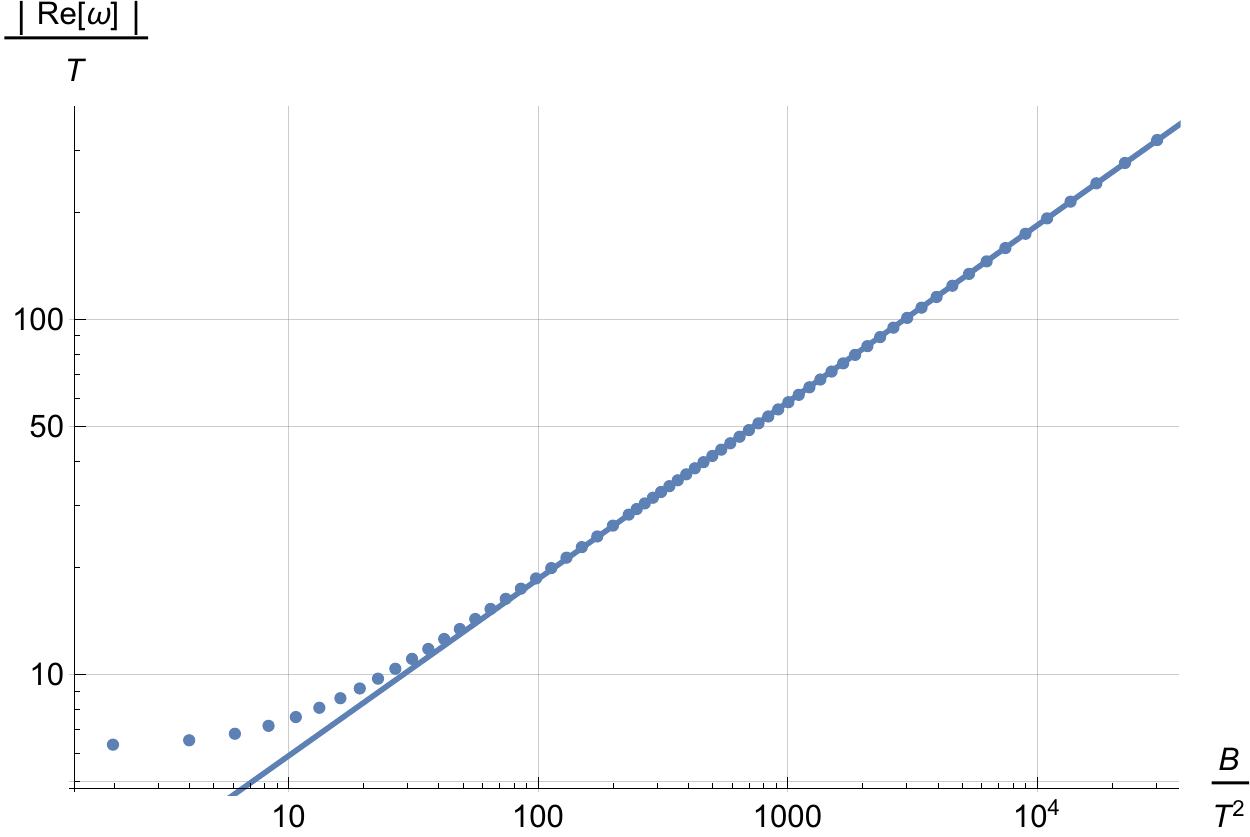}\hspace{10mm}
   \includegraphics[width = 0.4\textwidth]{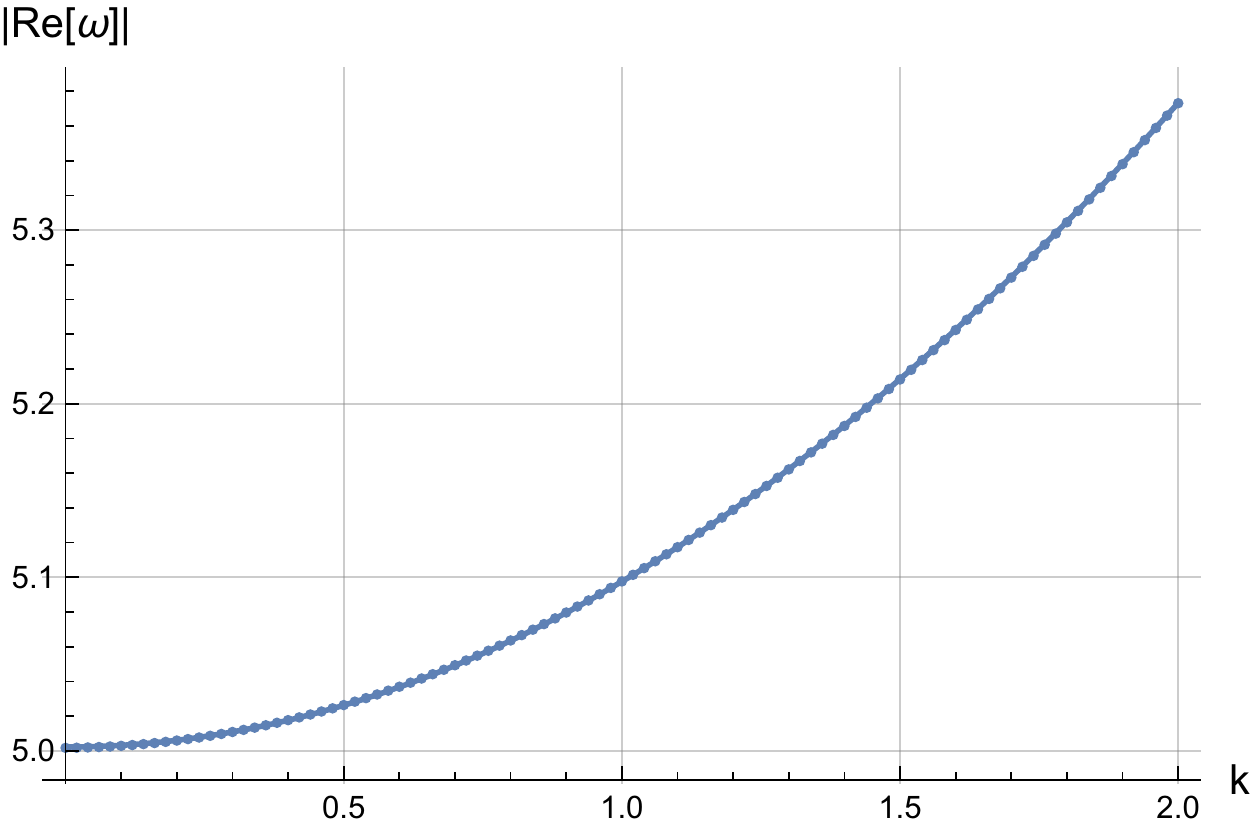}
	\caption{Left panel: real part of dispersion relation versus $\tilde{B}$ for $\mu = 0.01$. We can identify the expected $\sqrt{B}$ behaviour; Right panel: $k$ dependence of the real part of $\omega$ for $B =7.4$ for the next-to-lowest Landau level. The numerical data is fitted best by $\mbox{Re}(\omega) = 0.98 \sqrt{(5.1)^2+ k^2}$.}
	\label{fig:LandauRealSpin0}
\end{figure}

Besides the hydrodynamic modes, we find other modes which are long-lived in particular for large magnetic fields. These modes behave similarly to Landau levels, as discussed in \cite{Ammon:2016fru} for a similar model. As in the helicity-1 sector, we identify QNMs whose real part scales as $\sqrt{B}$ for large magnetic fields as can be seen from the left panel of Fig.~\ref{fig:LandauRealSpin0}. In particular, we find that for $\mu = 0.01$, the real part of the dispersion relation of the two next-to-lowest Landau levels may be fitted by $\textrm{Re}(\omega) \sim \pm 1.8 \, \sqrt{B}$ and $\textrm{Re}(\omega) \sim \pm 2.2 \, \sqrt{B}$ for large magnetic fields $B$. 

In the left panel of Fig.~\ref{fig:LandauRealSpin0} we investigated the $k$-dependence of the next-to-lowest Landau level. In particular, we fixed the magnetic field $B=7.4$ and the chemical potential $\mu = 0.01$, and plotted the real part of the dispersion relation as a function of $k$, the momentum along the magnetic field. The numerical data is best fitted by $\mbox{Re}(\omega) = 0.98 \sqrt{(5.1)^2+ k^2}$, the expected $k$-dependence for Landau levels if we extrapolate the dispersion-relation at weak coupling. 

\begin{figure}[ht]
	\centering
  \includegraphics[width = 0.4\textwidth]{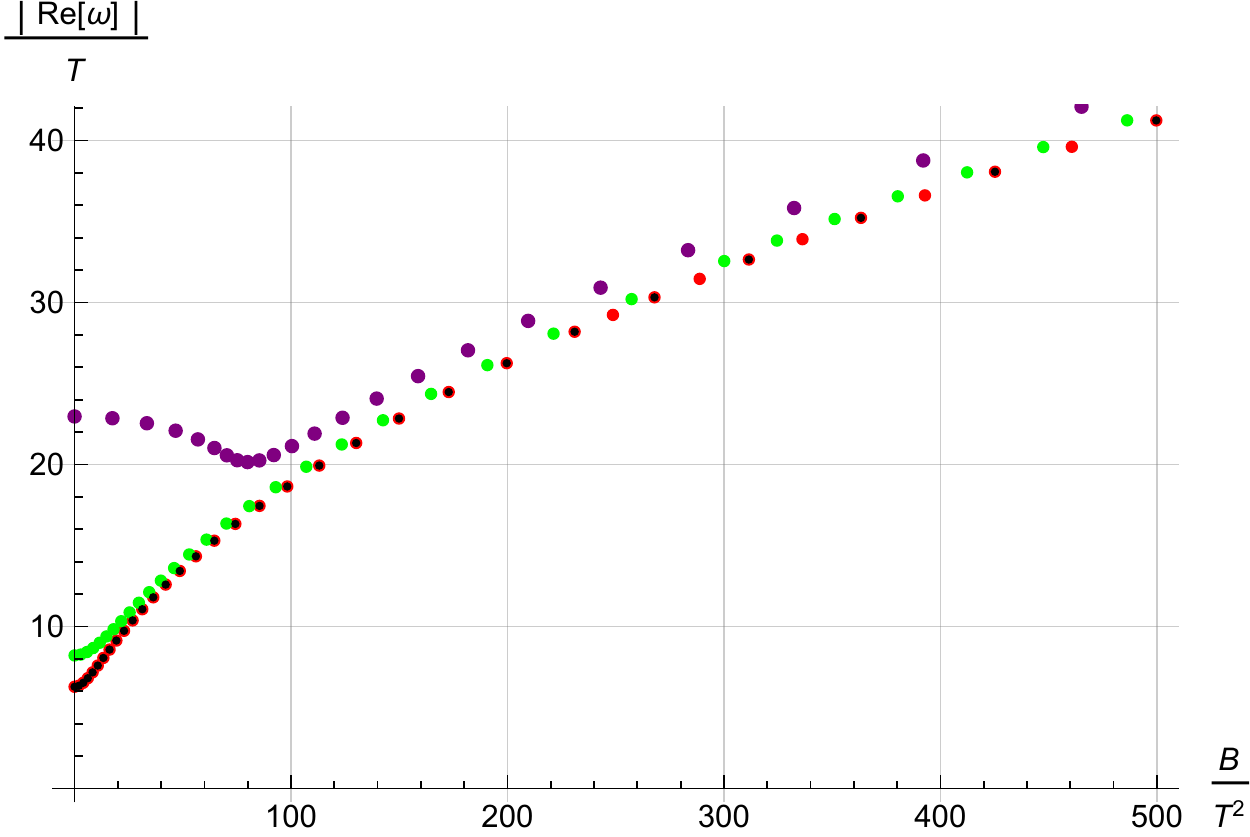} \hspace{10mm}
  \includegraphics[width = 0.4\textwidth]{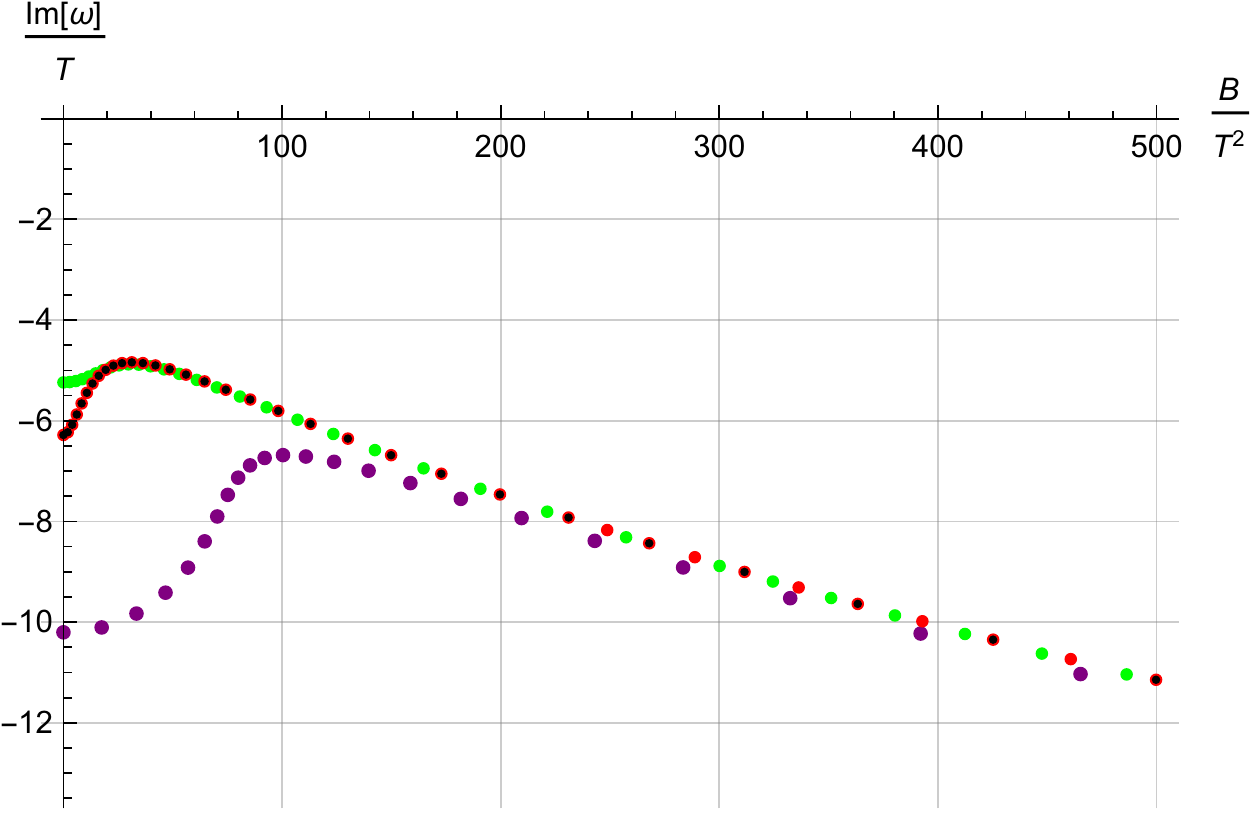}
	\caption{Left panel: Real part of dispersion relation versus $\tilde{B}$ for $\gamma=3/2$ and different chemical potential $\mu$: $\mu=0$ black dots, $\mu =0.01$ red dots and $\mu =1$ green dots, $\mu=2$ purple dots. Right panel: Imaginary part versus $\tilde{B}$ with the same colour coding. }
	\label{fig:LandauSpin0finitedensity}
\end{figure}

Moreover, we investigated the influence of finite charge density on the dispersion relation of the Landau levels. Again, the characteristic $\sqrt{B}$-behaviour of the real part of the frequency for large magnetic fields is unchanged. This is to be expected since for $\mu \ll \sqrt{B}$ the behaviour due to the magnetic field dominates.  However, as displayed in Fig.~\ref{fig:LandauSpin0finitedensity}, the chemical potential modifies the real and imaginary part of the dispersion relation for small magnetic fields. 

\begin{figure}[ht]
	\centering
  \includegraphics[width = 0.4\textwidth]{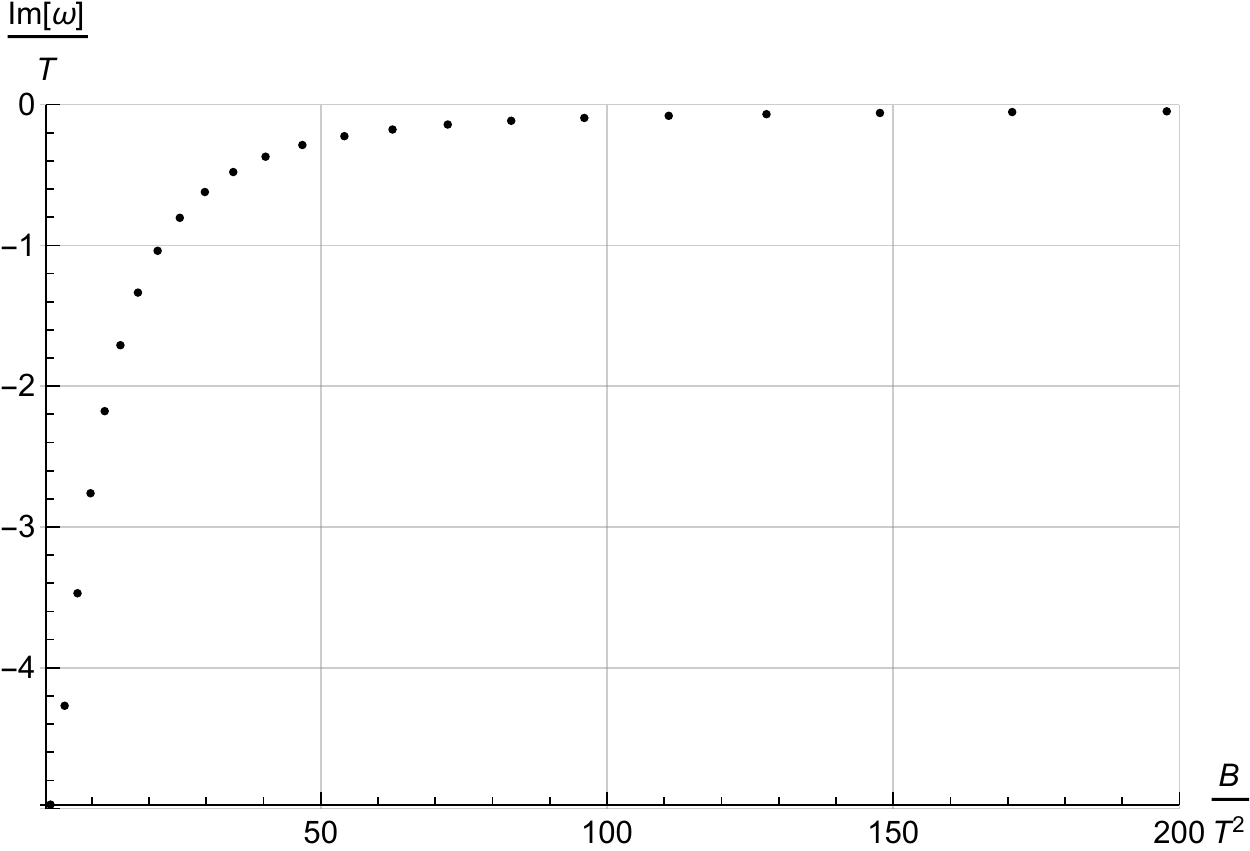}
	\caption{Imaginary part of next-to-lowest Landau level as a function of $\tilde{B}$ for $\gamma=5 > \gamma_c.$}
	\label{fig:LandauImSpin0}
\end{figure}

As in the case of the helicity-1 sector, the anomaly coefficient $\gamma$ plays an important role whether the modes are long-lived. For $\gamma < \gamma_c \approx 4$, the imaginary part seems to diverge, and hence the QNM is not long-lived. In particular, the imaginary part of the dispersion relation is fitted best by $\mbox{Im}(\omega) \sim - \sqrt{B}.$  However, as displayed in Fig.~\ref{fig:LandauImSpin0} for $\gamma > \gamma_c \approx 4$, the QNMs, corresponding to the next-to-lowest Landau levels, approach the real axis and hence are long-lived excitations for large magnetic fields.

\FloatBarrier

\section{Discussion}\label{sec:discussion}
In this paper, we have investigated transport and dissipation in a thermal plasma in a magnetic field of arbitrary strength at strong coupling. This plasma state is defined within a (3+1)-dimensional field theory with chiral anomaly. In the dual gravitational calculation we evaluate and analyze quasinormal modes (QNMs) of Einstein-Maxwell-Chern-Simons theory in the asymptotically $AdS$ charged magnetic black brane background. For a particular value of the Chern-Simons coupling, the action is a consistent truncation of type IIB supergravity, and the dual field theories are well-known, including $\mathcal{N}=4$ SYM.
We have investigated the effects of having a finite temperature $T$, chemical potential $\mu$, a non-zero external magnetic field $B$, on the QNMs. We have chosen the momentum of the QNMs, $k$, to be (anti-)parallel to $B$. 

In parallel to the holographic calculation, we have computed the weak $B$ hydrodynamic description of our system. The five (formerly) hydrodynamic poles were computed at nonzero $B$. We extend previous treatments  keeping our derivation general and providing expressions for the various velocities and attenuations. At leading order in derivatives, we have also worked out an expression for the energy momentun tensor and axial current containing polarization effects~\eqref{eq:Tmn0Hydro}. These hydrodynamic results are collected in Appendix~\ref{sec:hydrodynamics}.

In a systematic study of all metric and gauge field fluctuations propagating along the magnetic field, we investigated QNMs both within and outside the hydrodynamic regime. 
We find that some of the QNMs are long-lived and show characteristic behavior known from Landau levels, see Figs.~\ref{fig:LandauRealSpin0},~\ref{fig:LandauSpin0finitedensity} and~\ref{fig:LandauImSpin0}.
The helicity-2 QNMs do not exhibit any hydrodynamic modes. However, some of these QNMs appear to have vanishing real part in the limit of large $B$, see Fig.~\ref{fig:QNM2highB}. There are five hydrodynamic modes in total. Two of those modes at vanishing $B$ are momentum diffusion modes appearing in the helicity-1 sector. At nonzero $B$ these modes acquire a complex gap and can not be considered hydrodynamic modes according to our previous definition. However, they are still well described by the hydrodynamic dispersion relation~\eqref{eq:helicity1polesmain} at intermediate values of $B$ (modulo discrepancies stemming from third order corrections in the derivative expansion). The remaining three hydrodynamic modes appear in the helicity-0 sector and at $B=0$ they reduce to the charge diffusion pole and the two sound poles, see the dispersion relation~\eqref{eq:helicity0poles},~\eqref{eq:v0},~\eqref{eq:vPM},~\eqref{eq:GammaPM} and Fig.~\ref{spinkb}. These are hydrodynamic modes (ungapped) even at $B\neq 0$, 
as can be seen from the dispersion relations~\eqref{eq:helicity0poles}.
One of the helicity-0 modes (the former charge diffusion mode) turns into a chiral magnetic wave mode\footnote{We again refer to the clarification in footnote~\ref{fn:chiralMagWaveCaveat}.} with velocity $v_0$ and attenuation $D_0$ displayed from zero to large $B$ in Fig.~\ref{spinkb}. 
At large magnetic field the chiral magnetic wave velocity $v_0$ asymptotes to -1, i.e.~to the magnitude of the speed of light. This is particularly intriguing as that same chiral magnetic wave QNM shows Landau level behavior at large magnetic field, i.e.~the real part of the frequency is proportional to $\sqrt{B}$ and the imaginary part asymptotes to zero. This provides evidence at strong coupling for the relation between Landau level occupation and the large $B$ behavior of $v_0$. That relation was proposed by Kharzeev and Yee in~\cite{Kharzeev:2010gd} based on weak coupling reasoning. This can also be understood as supporting Kharzeev and Yee's conjecture that the form of the chiral magnetic velocity is valid at arbitrary $B$.  
In~\cite{Kharzeev:2010gd} a holographic and a field theory result for the chiral magnetic wave velocity, and a holographic result for its attenuation are provided. That holographic calculation was based on a probe-brane approach, which neglects backreaction by definition. In that sense, the present work extends \cite{Kharzeev:2010gd} to fully backreacted geometry and a solution valid at arbitrary $B$. We further provide a hydrodynamic and holographic calculation of the chiral magnetic wave attenuation $D_0$, as well as the velocities $v_\pm$ and attenuations $\Gamma_\pm$ of the former sound modes.  Remarkably, all of these velocities obtain contributions from the anomaly, as seen explicitly e.g.~in~\eqref{eq:v0} and~\eqref{eq:vPM}. 
This is also true for the two helicity-1 modes, see e.g.~\eqref{eq:helicity1poles}. 
Dispersion relations for all five hydrodynamic modes are provided in the Landau frame and in the thermodynamic frame~\cite{Jensen:2011xb,Jensen:2012jh}, see Appendix~\ref{sec:hydrodynamics}.

For the future, it would be interesting to relax the restriction of having the QNM momentum $k$ parallel to the external magnetic field $B$; see~\cite{Abbasi:2016rds} for a field theory discussion of a related system. This would allow many more modes to arise, and thus even richer phenomenology. 
It would also be interesting to have two U(1) gauge fields in the gravity theory~\cite{Jimenez-Alba:2014pea}, i.e.~introducing an axial and a conserved vector current in the dual field theory. This is relevant for testing some predictions for chiral magnetic waves and for Weyl semimetals and their surface states, see~\cite{Landsteiner:2015lsa,Landsteiner:2015pdh,Copetti:2016ewq,Ammon:2016mwa,Grignani:2016wyz}. 
Moreover, it would be interesting to include mixed gauge-gravity anomalies and study their effects on the QNMs, see for example~\cite{Liu:2016hqb} in which a mixed anomaly appears to lead to a phase transition. 
In this paper we have considered hydrodynamics in the regime of weak external gauge fields. It would be very interesting to work out the hydrodynamics in the case where the gauge fields are strong, i.e.~zeroth order in a derivative expansion, building on~\cite{Kovtun:2016lfw} and~\cite{Huang:2011dc}. 
Lastly, it would be desirable to extend our setup systematically to magnetohydrodynamics as viewed from a modern perspective~\cite{Grozdanov:2016tdf}. 


\paragraph*{Acknowledgements} 
We would like to thank P.~Kovtun for helpful discussions. JL acknowledges  financial support by Deutsche Forschungsgemeinschaft (DFG) GRK 1523/2.

\begin{appendix}

\section{Hydrodynamics}
\label{sec:hydrodynamics}

In this appendix, we derive expressions for the location of the poles of retarded correlation functions of the energy momentum tensor, $T^{\alpha\beta}$, and the axial current, $J^\mu$, in the hydrodynamic approximation. 
These poles are identified with the lowest lying QNMs of the corresponding gravitational fields via the gauge/gravity correspondence. Hence, this section will serve to predict the QNM frequencies at small frequency values and small momenta. 

\subsection{General framework and definitions}
Hydrodynamics can be considered as an effective field theory, or more precisely, as an expansion of $n$-point functions in terms of (small) gradients of the hydrodynamic variables. The defining relations for hydrodynamics are the constitutive equations (or 1-point functions) for the energy momentum tensor and the conserved current(s), as well as the corresponding conservation equations already defined in the main text in~\eqref{eq:hydroConservation}, see for example~\cite{Rangamani:2009xk,Kovtun:2012rj}. The constitutive equations express the conserved and hence long-lived quantities, i.e. the energy momentum tensor and the conserved current(s), in terms of temperature $T$, chemical $\mu$, and the fluid velocity, $u^\mu$. These hydrodynamic variables can be considered as fields in hydrodynamics when considered as an effective field theory. As such, they are only defined modulo field redefinitions leaving the physical quantities invariant. Fixing this freedom of field redefinitions is fixing a 
``hydrodynamic frame''. A well known example is the {\it Landau frame} in which the heat current vanishes, while we will be mostly working in the {\it thermodynamic frame}~\cite{Jensen:2011xb,Jensen:2012jh} defined below.  

We consider here a system in the presence of external sources, viz. the external gauge field, $A^\mu$, and the external metric, $g_{\mu\nu}$. We are interested here in a non-trivial gauge field background, 
$A_\mu = (\mu,\,-x_2 B/2,\,x_1 B/2,\,0)$, with a non-vanishing chemical potential $\mu$, and a magnetic field $B$ in the 
$x_3$-direction, plus a first order correction $a_\mu$. However, for the metric background we take it to be Minkowskian,
$\eta_{\mu\nu} = \text{diag}(-1,1,1,1)$, plus a first order correction $h_{\mu\nu}$.

The system will respond to the metric and gauge sources with corrections to the equilibrium values for all the hydrodynamic variables 
\begin{equation}\label{eq:Tumu}
T (t,{x_3}) = T_0 + \varepsilon\, T_1(t,{x_3})\, , \quad
u^\nu(t,{x_3}) = u^\nu_0 + \varepsilon\, u^\nu_1(t,{x_3})\, , \quad 
\mu (t,{x_3}) = \mu_0 + \varepsilon\, \mu_1(t,{x_3})\, ,
\end{equation}
with the expansion parameter $\varepsilon$ (not to be confused with the energy density $\epsilon$). Here we have choosen the particular case in which the momentum of the response is parallel to the magnetic field, i.e. $\vec{k}\,||\,\vec{B}$, where the vector arrow denotes the spatial part of the four-vectors.\footnote{Another interesting choice would be to consider the momentum of the response to be perpendicular to the magnetic field, see~\cite{Abbasi:2016rds} for such a choice in a system with an axial and a vector current.} 
Those corrections in the hydrodynamic variables will cause corrections in the thermodynamic quantities $\epsilon$, $P$, $n$, which we expand as
\begin{eqnarray}\label{eq:epsilonPn}
\epsilon(t,{x_3}) &=& \epsilon_0 + \varepsilon\, \frac{\partial \epsilon}{\partial T} T_1(t,{x_3}) + \varepsilon\, \frac{\partial \epsilon}{\partial \mu} \mu_1(t,{x_3})\, ,\\
P(t,{x_3}) &=& P_0 + \varepsilon\, \frac{\partial P}{\partial T} T_1(t,{x_3}) + \varepsilon\, \frac{\partial P}{\partial \mu} \mu_1(t,{x_3})\, ,\\
n(t,{x_3}) &=& n_0 + \varepsilon\, \frac{\partial n}{\partial T} T_1(t,{x_3}) + \varepsilon\, \frac{\partial n}{\partial \mu} \mu_1(t,{x_3}) \, ,
\end{eqnarray}
where here and in the remainder of this appendix partial derivatives with respect to $T$ are evaluated at fixed $\mu$ and vice versa, unless otherwise stated.

Below, we will make frequent use of the following thermodynamic relations:
\begin{eqnarray} 
\label{eq:de_dP}
d\epsilon = T ds +\mu d n\, ,\quad 
dP = s dT + n d\mu \, ,\quad \chi = \frac{\partial n}{\partial \mu} \, ,
\end{eqnarray}
and
\begin{eqnarray}
\label{eq:dedT_dedmu}
\frac{\partial \epsilon}{\partial T} &=& T_0 \frac{\partial s}{\partial T} + \mu_0 \frac{\partial n}{\partial T} \, , \qquad
\frac{\partial \epsilon}{\partial \mu} = T_0 \frac{\partial s}{\partial \mu} + \mu_0 \frac{\partial n}{\partial \mu} \,  .
\end{eqnarray}
Quantities of relevance to the constitutive equations used in the following subsections are the projector, electric field, the magnetic field and the vorticity:
\begin{equation}\label{eq:EBomega}
\Delta^{\mu\nu}= g^{\mu\nu}+u^\mu u^\nu\, , \quad 
E^\mu = F^{\mu\nu}u_\nu\, ,\quad 
B^\mu = \frac 1 2 \epsilon^{\mu\nu\rho\lambda}u_\nu F_{\rho\lambda}\, , \quad
\omega^\mu = \epsilon^{\mu\nu\rho\lambda}u_\nu\nabla_\rho u_\lambda \, .
\end{equation}
While the chiral transport coefficients in the thermodynamic frame are defined by
\begin{equation}
\label{eq:xisThermodynamicFrame}
\xi_V =  -3 C \mu^2 + \tilde{C} T^2,\quad 
\xi_B = -6 C \mu,\quad
\xi_3 = -2 C \mu^3 + 2 \tilde{C} \mu T^2,
\end{equation}
where the $T^2$ terms can be neglected in the large $N$ limit as they originate from supressed gauge-gravitational anomalies~\cite{Landsteiner:2011cp,Gynther:2010ed,Landsteiner:2011iq,Amado:2011zx}.

\subsection{Strong field thermodynamics ($B\sim\mathcal{O}(1)$)}
In this subsection, we consider strong magnetic field thermodynamics, more precisely zeroth order in derivatives hydrodynamics for time-independent quantities. A full hydrodynamic treatment of the strong magnetic field case would be interesting but is beyond the scope of this work.\footnote{For an attempt at this, see~\cite{Huang:2011dc}.}
By ``strong field'' we mean a background magnetic field of zeroth order in derivatives, i.e. the field strength 
$F\sim B\sim\mathcal{O}(1)$, with $B$ being the magnitude of the magnetic field. When a system is placed in a strong external field, strong field effects need to be taken into account, for example the polarization or magnetization of the medium~\cite{Israel:1978up,Kovtun:2016lfw}. Such polarization effects to zeroth order in a derivative expansion can be parametrized utilizing a polarization tensor $M^{\mu\nu}=2 \partial P/\partial F_{\mu\nu}$. Furthermore, effects of the chiral anomaly can contribute to thermodynamic quantities at zeroth order in derivatives~\cite{Jensen:2013kka}. Together, these effects at a strong external magnetic field $B$ can be expressed in terms of equilibrium constitutive equations for the energy momentum tensor
\begin{align}
\langle T^{\mu\nu}_\text{EFT} \rangle &=   
\epsilon_0 u^\mu u^\nu +P_0 \Delta^{\mu\nu} +q^\mu u^\nu +q^\nu u^\mu \notag \\
&\quad + M^{\mu\alpha} g_{\alpha\beta} F^{\beta\nu} + u^\mu u^\alpha \left (
M_{\alpha\beta} F^{\beta\nu} - F_{\alpha\beta} M^{\beta\nu}
\right ) +\mathcal{O}(\partial) \, ,
\end{align}
where
\begin{equation}
q^\mu = \xi_V B^\mu \,, \qquad 
M^{\mu\nu} = \chi_{BB} \epsilon^{\mu\nu\alpha\beta} B_\alpha u_\beta \, ,  
\end{equation}
and the axial current 
\begin{eqnarray}
\langle J^\mu_\text{EFT} \rangle & = & 
n_0 u^\mu + \xi_B B^\mu +\mathcal{O}(\partial) \,.
\end{eqnarray}
Note that all quantities above have to be evaluated at zeroth order in derivatives, e.g. $u^\mu = (1,0,0,0)$.
When the magnetic field is chosen to point in the $x_3$-direction, then these relations directly reduce to the equilibrium energy momentum tensor~\eqref{eq:Tmn0Hydro} and the equilibrium axial current~\eqref{eq:Jm0Hydro}. Note that in the Landau frame the heat current vanishes, $q^\mu =0$, hence the off-diagonal components in the energy momentum tensor vanish in Landau frame. Physically this simply means that in that frame the observer is traveling with the heat within the fluid. The interesting point is actually that such a heat current exists in equilibrium at strong magnetic fields and in the presence of an anomaly.

\subsection{Weak field hydrodynamics ($B\sim\mathcal{O}(\partial)$)}
By ``weak field'' we refer to a background magnetic field which is of first order in derivatives, i.e. the field strength is $F\sim B\sim\mathcal{O}(\partial)$. In other words, here we consider a background magnetic field which is of the same order as the spatial derivatives, and hence $B\sim k_3 = k$. So, for example, any term that contains one power of the magnetic field and one spatial derivative is of second order: $B \partial_3 (\dots)\sim \mathcal{O}(\partial^2)$. 

We start from the constitutive equations to first order in derivatives
\begin{eqnarray}\label{eq:constitutiveEquations}
 \langle T^{\mu\nu} \rangle &=&\epsilon u^\mu u^\nu +P \Delta^{\mu\nu} +u^\mu q^\nu+u^\nu q^\mu  + \tau^{\mu\nu} \,,\\
 \langle J^\mu \rangle &=& n u^\mu + \nu^\nu\,,
\end{eqnarray}
where
\begin{eqnarray}
\tau^{\mu\nu} &=& -\eta \Delta^{\mu\alpha} \Delta^{\nu\beta} \left( \nabla_\alpha u_\beta +\nabla_\beta u_\alpha - \frac 2 3 \nabla_\lambda u^\lambda g_{\alpha\beta} \right) - \zeta \Delta^{\mu\nu} \nabla_\lambda u^\lambda\,, \\
\nu^\mu &=& -\sigma T \Delta^{\mu\nu}\nabla_\nu\left(\frac{\mu}{T}\right) + \sigma E^\mu + \xi_{B} B^\mu + \xi_{V} \omega^\mu \, ,\\
q^\mu &=& \xi_V B^\mu + \xi_3 \omega^\mu \, ,
\end{eqnarray}
with $u_\mu \tau^{\mu\nu}=0$,  $ u_\mu \nu^\mu = 0$, and $u_\mu q^\mu=0$.
In order to determine the location of the poles of the hydrodynamic correlation functions, we have inserted the constitutive equations~\eqref{eq:constitutiveEquations} into the conservation equations~\eqref{eq:hydroConservation} and linearized in $\varepsilon$. We have then solved that system of linear equations for the corrections to the hydrodynamic variables, namely for $T_1$, $\mu_1$, and for the three spatial components of $u^\mu_1$, while the time component of $u_1^\mu$ is fixed by normalization $u^\mu u_\mu = -1$. The result of this exercise is provided in the following subsections sorted according to the helicity of the corrections and sources with respect to the $SO(2)$ rotation symmetry in the $(x_1, x_2)$-plane.\footnote{There are no hydrodynamic poles in the helicity-2 sector.}

\subsubsection{Poles in helicity-1 sector}
There are two poles in the correlation functions of the helicity-1 components of the energy momentum tensor $T^{\mu\nu}$ and the axial current $J^\mu$, located at
\begin{eqnarray}\label{eq:helicity1poles}
\omega &=& \mp \frac{B  n_0}{\epsilon_0 + P_0} - i k^2  \frac{\eta}{\epsilon_0+P_0} + k \frac{B n_0 \xi_3}{(\epsilon_0+P_0)^2} - \frac{i B^2\sigma}{\epsilon_0+P_0} \nonumber \\
&& \pm k^3 \frac{i \eta \xi_3}{(\epsilon_0+P_0)^2}
\mp k^2 B \frac{n_0 \xi_3^2}{(\epsilon_0+P_0)^3}
\pm k B^2 \frac{i \sigma \xi_3}{(\epsilon_0+P_0)^2}+ \mathcal{O}(\partial^3) \, ,
\end{eqnarray}
which originates from the two solutions 
\begin{equation}
\omega= -\frac{\pm B n_0 + i k^2 \eta + i B^2 \sigma}{
 \epsilon_0 + P_0 \pm k \xi_3} \, ,
\end{equation}
when expanded in $k$ and $B$. The second line in \eqref{eq:helicity1poles} is third order in $B$ and $k$, and it will receive corrections from second order contributions to the constitutive equations. It is remarkable that in the limit of $B=0$, explicit checks confirm that the $k^3$-term in the pole does not appear to receive any corrections from terms of second order in the constitutive relations in the Reissner-Nordstr\"om case. In that case \eqref{eq:helicity1poles} reduces to 
\begin{eqnarray}\label{eq:helicity1polesRN}
\omega &=& - i k^2  \frac{\eta}{\epsilon_0+P_0} 
\pm k^3 \frac{i \eta \xi_3}{(\epsilon_0+P_0)^2}
+ \mathcal{O}(\partial^3) \, ,
\end{eqnarray}
which is the standard charge diffusion $k^2$ term plus a contribution from the anomaly. This expression is in agreement with the holographic result~\cite{Sahoo:2009yq,Matsuo:2009xn}. Note that this $k^3$-term in equation~\eqref{eq:helicity1poles} is not present in Landau frame at first order in the derivative expansion as it originates from the heat current contribution $q^\mu$ in the energy momentum tensor in thermodynamic frame. It was shown in~\cite{Kharzeev:2011ds} that this $k^3$ term appears when working to second order in the constitutive equations in Landau frame.

\subsubsection{Poles in helicity-0 sector}
For convenience, we define first some useful notation: $w_0 = \epsilon_0 + P_0$ is the enthalpy density,
$\mathfrak{s}_0 = s_0/n_0$ is the entropy per particle, $c_s^2 = (\partial P/\partial\epsilon)_\mathfrak{s}$ is the speed of sound, $\tilde{c}_n = T_0(\partial\mathfrak{s}/\partial T)_n$ and $\tilde{c}_P = T_0(\partial\mathfrak{s}/\partial T)_P$ are the specific heats at constant density and pressure respectively\footnote{Note that this definition of the specific heat at constant pressure is different from that in~\cite{Kalaydzhyan:2016dyr} by a factor of the charge density $n_0$.}, 
and $\alpha_P = -(1/n_0)(\partial n/\partial T)_P$ is the thermal expansivity at constant pressure.

There are three poles in the helicity-0 sector: 
\begin{align}\label{eq:helicity0poles}
\omega_0 &= v_0\,k - i D_0\,k^2  + \mathcal{O}(\partial^3) \,, \\ 
\omega_+ &=  v_+\,k - i \Gamma_+\,k^2  + \mathcal{O}(\partial^3) \,, \\
\omega_- &=  v_-\,k - i \Gamma_-\,k^2 + \mathcal{O}(\partial^3) \,.
\end{align}
For the generalized diffusion pole, which we refer to as a chiral magnetic wave pole from now on, we have
\begin{equation}\label{eq:v0}
v_0 = 
\frac{2B\,T_0}{\tilde{c}_P n_0}\left(\tilde{C} - 3C\mathfrak{s}_0^2\right) \,, \qquad
D_0 = \frac{w_0^2\,\sigma}{\tilde{c}_P n_0^3 T_0} \,,
\end{equation}
where $D_0$ is the generalized diffusion coefficient. 
For the generalized sound poles, we have
\begin{align}
\label{eq:vPM}
v_\pm &= \pm c_s 
- B\,\frac{c_s^2}{n_0}\!\left(1 - \frac{\alpha_P w_0}{\tilde{c}_P n_0}\right)
\left[3C T_0\mathfrak{s}_0 + \frac{\alpha_P T_0^2}{\tilde{c}_P}(\tilde{C} - 3C\mathfrak{s}_0^2) 
+ \frac{1}{2}\xi_B^{(0)} - \frac{n_0}{w_0}\xi_V^{(0)}\right] \notag \\
&\qquad\quad  
+B\,\frac{1 - c_s^2}{w_0}\,\xi_V^{(0)} \,, \\
\label{eq:GammaPM}
\Gamma_\pm &= \frac{3\zeta + 4\eta}{6w_0} 
+ c_s^2\,\frac{w_0\,\sigma}{2n_0^2}\!\left(1 - \frac{\alpha_P w_0}{\tilde{c}_P n_0}\right)^2 \,.
\end{align}

In the RN limit, $B = 0$ and $\zeta = 0$~\footnote{Conformal symmetry is not broken explicitly in the dual field theory. It is  only broken at the level of the state.}. We have thus 
\begin{equation}\label{eq:RNlim}
v_0 = 0 \,, \qquad D_0 = \frac{w_0^2\,\sigma}{\tilde{c}_P n_0^3 T_0} \,, \qquad
v_\pm = \pm c_s \,, \qquad \Gamma_\pm = \frac{2\eta}{3w_0} \,.
\end{equation}
In particular, in the RN limit we have the relation
\begin{equation}
\tilde{c}_P n_0 = \alpha_P w_0 \,.
\end{equation}
This relation actually holds more generally than just in the RN case. It is true whenever $P_0\propto\epsilon_0$, and this can be seen as follows. By using Eq.~\eqref{eq:dedT_dedmu}, it can be shown that 
\begin{equation}\label{eq:cPn02aPw0}
\tilde{c}_P n_0 - \alpha_P w_0 = \left(\frac{\partial\epsilon}{\partial T}\right)_\mu 
- \mathfrak{s}_0\left(\frac{\partial\epsilon}{\partial\mu}\right)_T \,. 
\end{equation}
Then by Eq.~\eqref{eq:de_dP}, we have for constant $P_0$ and thus $\epsilon_0$
\begin{equation}
\mathfrak{s}_0 
= -\left(\frac{\partial\mu}{\partial T}\right)_P = -\left(\frac{\partial\mu}{\partial T}\right)_\epsilon
= \frac{(\partial\epsilon/\partial T)_\mu}{(\partial\epsilon/\partial\mu)_T} \,,
\end{equation}
and thus the desired result.

In the limit of vanishing charge density, $n_0=0$, the velocity and attenuation of the generalized diffusion pole are given by
\begin{equation} \label{eq:n020lim}
v_0 = 
-\frac{6 B C}{\chi} \,, \qquad 
D_0 = \frac{\sigma}{\chi} \,. 
\end{equation}
For the sound poles at $n_0=0$, with the additional simplifying assumption of $P_0\propto\epsilon_0$, velocities and attenuations are given by
\begin{equation}
v_\pm = \pm c_s  
+ B\,\frac{1 - c_s^2}{w_0}\,\xi_V^{(0)} \,, \qquad \Gamma_\pm = \frac{3\zeta + 4\eta}{6w_0} \,.
\end{equation}
Note that when $n_0=0$, $d\epsilon = T ds$ and $dP = s dT$, hence 
$c_s^2 = \partial P/\partial\epsilon = (\partial P/\partial T)/(\partial\epsilon/\partial T)$.

\subsubsection{Comparison to Landau frame results} \label{sec:LandauFrame}
A field theory calculation~\cite{Kalaydzhyan:2016dyr} similar to ours has been carried out while this paper was in preparation. Whereas our holographic model prompted us to work in the thermodynamic frame, Landau frame was chosen in~\cite{Kalaydzhyan:2016dyr}. Here we repeat our calculation in the Landau frame for comparison, and we point out differences between the two hydrodynamic frames.~\footnote{We find full agreement between those results provided explicitly in~\cite{Kalaydzhyan:2016dyr} and ours. We find only partial agreement between our results and the results of~\cite{Abbasi:2015saa}. In particular for the spin-0 modes at vanishing chemical potential and momentum we were able to find agreement with what the authors call $\omega_1, \, \omega_2,\, \omega_3$. The other results reported in~\cite{Abbasi:2015saa} we can not reproduce unless we make further assumptions, e.g. about the relation between the energy density and the pressure.}

In the Landau frame, the chiral conductivities are given by
\begin{eqnarray}
\xi_V^{L}& = &  \frac{1}{2} C^L \mu^2 \left (1-\frac{2}{3} \frac{n\mu}{\epsilon + P} \right) + \frac{1}{2}\tilde{C}^L T^2 \left( 1-\frac{2}{3} \frac{n\mu}{\epsilon+P} \right)\, , \\
\xi_B^{L} & = & C^L \mu \left ( 1 - \frac{1}{2} \frac{n \mu}{\epsilon+P}\right ) -\frac{\tilde{C}^L}{2} T^2 \frac{n}{\epsilon + P}\, .
\end{eqnarray}
Note the relation between the anomaly coefficients defined in the two frames:
\begin{equation}
 C^L=-6C \,, \qquad \tilde{C}^L = \tilde{C} \,.
\end{equation}
The chiral magnetic wave pole is given by
\begin{equation}\label{eq:v0D0Landau}
v_0^L = 
-B \left(\frac{C^L \mu_0^2+\tilde{C}^L T_0^2}{2w_0} 
- T_0\frac{C^L\mathfrak{s}_0^2 + \tilde{C}^L}{\tilde{c}_P n_0}\right) \,, \qquad
D_0^L = D_0 \,,
\end{equation}
and the generalized sound poles by
\begin{align}
v_\pm^L &= \pm c_s - B\,\frac{c_s^2}{2n_0}\!\left(1 - \frac{\alpha_P w_0}{\tilde{c}_P n_0}\right)\times \notag \\
&\qquad\qquad\quad
\bigg[C^L\!\left(\frac{\alpha_P T_0^2\mathfrak{s}_0^2}{\tilde{c}_P} - \frac{w_0^2 + T_0^2 s_0^2}{2w_0 n_0}\right) 
+ \tilde{C}^L\frac{T_0 n_0}{2w_0}\!\left(1 - \frac{2\alpha_P w_0}{\tilde{c}_P n_0}\right) + \xi_B^{(0)}\bigg] \notag \\
\Gamma_\pm^L &= \Gamma_\pm \,. 
\end{align}
Note that only the ``velocities'' $v_0^L$ and $v_\pm^L$ are different compared to their counterparts in the thermodynamic frame, while the ``diffusion coefficient'' and ``attenuation coefficients'' are identical in the two frames. 
Lastly, the RN limit in the Landau frame is the same as in the thermodynamic frame, while in the $n_0=0$ limit,
\begin{equation}\label{eq:zeroChargev0D0vPMGammaPMLandau}
v_0^L = B\left[C^L\!\left(\frac{1}{\chi} - \frac{\mu_0^2}{2w_0}\right) - \tilde{C}^L\frac{T_0^2}{2 w_0}\right] \,.
\end{equation}

We point out that the expressions for $v_0^L,\, v_\pm^L,\, D_0$ and $\Gamma_\pm$ agree with the results of~\cite{Kalaydzhyan:2016dyr}. 
In particular, when $n_0 = 0$, $v_0$ and $D_0$ from~\eqref{eq:zeroChargev0D0vPMGammaPMLandau} and~\eqref{eq:n020lim} respectively are identical to $v_\chi$ and $D_\chi$ defined in (42) of~\cite{Kalaydzhyan:2016dyr}\footnote{Note the definition in~\cite{Kalaydzhyan:2016dyr} is given at vanishing chemical potential, whereas we allow for $\mu_0\neq0$.}. 
We can also compare the expressions for $v_\chi$ and $D_\chi$ at $n_0 \neq 0$ in~\cite{Kalaydzhyan:2016dyr} to our Landau frame results, \eqref{eq:v0D0Landau}. Note that our calculation is more general, as we do not impose any restrictions such as 
$\delta p=0$, which was imposed in~\cite{Kalaydzhyan:2016dyr}. The comparison shows agreement between the velocity $v_0^L$ in our~\eqref{eq:v0D0Landau} and $v_\chi^T$ given in (45) of~\cite{Kalaydzhyan:2016dyr}. 
We also provide the attenuation coefficient, $D_0^L$, in~\eqref{eq:v0D0Landau} and~\eqref{eq:v0}, which is left undetermined as ``$D_T$'' in~\cite{Kalaydzhyan:2016dyr}. 

The particular collective excitation with the velocity $v_0^L$ given in~\eqref{eq:v0D0Landau} was coined ``thermal chiral magnetic wave'' in~\cite{Kalaydzhyan:2016dyr}. But the classification according to the $SO(2)$ rotation symmetry, seems to support the claim that~\eqref{eq:v0D0Landau} is merely a generalized version of the chiral magnetic wave~\cite{Kharzeev:2010gd}. In addition to a background magnetic field, a background vorticity has been considered in~\cite{Kalaydzhyan:2016dyr}, and additional effects such as a ``thermal chiral vortical wave'' and two modified sound modes with a vorticity-dependent velocity were reported. Certainly these three modes bear intriguing effects, we would, however, still refer to them simply as the hydrodynamic modes of the helicity-0 sector, one of them being the chiral magnetic wave mode.

The helicity-1 poles in Landau frame are given by
\begin{equation}
\omega = \, \pm B\frac{n_0}{w_0} - i k^2 \frac{\eta}{w_0}-k\, B \frac{\xi^{L, (0)}_V}{w_0} -i B^2\frac{\sigma}{w_0} +O(\partial^3).
\end{equation}

It is important to recall that we are working with a single axial current and an energy momentum tensor here, in contrast to QCD, which has an axial and a vector current. Hence, there are more hydrodynamic modes to be expected in QCD due to the interplay between its axial current and its vector current. The weak field hydrodynamic description of such a system has been considered in~\cite{Abbasi:2016rds}.

\section{Numerical methods. Convergence tests}
\label{sec:appendixConv}
The equations of motion for the background and the equations of motion for the fluctuations on such a background are given by \eqref{eq:EOM1} and \eqref{eq:EOM2} using the ansatz \eqref{eq:ansatzmetric_ef},\eqref{eq:ansatzA_ef} or \eqref{eq:ansatzqnm},\eqref{eq:fouriertrafo} respectively. The equations obtained are solved numerically with a spectral method. In this section, we discuss the numerics for the nonlinear background equations as well as the nonlinear generalized eigenvalue problem for the QNMs in detail. The whole numerics is set up in Mathematica. This enforces to use several performance optimizations, which we will discuss for the background and the QNMs.  

Spectral methods as a high accuracy method have the benefit of exponential convergence for analytic solutions. Yet the background magnetic field $B$ introduces logarithmic terms, giving an algebraic convergence rate. Additionally spectral methods allow to include the singular points on the boundary, that appear in holographic models. Therefore the background equations and the eigenvalue problem can be solved on the whole domain, including both the conformal boundary and the horizon.

\subsection{Background}
The equations of motion \eqref{eq:EOM1} and \eqref{eq:EOM2} for the background are given by six nonlinear ordinary differential equations and one constraint equation for the six unknown functions, $u(z), v(z), w(z), c(z)$ and $ A_v(z), P(z)$. Expanding the unknown functions in Chebyshev Polynomials
\begin{equation}
f(z) = \sum\limits^{N}_{i=0} c_i T_i(z)
\end{equation}
evaluated on a Chebyshev-Lobatto grid
\begin{equation}
 z_j = \frac{1}{2}\left( 1+\cos \left( \pi \frac{j}{N} \right) \right), \qquad j=0,...,N
\end{equation}
gives a nonlinear algebraic problem. This nonlinear algebraic problem is solved by means of a Newton-Raphson Method providing an initial guess. The constraint is incorporated in the boundary conditions. For this project we also subtract the leading logarithms as in \cite{Ammon:2016szz} and calculate all background data with 100 gridpoints, which we justify in the following discussion on QNMs. A detailed description of both the nonlinear system and the subtraction of leading logarithms can be found in the more general setup of \cite{Ammon:2016szz}.

To be able to use large gridsizes and to get high precision results we use the following advanced techniques in our numerical code. In the first step we carefully make sure to keep the equations of motion in a very short form, in particular after extracting the logarithms, terms like e.g. $u(z)^2 = \left(1 + z^4  u_4  + z^4\ln(z) \frac{B^2}{6} \right)^2$ should not be expanded, which can be the case when using simplify.

We than replace the functions $f(z) = \{ u(z), v(z), w(z), c(z), A_v(z), P(z) \}$ by symbolic quantities by $f = \{u, v, w, c, A_v, P \}$. Instead of replacing derivatives of $f$ by their spectral representation, we also replace $f'(z)$ by $df = \{du, dv, dw, dc, dA_v, dP \}$ and their second derivatives by $d^2f$. This has the advantage, that numerical values for the derivatives are calculated only once in each step using spectral differentiation matrices, such that e.g. the numerical value  of the symbol $du$ is calculated once and then replaced everywhere it appears in the equations of motion or in the Jacobian. The replacement is done with a faster Dispatch-rule in Mathematica instead of a trivial replacement-rule.

In addition we calculate the Jacobian only in the first step of the Newton-Raphson iteration symbolically and store it for the further iterations. The Jacobian is calculated using the chain rule, i.e. calculating 
\begin{equation}
\frac{\partial (EoM)_i}{\partial (f)_j} + \frac{\partial (EoM)_i}{\partial (df)_l} \frac{\partial (df)_l}{\partial (f)_j}  + \frac{\partial (EoM)_i}{\partial (d^2f)_l} \frac{\partial (d^2f)_l}{\partial (f)_j}\, ,
\end{equation}
where $\frac{\partial (df)_l}{\partial (f)_j}$ and $\frac{\partial (d^2f)_l}{\partial (f)_j}$ are given by a block matrix of the spectral differentiation matrix.

Moreover to calculate backgrounds with constant $\bar{T}$ and varying $\bar{B}$ or constant $\bar{B}$ and varying $\bar{T}$, we replace the boundary condition  $A_v(0) = \mu$ at the conformal boundary by a condition for the temperature $u'(1) = 4 \pi \mu \bar{T}$. This is equivalent to promoting $\mu$ to an extra parameter and implementing $u'(1) = 4 \pi \mu \bar{T}$ as an extra condition.

\subsection{Quasinormal Modes}
The fluctuation equations are discretized in the same way, where we now have to insert the numerical values for the previously calculated background. We use the second order differential equations to formulate the eigenvalue problem and include the constraints in the boundary conditions. The QNMs are determined as solution of a generalized eigenvalue problem
\begin{equation}
\left( A - \omega\, B \right ) \vec{x} = 0\, ,
\end{equation}
with respect to the right boundary conditions, see Sec.~\ref{subsec:QNMandnumericaldetails}. The vector $\vec{x}$ contains the values of all fields evaluated at the gridpoints, which represents our numerical solution for the QNM functions.
We force the boundary value of all fluctuations to be zero by factorizing out proper powers of $z$ 
\begin{equation}
 h_{ij}(z) = z^4 \tilde{h}_{ij}(z) \,, \qquad  a_{i}(z) = z^2 \tilde{a}_{i}(z) \, .
\end{equation}
Despite being in Eddington Finkelstein coordinates, the helicity-0 sector has a quadratic $\omega$ dependence, giving a nonlinear eigenvalue problem. This arises from additional $\omega$ contributions in the determinant of the metric. The helicity-1 and helicity-2 sectors are still linear in $\omega$. The quadratic eigenvalue problem can be reduced to a linear eigenvalue problem by doubling the number of fields, introducing $ h^{new}_{ij}(z) = \omega\, h_{ij}(z)$.

In addition to the numerical techniques for the background, we replace the derivatives of the background and the fluctuations again symbolically and calculate their values only once. Moreover we use the background equations of motion to eliminate all second derivatives of background fields in our QNM equations. To interpolate we use the fast and more accurate Clenshaw algorithm. The linear eigenvalue problems are solved with Mathematica Eigenvalue/Eigensystem, to get the QNM and the eigenfunctions respectively. All QNM data in this project are calculated also with 100 gridpoints and 60 digit precision, which means solving a $100\times100$ matrix for helicty-2, two $300\times300$ matrices for helicty-1 and a $800\times800$ matrix for helicty-0 sector.

By plugging back the eigenvalue and the eigenfunctions into the equation of motion, we check if we have a true solution of the full system including constraints.
In our analysis QNM appear, that converge and fulfil the equations of motion, but not the constraint(s). It happens that those QNM additionally do not change with k. Accordingly these QNM are identified as numerical artefacts and are rejected for the analysis. We refer them as fake QNM which also show up in the continued fraction method \cite{Starinets:2002br, Janiszewski:2011sx}. In particular, the fake QNM may be traced back to a degeneracy of the ingoing/outgoing horizon conditions in the Poicare chart. At the horizon, ingoing and outgoing modes have a term of the form $(1-z)^{\pm i \omega / u_1}$, which for $\omega = -i \, n \, u_1 / 2$, with $n\in \mathbb{N}$, is degnerate from the power series in $1-u.$ Taking into account $u_1 = 4\pi T$, the fake QNMs lie at $\omega =-2 \pi i\, T\, n$ with $n \in \mathbb{N}$, as expected.

\subsection{Helicity-2 sector}
\noindent
The helicity-2 sector is described by two decoupled equations of motion.
The convergence in helicity-2 sector is slow due to logarithms in the background and logarithms in the QNM eigenfunctions.
It further shows an even/odd gridsize oscillation, meaning that the QNM value for even/odd gridsizes approaches the correct value from above and below respectively.

Therefore we have to introduce a $z \mapsto z^2$ mapping. This moves the logarithms to higher orders
\begin{equation}
z^n \log(z) \mapsto 2\,z^{2n}\log(z) \,,
\end{equation}
which improves the convergence. Interestingly we find that this allows to use simplified boundary conditions at $z=0$. In principle one has to specify the new boundary condition $f'(z) = 0$ for all fields $f$ as well as the boundary condition before the mapping. In our case the condition $f'(z) = 0$ was sufficient to reproduce the results before the mapping, since our boundary conditions are solely given by the equations of motion.
All calculations, and in particular the convergence plots shown below, are obtained with this mapping.

\begin{figure}[ht]
	\centering
 \includegraphics[width = 0.3\textwidth]{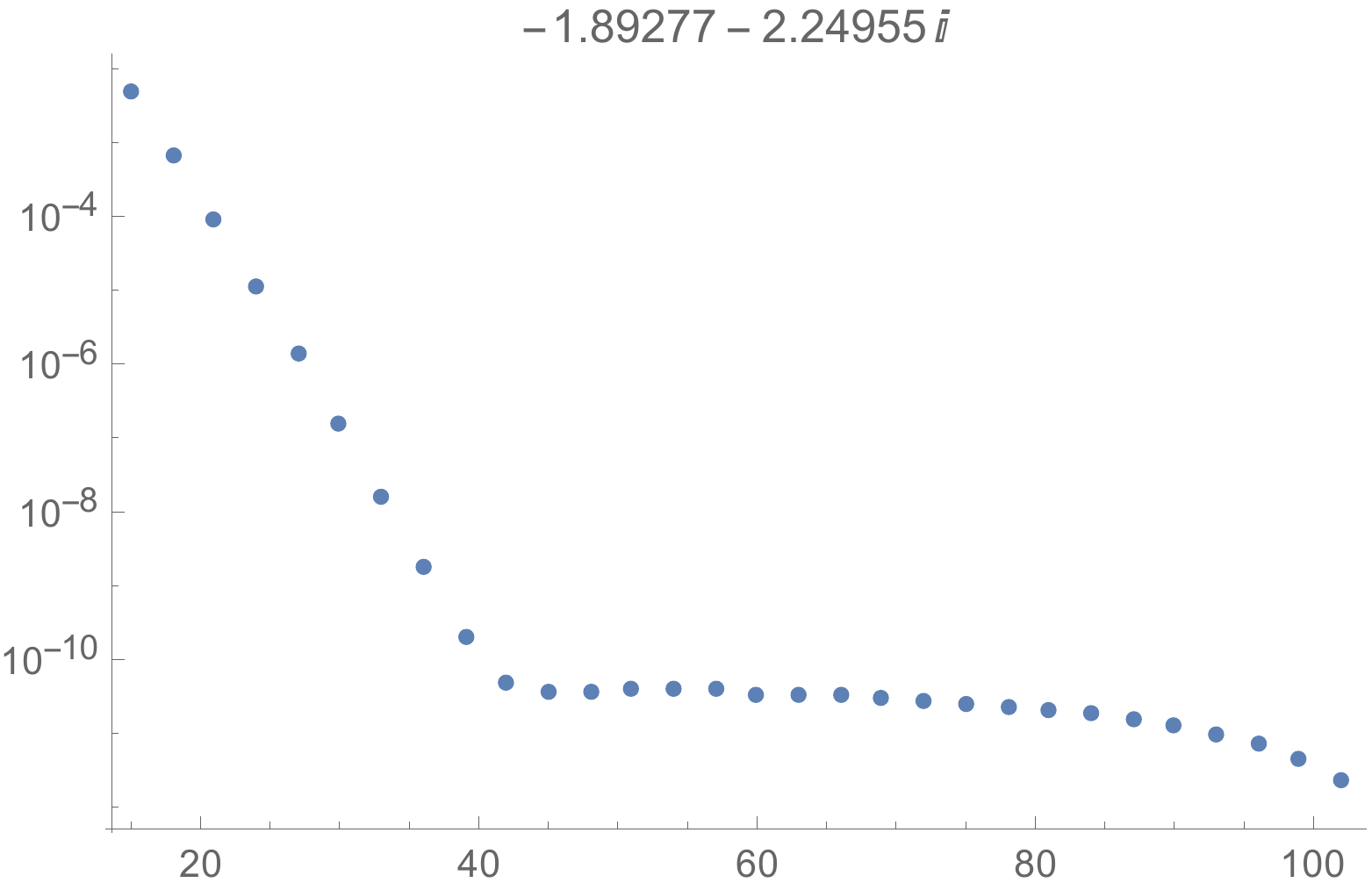}
 \includegraphics[width = 0.3\textwidth]{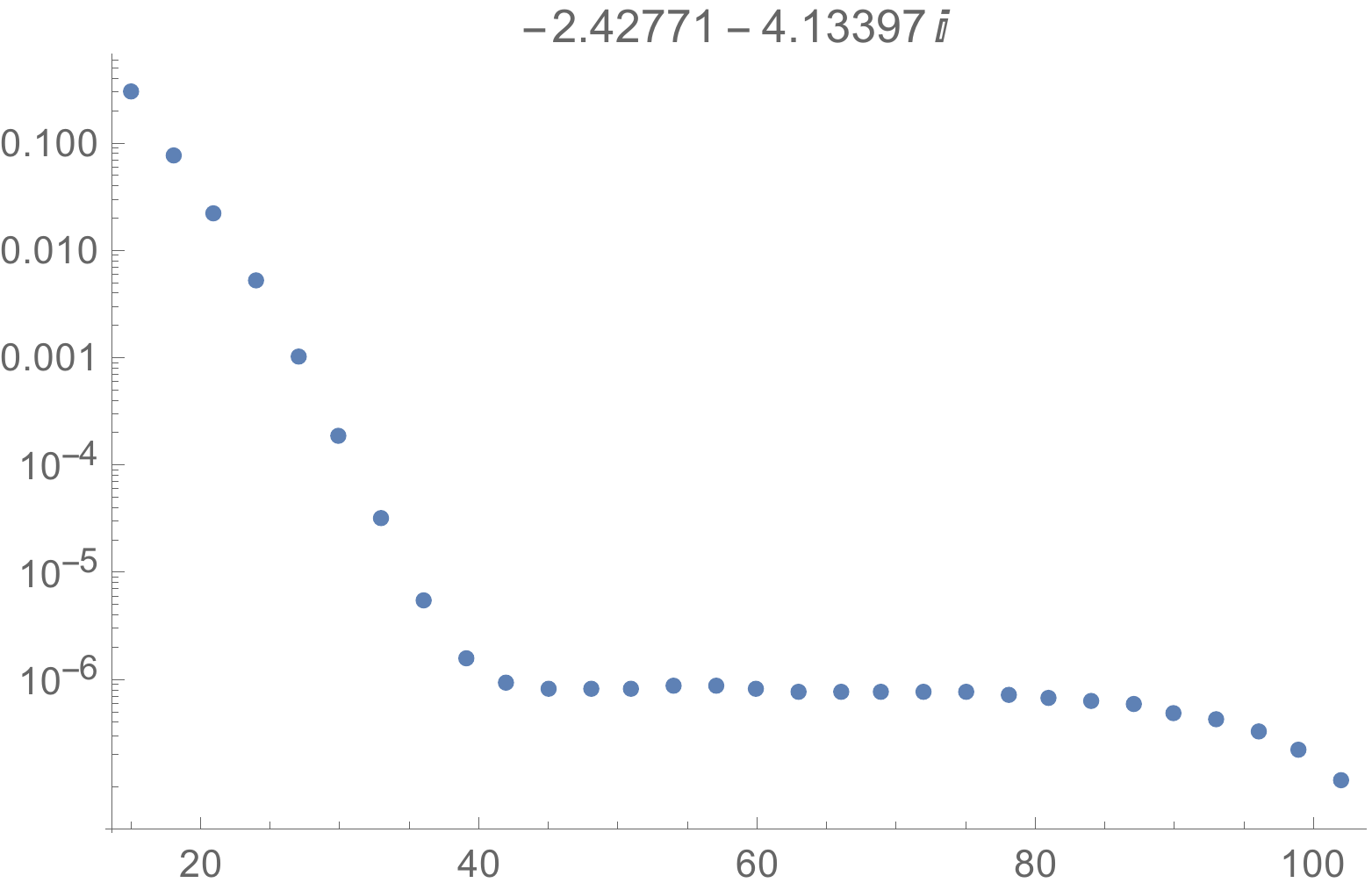}
  \includegraphics[width = 0.3\textwidth]{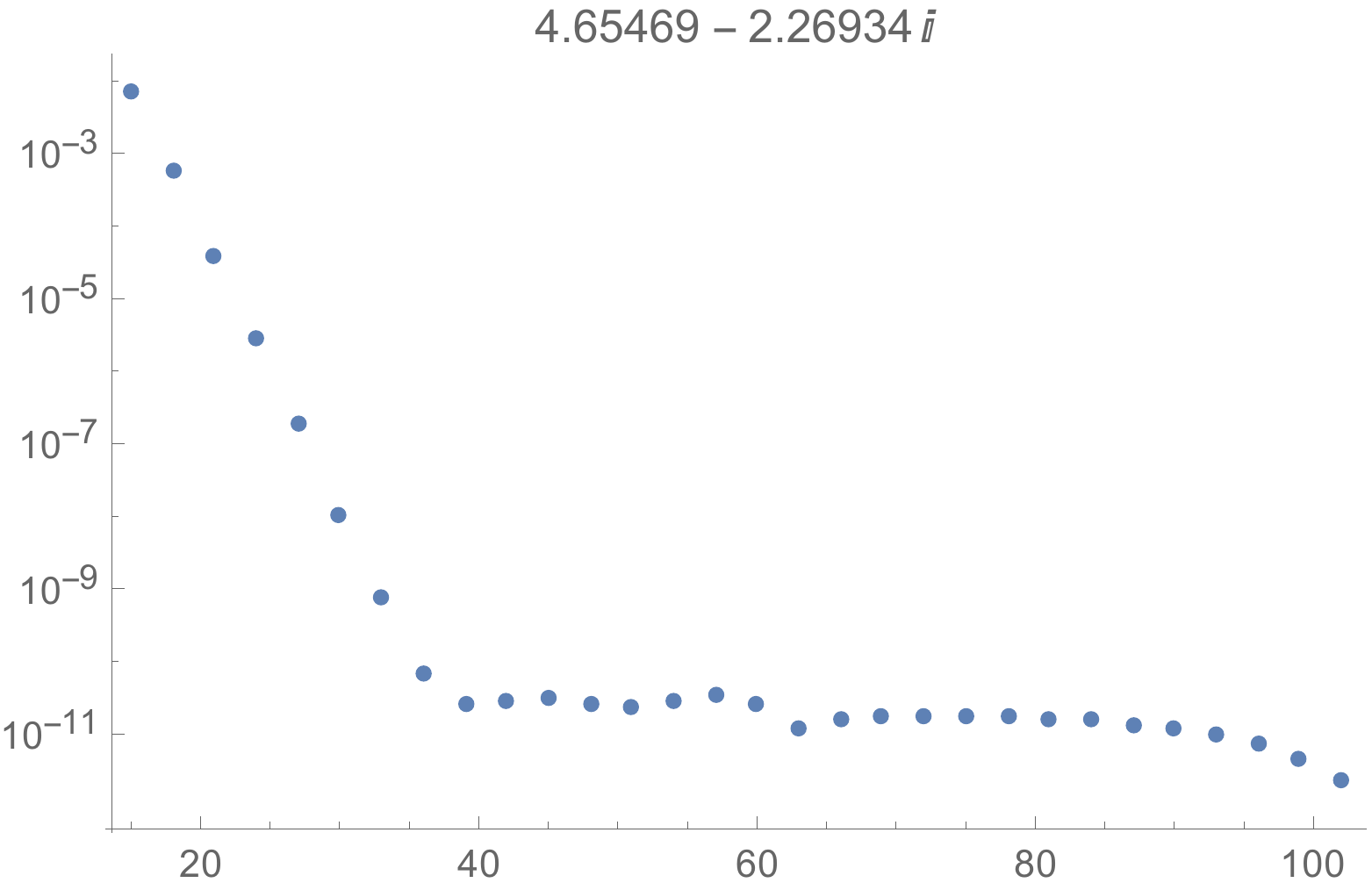}
	\caption{convergence test helicity-2 for $\tilde{k}=20$, $\tilde{B}=65$, $\bar{T} = 0.1$}
	\label{convergencespin2kmax}
\end{figure}

With the $z^2$ mapping we already obtain good results with 40 gridpoints, see \ref{convergencespin2kmax}. 

\subsection{Helicity-1 and helicity-0 sector}
\begin{figure}[ht]
	\centering
  \includegraphics[width = 0.3\textwidth]{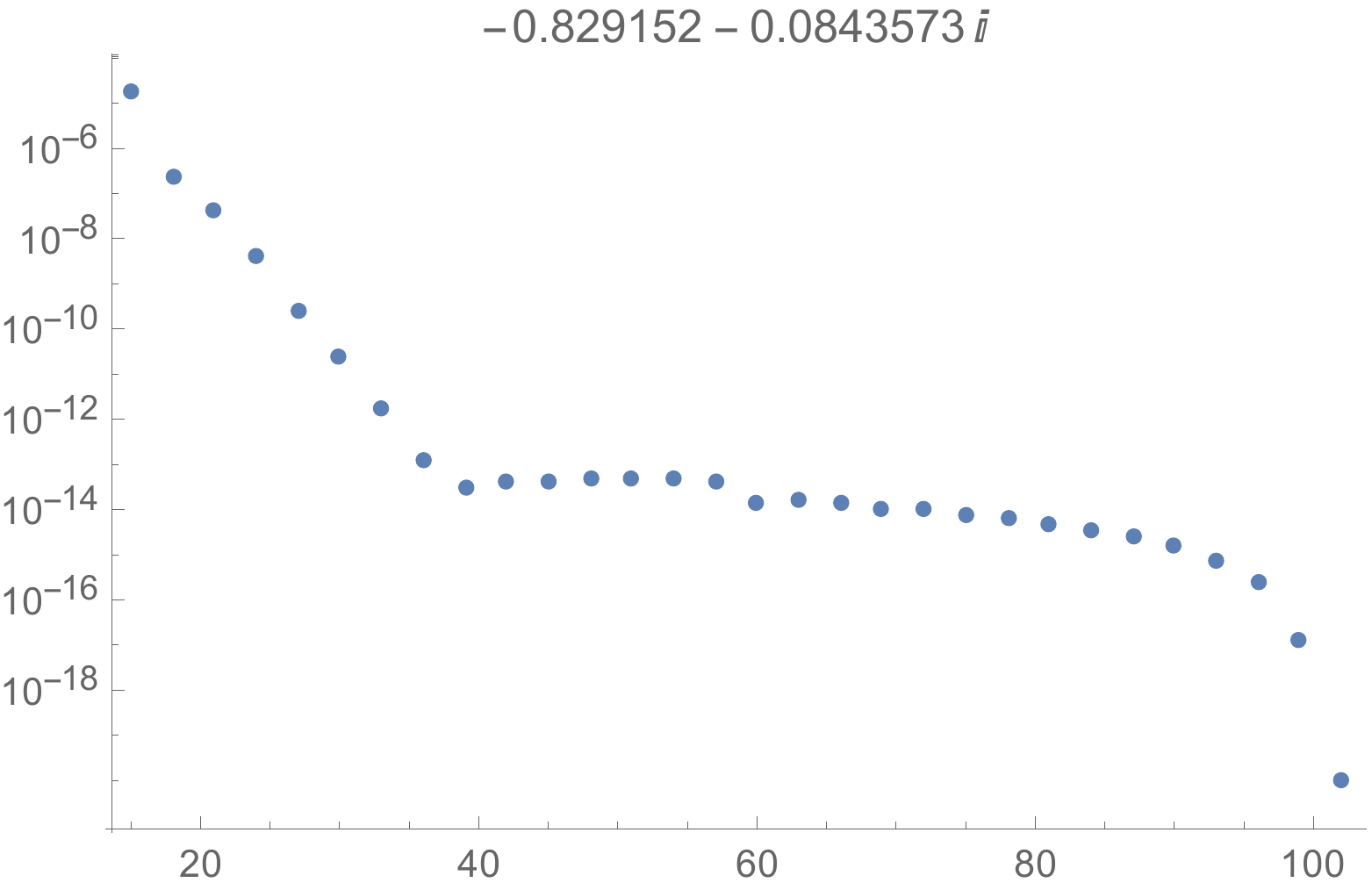}
  \includegraphics[width = 0.3\textwidth]{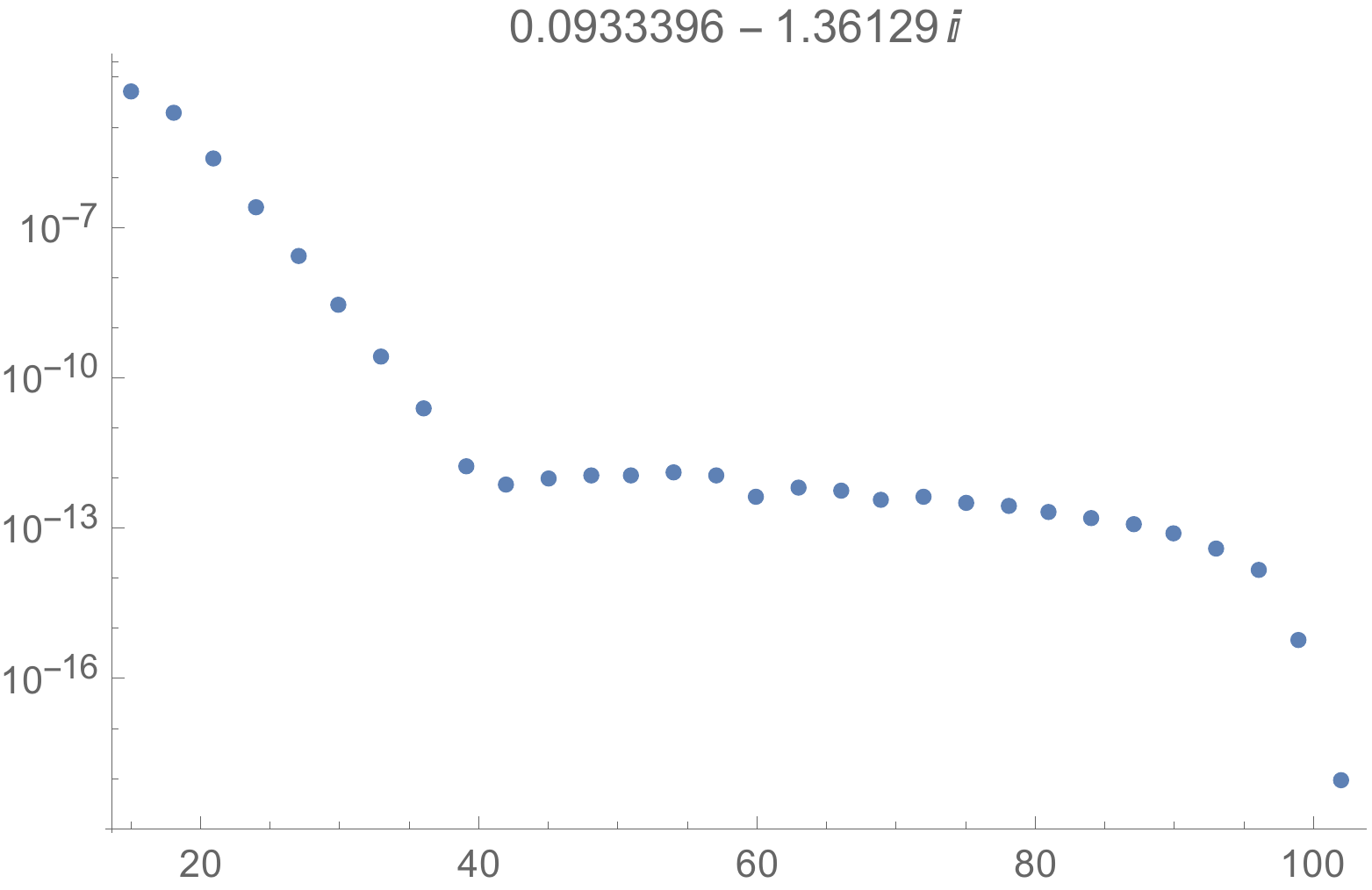}
  \includegraphics[width = 0.3\textwidth]{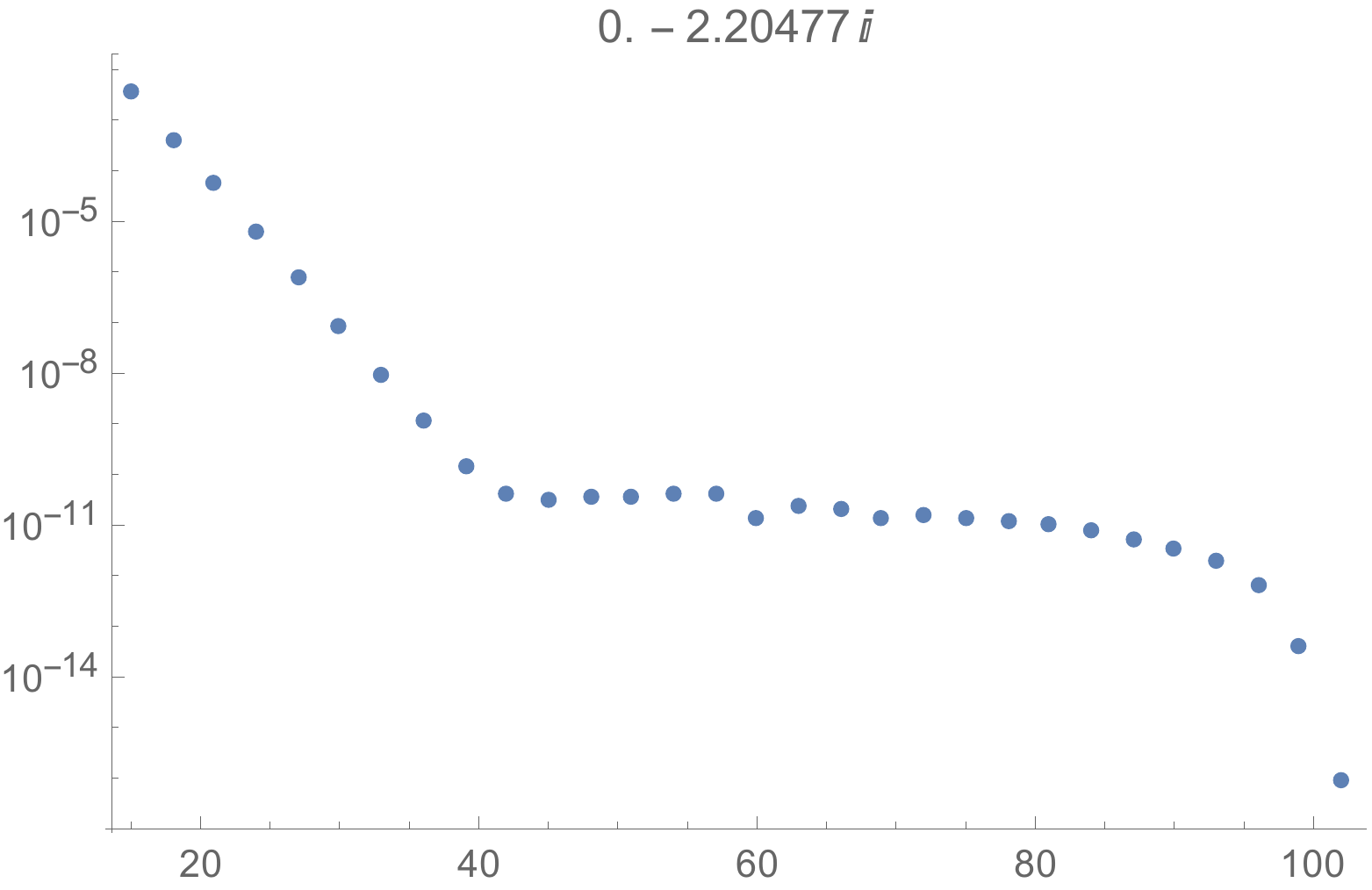}
	\caption{Convergence test helicity-$1^+$ for $\tilde{k}=0$, $\tilde{B}=65$, $\bar{T} = 0.1$}
	\label{convergencespin1pk0}
\end{figure}

\begin{figure}[ht]
	\centering
 \includegraphics[width = 0.3\textwidth]{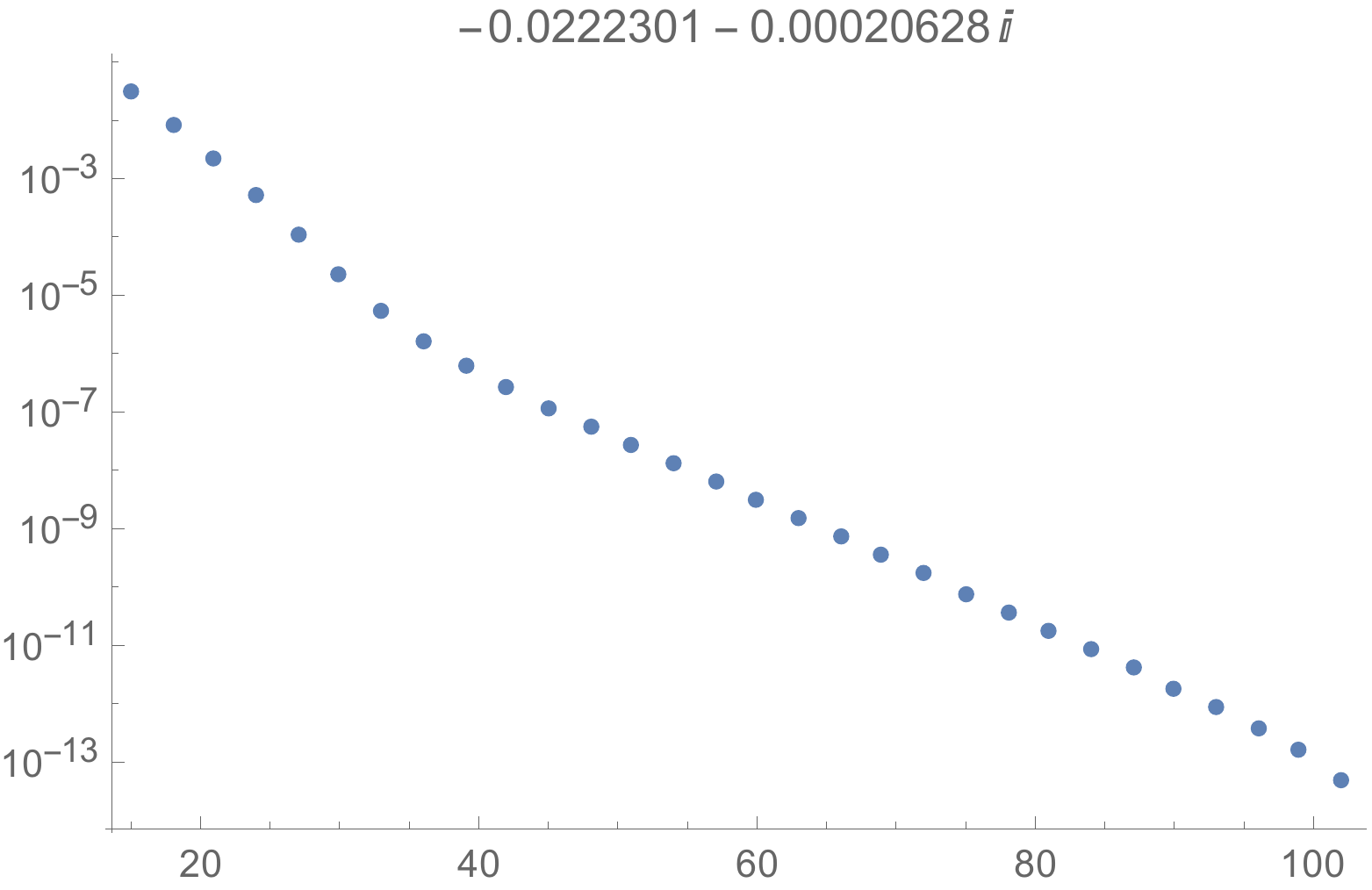}
 \includegraphics[width = 0.3\textwidth]{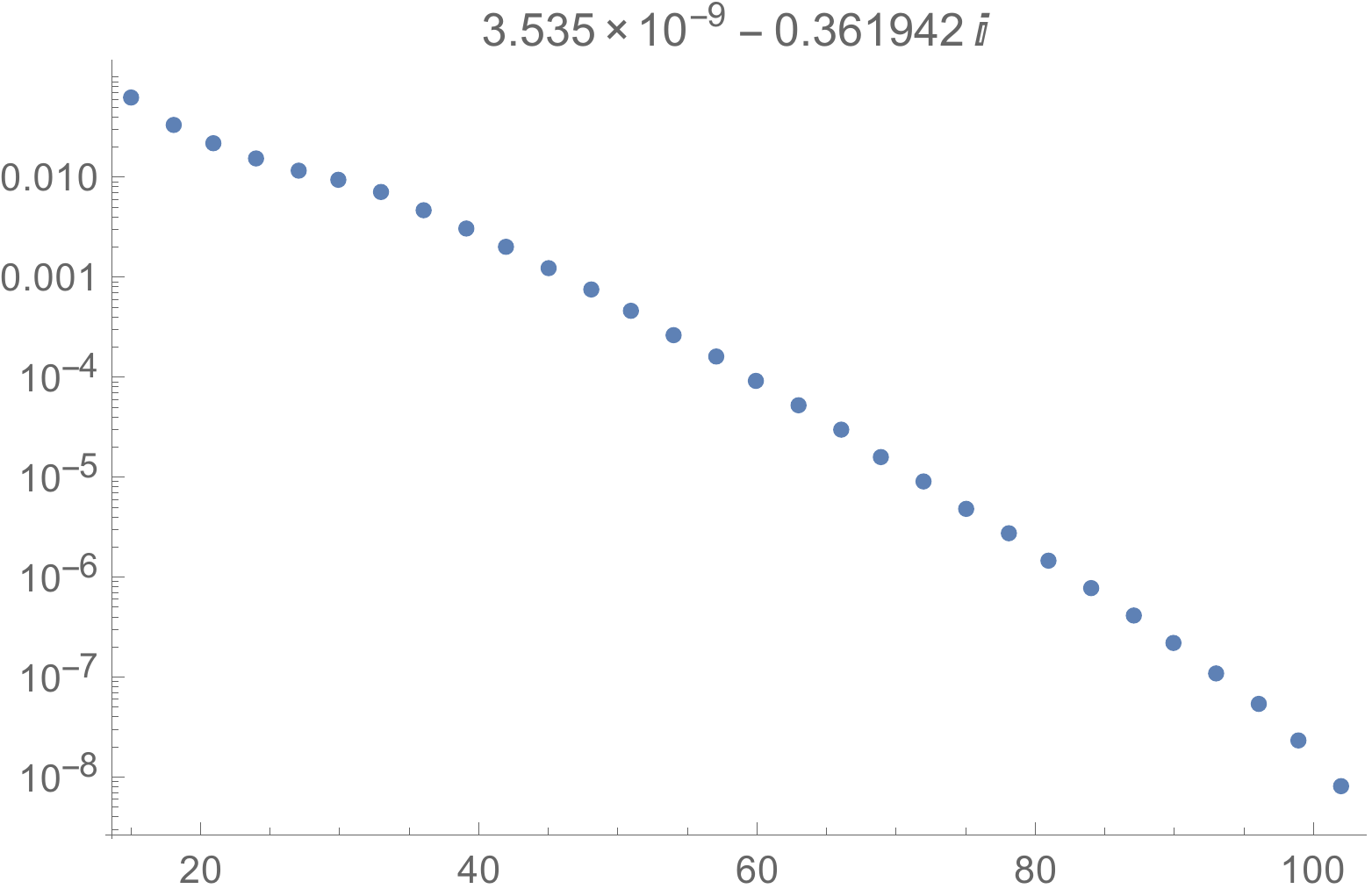}
 \includegraphics[width = 0.3\textwidth]{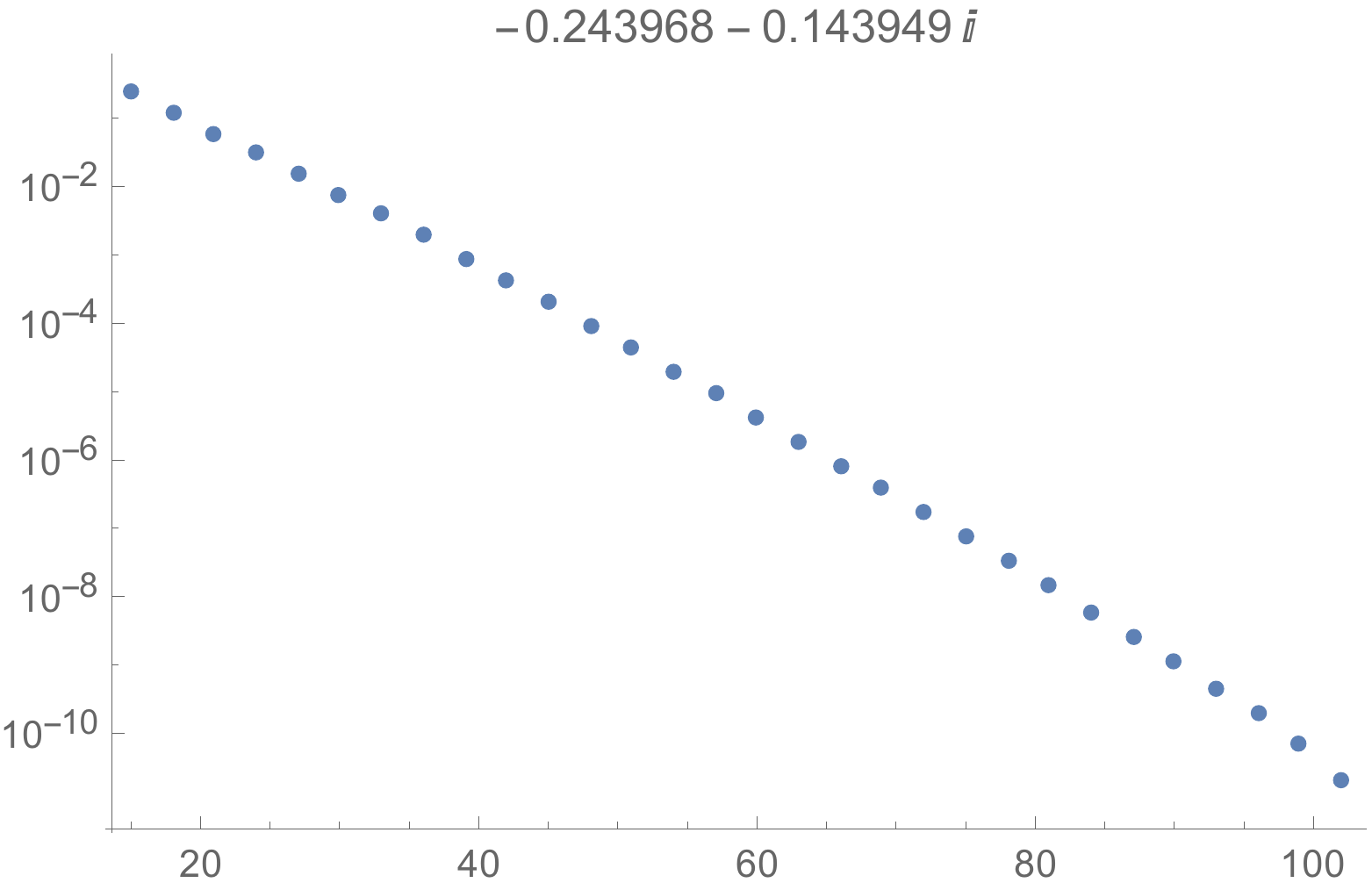}	\caption{Convergence test helicity-$1^+$ for $\tilde{k}=20$, $\tilde{B}=65$, $\bar{T} = 0.1$}
	\label{convergencespin1pkmax}
\end{figure}

Helicity-1 is given by two decoupled subsectors, helicity-$1^\pm$s. Each subsector consists of three equations of motion and one constraint and has to be solved separately. We exemplary show the lowest eigenmodes and the error in the constraint in table \ref{constraintcheck} for helicity-$1^-$. The error is obtained by plugging the corresponding eigenvector, which is the numerical solution for the fluctuations evaluated at the gridpoints, into the constraint equations and taking the norm. For example the QNM $ 0.  - 2.20477 \,i$ in table \ref{constraintcheck} does not fulfil the constraint and does also not change from $\tilde{k}=0$ to $\tilde{k}=20$. This mode is identified as a fake QNM.
\begin{table}
\begin{tabular}{ r || r }
  QNM $\omega$, $\tilde{k}=0$\, & $\Delta$ constraint  \\
  \hline
  0.829152 - 0.0843573 i \,& $3. \times 10^{-14}$  \\
 -0.0933396 - 1.36129 i \,& $4. \times 10^{-12}$  \\
  0.  - 2.20477 i \,& 14.  \\
  1.84644 - 1.34191 i \,&$7. \times 10^{-12}$ \\
  -0.267106 - 3.58127 i\,&$5. \times 10^{-8}$ \\
\end{tabular}
\qquad
\begin{tabular}{ r || r }
  QNM $\omega$, $\tilde{k}=20$\, & $\Delta$ constraint  \\
  \hline
  0. - 2.20477 i \,& 31.  \\
 -1.68677 - 1.54155 i \,& $4.\times10^{-11}$  \\
  3.41168 - 0.845203 i \,& $3.\times10^{-12}$  \\
 -2.23184 - 3.36554 i \,& $2.\times10^{-8}$  \\
 0. -4.40955 i \,& 43. \\
\end{tabular}
\caption{Constraint for $k=0$ and $\tilde{k}= 20$, $\tilde{B}=65$, $\bar{T} = 0.1$ in the helicity-$1^-$ sector}
\label{constraintcheck}
\end{table}
This mode also appears in helicty-$1^+$. Despite being a fake mode, it converges nicely, as shown in fig. \ref{convergencespin1pk0} 
In fig \ref{convergencespin1pkmax} we show the slower convergence for larger $\tilde{B}$ and $\tilde{k}$.
Comparing the convergence at the same point in parameter space $\tilde{B}=65$, $\bar{T} = 0.1$ for $\tilde{k}=0$, shown in fig. \ref{convergencespin1pk0}, and $\tilde{k}=20$, shown in fig. \ref{convergencespin1pkmax}, reveals the slowing down of the convergence for larger $\tilde{k}$,  which requires to use large gridsizes.


The helicity-0 sector is given by six equations of motion and four constraint.
It is in addition quadratically in $\omega$, therefore we introduce two additional fields,known as doubling trick, to get a linear system again. Convergence does not pose a problem in the helicity-0 sector, however calculations are time consuming compared to helicity-1 and helicity-2.

\subsection{Instabilities}
\begin{figure}[ht]
	\centering
  \includegraphics[width = 0.45\textwidth]{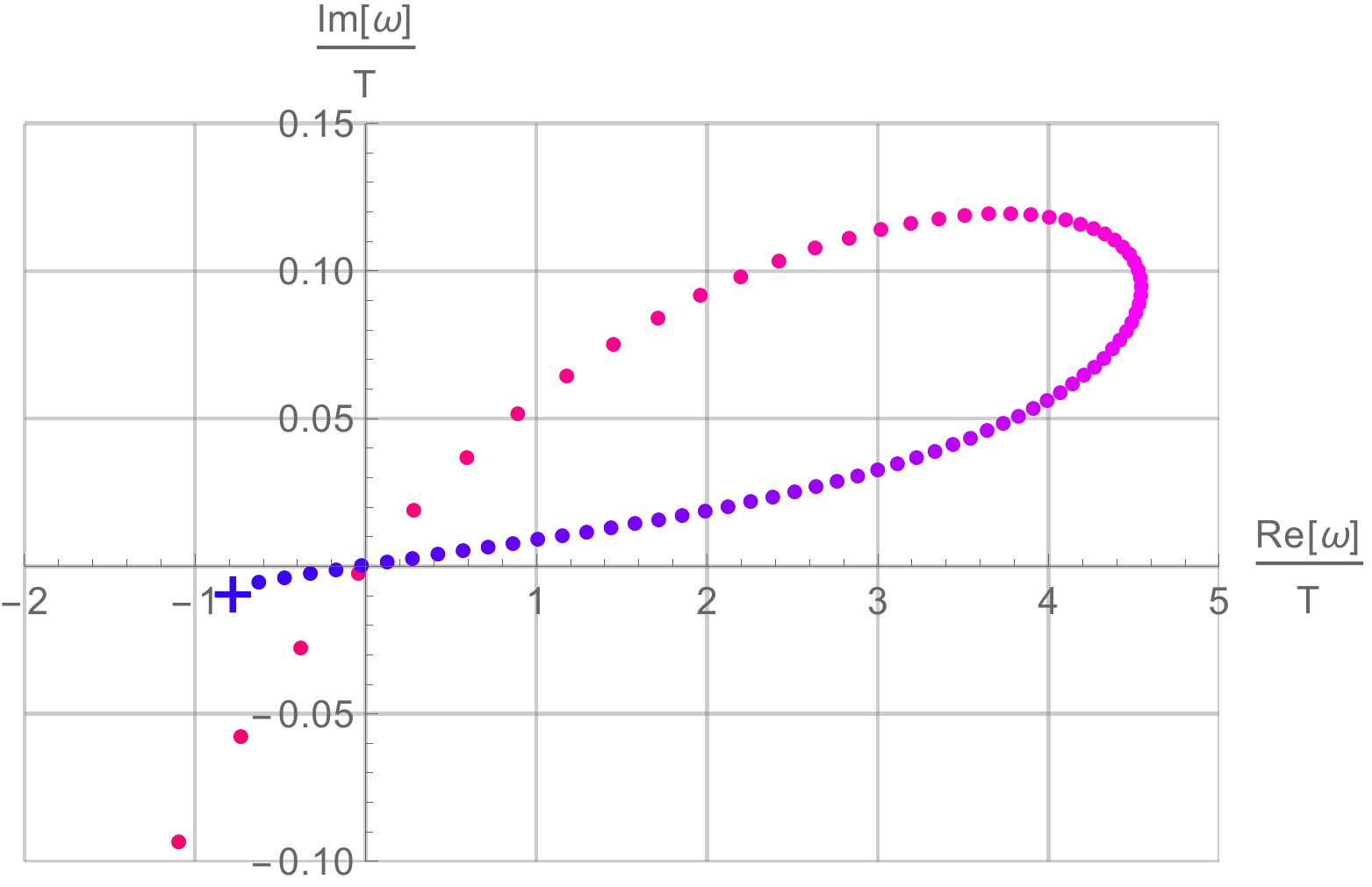}
  \includegraphics[width = 0.45\textwidth]{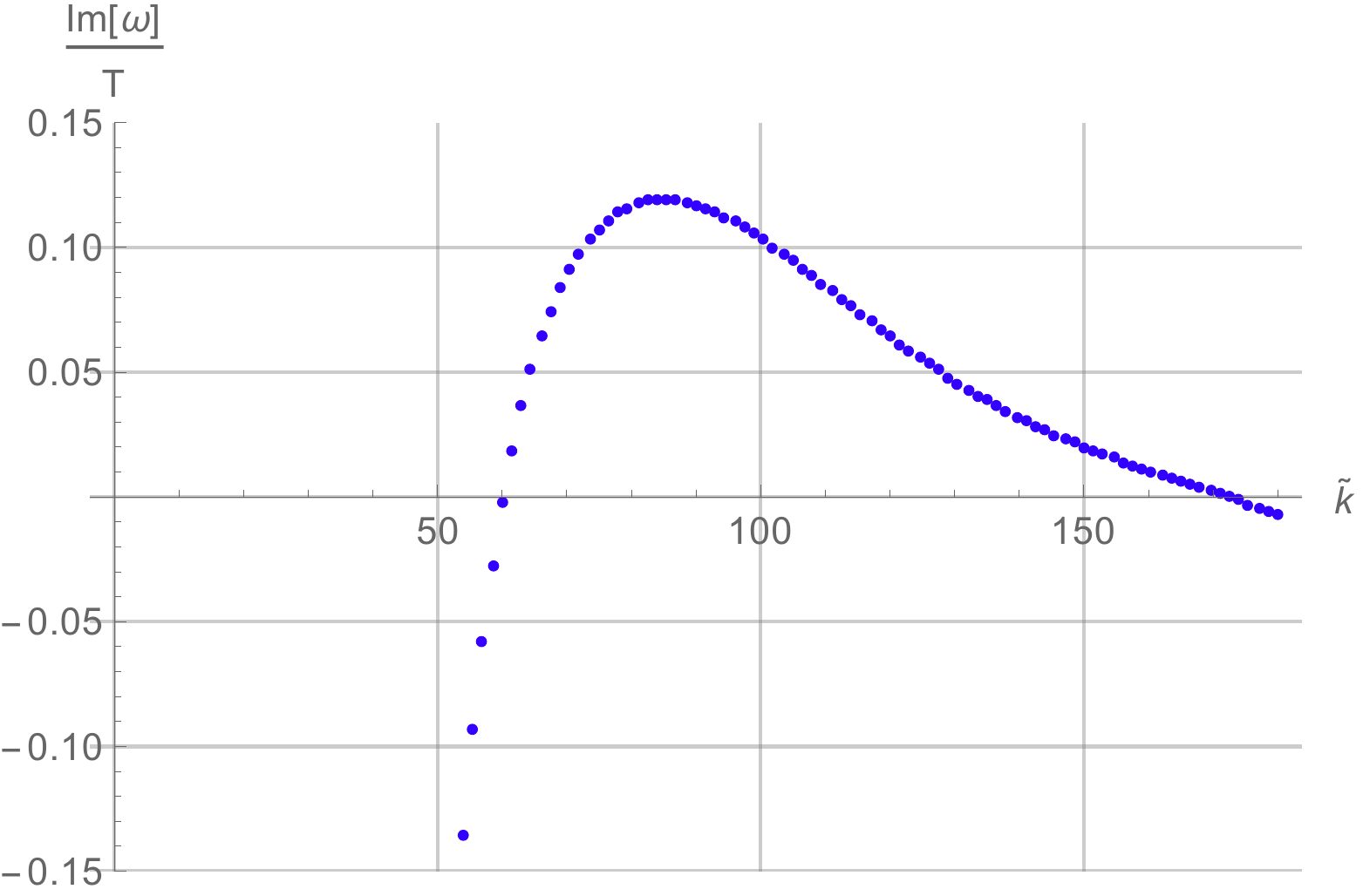}
	\caption{helical magnetic instabilty at $\tilde{k}=60-173$, $\bar{T}=0.01$, $B=2, \Tilde{B}= 69.44$, $\tilde{k}_{max} = 180$}
	\label{instability}
\end{figure}

We find only one QNM with positive imaginary part for very low temperatures $\bar{T}=0.01$ in the helicity-1 sector for finite $k$ \cite{Nakamura:2009tf}. This mode corresponds to the helical charged magnetic phase \citep{Donos:2012wi}, \cite{Ammon:2016szz}. We find the instability exactly in the $k$-range where the helical phase exists, see fig. \ref{instability}. We have not found any further instabilities in our numerical analysis.

\FloatBarrier


\end{appendix}


\bibliographystyle{JHEP}
\bibliography{EMBranes.v2}

\providecommand{\href}[2]{#2}\begingroup\raggedright\begin{thebibliography}{10}

\bibitem{Kharzeev:2004ey}
D.~Kharzeev, {\it {Parity violation in hot QCD: Why it can happen, and how to
  look for it}},  {\em Phys.Lett.} {\bf B633} (2006) 260--264,
  [\href{http://arxiv.org/abs/hep-ph/0406125}{{\tt hep-ph/0406125}}].

\bibitem{Fukushima:2008xe}
K.~Fukushima, D.~E. Kharzeev, and H.~J. Warringa, {\it {The Chiral Magnetic
  Effect}},  {\em Phys. Rev.} {\bf D78} (2008) 074033,
  [\href{http://arxiv.org/abs/0808.3382}{{\tt arXiv:0808.3382}}].

\bibitem{Son:2009tf}
D.~T. Son and P.~Surowka, {\it {Hydrodynamics with Triangle Anomalies}},  {\em
  Phys. Rev. Lett.} {\bf 103} (2009) 191601,
  [\href{http://arxiv.org/abs/0906.5044}{{\tt arXiv:0906.5044}}].

\bibitem{Erdmenger:2008rm}
J.~Erdmenger, M.~Haack, M.~Kaminski, and A.~Yarom, {\it {Fluid dynamics of
  R-charged black holes}},  {\em JHEP} {\bf 01} (2009) 055,
  [\href{http://arxiv.org/abs/0809.2488}{{\tt arXiv:0809.2488}}].

\bibitem{Banerjee:2008th}
N.~Banerjee et~al., {\it {Hydrodynamics from charged black branes}},  {\em
  JHEP} {\bf 01} (2011) 094, [\href{http://arxiv.org/abs/0809.2596}{{\tt
  arXiv:0809.2596}}].

\bibitem{Vilenkin:1978is}
A.~Vilenkin, {\it {Parity Nonconservation and Rotating Black Holes}},  {\em
  Phys.Rev.Lett.} {\bf 41} (1978) 1575--1577.

\bibitem{PhysRevD.22.3080}
A.~Vilenkin, {\it Equilibrium parity-violating current in a magnetic field},
  {\em Phys. Rev. D} {\bf 22} (Dec, 1980) 3080--3084.

\bibitem{Jensen:2013vta}
K.~Jensen, P.~Kovtun, and A.~Ritz, {\it {Chiral conductivities and effective
  field theory}},  {\em JHEP} {\bf 10} (2013) 186,
  [\href{http://arxiv.org/abs/1307.3234}{{\tt arXiv:1307.3234}}].

\bibitem{Kharzeev:2015znc}
D.~E. Kharzeev, J.~Liao, S.~A. Voloshin, and G.~Wang, {\it {Chiral magnetic and
  vortical effects in high-energy nuclear collisionsÑA status report}},  {\em
  Prog. Part. Nucl. Phys.} {\bf 88} (2016) 1--28,
  [\href{http://arxiv.org/abs/1511.04050}{{\tt arXiv:1511.04050}}].

\bibitem{2014arXiv1412.6543L}
Q.~{Li}, D.~E. {Kharzeev}, C.~{Zhang}, Y.~{Huang}, I.~{Pletikosic}, A.~V.
  {Fedorov}, R.~D. {Zhong}, J.~A. {Schneeloch}, G.~D. {Gu}, and T.~{Valla},
  {\it {Observation of the chiral magnetic effect in ZrTe5}},  {\em ArXiv
  e-prints} (Dec., 2014) [\href{http://arxiv.org/abs/1412.6543}{{\tt
  arXiv:1412.6543}}].

\bibitem{2015arXiv150606577S}
C.~{Shekhar}, F.~{Arnold}, S.-C. {Wu}, Y.~{Sun}, M.~{Schmidt}, N.~{Kumar},
  A.~G. {Grushin}, J.~H. {Bardarson}, R.~{Donizeth dos Reis}, M.~{Naumann},
  M.~{Baenitz}, H.~{Borrmann}, M.~{Nicklas}, E.~{Hassinger}, C.~{Felser}, and
  B.~{Yan}, {\it {Large and unsaturated negative magnetoresistance induced by
  the chiral anomaly in the Weyl semimetal TaP}},  {\em ArXiv e-prints} (June,
  2015) [\href{http://arxiv.org/abs/1506.06577}{{\tt arXiv:1506.06577}}].

\bibitem{2015arXiv150407698Z}
C.~{Zhang}, E.~{Zhang}, Y.~{Liu}, Z.-G. {Chen}, S.~{Liang}, J.~{Cao},
  X.~{Yuan}, L.~{Tang}, Q.~{Li}, T.~{Gu}, Y.~{Wu}, J.~{Zou}, and F.~{Xiu}, {\it
  {Detection of chiral anomaly and valley transport in Dirac semimetals}},
  {\em ArXiv e-prints} (Apr., 2015)
  [\href{http://arxiv.org/abs/1504.07698}{{\tt arXiv:1504.07698}}].

\bibitem{2015arXiv150600924W}
Z.~{Wang}, Y.~{Zheng}, Z.~{Shen}, Y.~{Zhou}, X.~{Yang}, Y.~{Li}, C.~{Feng}, and
  Z.-A. {Xu}, {\it {Helicity protected ultrahigh mobility Weyl fermions in
  NbP}},  {\em ArXiv e-prints} (June, 2015)
  [\href{http://arxiv.org/abs/1506.00924}{{\tt arXiv:1506.00924}}].

\bibitem{2015arXiv150603190Y}
X.~{Yang}, Y.~{Liu}, Z.~{Wang}, Y.~{Zheng}, and Z.-a. {Xu}, {\it {Chiral
  anomaly induced negative magnetoresistance in topological Weyl semimetal
  NbAs}},  {\em ArXiv e-prints} (June, 2015)
  [\href{http://arxiv.org/abs/1506.03190}{{\tt arXiv:1506.03190}}].

\bibitem{Kharzeev:2010gr}
D.~E. Kharzeev and D.~T. Son, {\it {Testing the chiral magnetic and chiral
  vortical effects in heavy ion collisions}},  {\em Phys.Rev.Lett.} {\bf 106}
  (2011) 062301, [\href{http://arxiv.org/abs/1010.0038}{{\tt
  arXiv:1010.0038}}].

\bibitem{Khachatryan:2016got}
{\bf CMS} Collaboration, V.~Khachatryan et~al., {\it {Observation of
  charge-dependent azimuthal correlations in pPb collisions and its implication
  for the search for the chiral magnetic effect}},  {\em Submitted to: Phys.
  Rev. Lett} (2016) [\href{http://arxiv.org/abs/1610.00263}{{\tt
  arXiv:1610.00263}}].

\bibitem{Abelev:2009ac}
{\bf STAR} Collaboration, B.~I. Abelev et~al., {\it {Azimuthal Charged-Particle
  Correlations and Possible Local Strong Parity Violation}},  {\em Phys. Rev.
  Lett.} {\bf 103} (2009) 251601, [\href{http://arxiv.org/abs/0909.1739}{{\tt
  arXiv:0909.1739}}].

\bibitem{Abelev:2009ad}
{\bf STAR} Collaboration, B.~I. Abelev et~al., {\it {Observation of
  charge-dependent azimuthal correlations and possible local strong parity
  violation in heavy ion collisions}},  {\em Phys. Rev.} {\bf C81} (2010)
  054908, [\href{http://arxiv.org/abs/0909.1717}{{\tt arXiv:0909.1717}}].

\bibitem{Adamczyk:2013kcb}
{\bf STAR} Collaboration, L.~Adamczyk et~al., {\it {Measurement of charge
  multiplicity asymmetry correlations in high-energy nucleus-nucleus collisions
  at $\sqrt{{s}_{NN}} =$ 200 GeV}},  {\em Phys. Rev.} {\bf C89} (2014), no.~4
  044908, [\href{http://arxiv.org/abs/1303.0901}{{\tt arXiv:1303.0901}}].

\bibitem{Adamczyk:2014mzf}
{\bf STAR} Collaboration, L.~Adamczyk et~al., {\it {Beam-energy dependence of
  charge separation along the magnetic field in Au+Au collisions at RHIC}},
  {\em Phys. Rev. Lett.} {\bf 113} (2014) 052302,
  [\href{http://arxiv.org/abs/1404.1433}{{\tt arXiv:1404.1433}}].

\bibitem{Adamczyk:2013hsi}
{\bf STAR} Collaboration, L.~Adamczyk et~al., {\it {Fluctuations of charge
  separation perpendicular to the event plane and local parity violation in
  $\sqrt{s_{NN}}=200$ GeV Au+Au collisions at the BNL Relativistic Heavy Ion
  Collider}},  {\em Phys. Rev.} {\bf C88} (2013), no.~6 064911,
  [\href{http://arxiv.org/abs/1302.3802}{{\tt arXiv:1302.3802}}].

\bibitem{Abelev:2012pa}
{\bf ALICE} Collaboration, B.~Abelev et~al., {\it {Charge separation relative
  to the reaction plane in Pb-Pb collisions at $\sqrt{s_{NN}}= 2.76$ TeV}},
  {\em Phys. Rev. Lett.} {\bf 110} (2013), no.~1 012301,
  [\href{http://arxiv.org/abs/1207.0900}{{\tt arXiv:1207.0900}}].

\bibitem{Wang:2009kd}
F.~Wang, {\it {Effects of Cluster Particle Correlations on Local Parity
  Violation Observables}},  {\em Phys. Rev.} {\bf C81} (2010) 064902,
  [\href{http://arxiv.org/abs/0911.1482}{{\tt arXiv:0911.1482}}].

\bibitem{Bzdak:2010fd}
A.~Bzdak, V.~Koch, and J.~Liao, {\it {Azimuthal correlations from transverse
  momentum conservation and possible local parity violation}},  {\em Phys.
  Rev.} {\bf C83} (2011) 014905, [\href{http://arxiv.org/abs/1008.4919}{{\tt
  arXiv:1008.4919}}].

\bibitem{Schlichting:2010qia}
S.~Schlichting and S.~Pratt, {\it {Charge conservation at energies available at
  the BNL Relativistic Heavy Ion Collider and contributions to local parity
  violation observables}},  {\em Phys. Rev.} {\bf C83} (2011) 014913,
  [\href{http://arxiv.org/abs/1009.4283}{{\tt arXiv:1009.4283}}].

\bibitem{Abbasi:2015saa}
N.~Abbasi, A.~Davody, K.~Hejazi, and Z.~Rezaei, {\it {Hydrodynamic Waves in an
  Anomalous Charged Fluid}},  {\em Phys. Lett.} {\bf B762} (2016) 23--32,
  [\href{http://arxiv.org/abs/1509.08878}{{\tt arXiv:1509.08878}}].

\bibitem{Kalaydzhyan:2016dyr}
T.~Kalaydzhyan and E.~Murchikova, {\it {Thermal chiral vortical and magnetic
  waves: new excitation modes in chiral fluids}},
  \href{http://arxiv.org/abs/1609.00024}{{\tt arXiv:1609.00024}}.

\bibitem{Abbasi:2016rds}
N.~Abbasi, D.~Allahbakhshi, A.~Davody, and S.~F. Taghavi, {\it {Collective
  Excitations in QCD Plasma}},  \href{http://arxiv.org/abs/1612.08614}{{\tt
  arXiv:1612.08614}}.

\bibitem{Huang:2011dc}
X.-G. Huang, A.~Sedrakian, and D.~H. Rischke, {\it {Kubo formulae for
  relativistic fluids in strong magnetic fields}},  {\em Annals Phys.} {\bf
  326} (2011) 3075--3094, [\href{http://arxiv.org/abs/1108.0602}{{\tt
  arXiv:1108.0602}}].

\bibitem{Kovtun:2016lfw}
P.~Kovtun, {\it {Thermodynamics of polarized relativistic matter}},  {\em JHEP}
  {\bf 07} (2016) 028, [\href{http://arxiv.org/abs/1606.01226}{{\tt
  arXiv:1606.01226}}].

\bibitem{Israel:1978up}
W.~Israel, {\it {The Dynamics of Polarization}},  {\em Gen. Rel. Grav.} {\bf 9}
  (1978) 451--468.

\bibitem{Kharzeev:2010gd}
D.~E. Kharzeev and H.-U. Yee, {\it {Chiral Magnetic Wave}},  {\em Phys. Rev.}
  {\bf D83} (2011) 085007, [\href{http://arxiv.org/abs/1012.6026}{{\tt
  arXiv:1012.6026}}].

\bibitem{Maldacena:1997re}
J.~M. Maldacena, {\it {The Large N limit of superconformal field theories and
  supergravity}},  {\em Int. J. Theor. Phys.} {\bf 38} (1999) 1113--1133,
  [\href{http://arxiv.org/abs/hep-th/9711200}{{\tt hep-th/9711200}}]. [Adv.
  Theor. Math. Phys.2,231(1998)].

\bibitem{Ammon:2015wua}
M.~Ammon and J.~Erdmenger, {\em {Gauge/gravity duality}}.
\newblock Cambridge Univ. Pr., Cambridge, UK, 2015.

\bibitem{Nastase}
H.~Nastase, {\em {Introduction to the AdS/CFT Correspondence}}.
\newblock Cambridge Univ. Pr., Cambridge, UK, 2015.

\bibitem{Schalm}
J.~Zaanen, Y.~Liu, Y.-W. Sun, and K.~Schalm, {\em {Holographic Duality in
  Condensed Matter Physics}}.
\newblock Cambridge Univ. Pr., Cambridge, UK, 2016.

\bibitem{Heinz:2008tv}
U.~W. Heinz, {\it {The Strongly coupled quark-gluon plasma created at RHIC}},
  {\em J. Phys.} {\bf A42} (2009) 214003,
  [\href{http://arxiv.org/abs/0810.5529}{{\tt arXiv:0810.5529}}].

\bibitem{Landsteiner:2015lsa}
K.~Landsteiner and Y.~Liu, {\it {The holographic Weyl semi-metal}},  {\em Phys.
  Lett.} {\bf B753} (2016) 453--457,
  [\href{http://arxiv.org/abs/1505.04772}{{\tt arXiv:1505.04772}}].

\bibitem{Landsteiner:2015pdh}
K.~Landsteiner, Y.~Liu, and Y.-W. Sun, {\it {Quantum phase transition between a
  topological and a trivial semimetal from holography}},  {\em Phys. Rev.
  Lett.} {\bf 116} (2016), no.~8 081602,
  [\href{http://arxiv.org/abs/1511.05505}{{\tt arXiv:1511.05505}}].

\bibitem{Copetti:2016ewq}
C.~Copetti, J.~Fern‡ndez-Pend‡s, and K.~Landsteiner, {\it {Axial Hall effect
  and universality of holographic Weyl semi-metals}},
  \href{http://arxiv.org/abs/1611.08125}{{\tt arXiv:1611.08125}}.

\bibitem{Ammon:2016mwa}
M.~Ammon, M.~Heinrich, A.~JimŽnez-Alba, and S.~Moeckel, {\it {Surface States in
  Holographic Weyl Semimetals}},  \href{http://arxiv.org/abs/1612.00836}{{\tt
  arXiv:1612.00836}}.

\bibitem{Grignani:2016wyz}
G.~Grignani, A.~Marini, F.~Pena-Benitez, and S.~Speziali, {\it {AC conductivity
  for a holographic Weyl Semimetal}},
  \href{http://arxiv.org/abs/1612.00486}{{\tt arXiv:1612.00486}}.

\bibitem{Kharzeev:2016mvi}
D.~Kharzeev, Y.~Kikuchi, and R.~Meyer, {\it {Chiral magnetic effect without
  chirality source in asymmetric Weyl semimetals}},
  \href{http://arxiv.org/abs/1610.08986}{{\tt arXiv:1610.08986}}.

\bibitem{Berti:2009kk}
E.~Berti, V.~Cardoso, and A.~O. Starinets, {\it {Quasinormal modes of black
  holes and black branes}},  {\em Class.Quant.Grav.} {\bf 26} (2009) 163001,
  [\href{http://arxiv.org/abs/0905.2975}{{\tt arXiv:0905.2975}}].

\bibitem{Konoplya:2011qq}
R.~A. Konoplya and A.~Zhidenko, {\it {Quasinormal modes of black holes: From
  astrophysics to string theory}},  {\em Rev. Mod. Phys.} {\bf 83} (2011)
  793--836, [\href{http://arxiv.org/abs/1102.4014}{{\tt arXiv:1102.4014}}].

\bibitem{Bu:2016vum}
Y.~Bu, M.~Lublinsky, and A.~Sharon, {\it {Anomalous transport from holography:
  Part II}},  \href{http://arxiv.org/abs/1609.09054}{{\tt arXiv:1609.09054}}.

\bibitem{Bu:2016oba}
Y.~Bu, M.~Lublinsky, and A.~Sharon, {\it {Anomalous transport from holography:
  Part I}},  {\em JHEP} {\bf 11} (2016) 093,
  [\href{http://arxiv.org/abs/1608.08595}{{\tt arXiv:1608.08595}}].

\bibitem{Heller:2013oxa}
M.~P. Heller, D.~Mateos, W.~van~der Schee, and M.~Triana, {\it {Holographic
  isotropization linearized}},  {\em JHEP} {\bf 09} (2013) 026,
  [\href{http://arxiv.org/abs/1304.5172}{{\tt arXiv:1304.5172}}].

\bibitem{Buchel:2015saa}
A.~Buchel, M.~P. Heller, and R.~C. Myers, {\it {Equilibration rates in a
  strongly coupled nonconformal quark-gluon plasma}},  {\em Phys. Rev. Lett.}
  {\bf 114} (2015), no.~25 251601, [\href{http://arxiv.org/abs/1503.07114}{{\tt
  arXiv:1503.07114}}].

\bibitem{Fuini:2015hba}
J.~F. Fuini and L.~G. Yaffe, {\it {Far-from-equilibrium dynamics of a strongly
  coupled non-Abelian plasma with non-zero charge density or external magnetic
  field}},  {\em JHEP} {\bf 07} (2015) 116,
  [\href{http://arxiv.org/abs/1503.07148}{{\tt arXiv:1503.07148}}].

\bibitem{Janiszewski:2015ura}
S.~Janiszewski and M.~Kaminski, {\it {Quasinormal modes of magnetic and
  electric black branes versus far from equilibrium anisotropic fluids}},  {\em
  Phys. Rev.} {\bf D93} (2016), no.~2 025006,
  [\href{http://arxiv.org/abs/1508.06993}{{\tt arXiv:1508.06993}}].

\bibitem{Janik:2016btb}
R.~A. Janik, J.~Jankowski, and H.~Soltanpanahi, {\it {Quasinormal modes and the
  phase structure of strongly coupled matter}},  {\em JHEP} {\bf 06} (2016)
  047, [\href{http://arxiv.org/abs/1603.05950}{{\tt arXiv:1603.05950}}].

\bibitem{Attems:2016ugt}
M.~Attems, J.~Casalderrey-Solana, D.~Mateos, I.~Papadimitriou,
  D.~Santos-Oliván, C.~F. Sopuerta, M.~Triana, and M.~Zilhão, {\it
  {Thermodynamics, transport and relaxation in non-conformal theories}},  {\em
  JHEP} {\bf 10} (2016) 155, [\href{http://arxiv.org/abs/1603.01254}{{\tt
  arXiv:1603.01254}}].

\bibitem{Son:2002sd}
D.~T. Son and A.~O. Starinets, {\it {Minkowski space correlators in AdS / CFT
  correspondence: Recipe and applications}},  {\em JHEP} {\bf 09} (2002) 042,
  [\href{http://arxiv.org/abs/hep-th/0205051}{{\tt hep-th/0205051}}].

\bibitem{Kovtun:2005ev}
P.~K. Kovtun and A.~O. Starinets, {\it {Quasinormal modes and holography}},
  {\em Phys. Rev.} {\bf D72} (2005) 086009,
  [\href{http://arxiv.org/abs/hep-th/0506184}{{\tt hep-th/0506184}}].

\bibitem{Nunez:2003eq}
A.~Nunez and A.~O. Starinets, {\it {AdS / CFT correspondence, quasinormal
  modes, and thermal correlators in N=4 SYM}},  {\em Phys. Rev.} {\bf D67}
  (2003) 124013, [\href{http://arxiv.org/abs/hep-th/0302026}{{\tt
  hep-th/0302026}}].

\bibitem{Sahoo:2009yq}
B.~Sahoo and H.-U. Yee, {\it {Holographic chiral shear waves from anomaly}},
  {\em Phys. Lett.} {\bf B689} (2010) 206--212,
  [\href{http://arxiv.org/abs/0910.5915}{{\tt arXiv:0910.5915}}].

\bibitem{Matsuo:2009xn}
Y.~Matsuo, S.-J. Sin, S.~Takeuchi, and T.~Tsukioka, {\it {Magnetic conductivity
  and Chern-Simons Term in Holographic Hydrodynamics of Charged AdS Black
  Hole}},  {\em JHEP} {\bf 04} (2010) 071,
  [\href{http://arxiv.org/abs/0910.3722}{{\tt arXiv:0910.3722}}].

\bibitem{Hur:2008tq}
J.~Hur, K.~K. Kim, and S.-J. Sin, {\it {Hydrodynamics with conserved current
  from the gravity dual}},  {\em JHEP} {\bf 0903} (2009) 036,
  [\href{http://arxiv.org/abs/0809.4541}{{\tt arXiv:0809.4541}}].

\bibitem{Matsuo:2009yu}
Y.~Matsuo, S.-J. Sin, S.~Takeuchi, T.~Tsukioka, and C.-M. Yoo, {\it {Sound
  Modes in Holographic Hydrodynamics for Charged AdS Black Hole}},  {\em Nucl.
  Phys.} {\bf B820} (2009) 593--619,
  [\href{http://arxiv.org/abs/0901.0610}{{\tt arXiv:0901.0610}}].

\bibitem{DHoker:2009ixq}
E.~D'Hoker and P.~Kraus, {\it {Charged Magnetic Brane Solutions in AdS (5) and
  the fate of the third law of thermodynamics}},  {\em JHEP} {\bf 03} (2010)
  095, [\href{http://arxiv.org/abs/0911.4518}{{\tt arXiv:0911.4518}}].

\bibitem{Buchel:2006gb}
A.~Buchel and J.~T. Liu, {\it {Gauged supergravity from type IIB string theory
  on Y**p,q manifolds}},  {\em Nucl. Phys.} {\bf B771} (2007) 93--112,
  [\href{http://arxiv.org/abs/hep-th/0608002}{{\tt hep-th/0608002}}].

\bibitem{Gauntlett:2006ai}
J.~P. Gauntlett, E.~O~Colgain, and O.~Varela, {\it {Properties of some
  conformal field theories with M-theory duals}},  {\em JHEP} {\bf 02} (2007)
  049, [\href{http://arxiv.org/abs/hep-th/0611219}{{\tt hep-th/0611219}}].

\bibitem{Gauntlett:2007ma}
J.~P. Gauntlett and O.~Varela, {\it {Consistent Kaluza-Klein reductions for
  general supersymmetric AdS solutions}},  {\em Phys. Rev.} {\bf D76} (2007)
  126007, [\href{http://arxiv.org/abs/0707.2315}{{\tt arXiv:0707.2315}}].

\bibitem{Colgain:2014pha}
E.~O. Colg\'{a}in, M.~M. Sheikh-Jabbari, J.~F. V\'{a}zquez-Poritz,
  H.~Yavartanoo, and Z.~Zhang, {\it {Warped Ricci-flat reductions}},  {\em
  Phys. Rev.} {\bf D90} (2014), no.~4 045013,
  [\href{http://arxiv.org/abs/1406.6354}{{\tt arXiv:1406.6354}}].

\bibitem{Finazzo:2016mhm}
S.~I. Finazzo, R.~Critelli, R.~Rougemont, and J.~Noronha, {\it {Momentum
  transport in strongly coupled anisotropic plasmas in the presence of strong
  magnetic fields}},  {\em Phys. Rev.} {\bf D94} (2016), no.~5 054020,
  [\href{http://arxiv.org/abs/1605.06061}{{\tt arXiv:1605.06061}}].

\bibitem{Critelli:2014kra}
R.~Critelli, S.~I. Finazzo, M.~Zaniboni, and J.~Noronha, {\it {Anisotropic
  shear viscosity of a strongly coupled non-Abelian plasma from magnetic
  branes}},  {\em Phys. Rev.} {\bf D90} (2014), no.~6 066006,
  [\href{http://arxiv.org/abs/1406.6019}{{\tt arXiv:1406.6019}}].

\bibitem{Ammon:2016szz}
M.~Ammon, J.~Leiber, and R.~P. Macedo, {\it {Phase diagram of 4D field theories
  with chiral anomaly from holography}},  {\em JHEP} {\bf 03} (2016) 164,
  [\href{http://arxiv.org/abs/1601.02125}{{\tt arXiv:1601.02125}}].

\bibitem{Adler:1969gk}
S.~L. Adler, {\it {Axial vector vertex in spinor electrodynamics}},  {\em Phys.
  Rev.} {\bf 177} (1969) 2426--2438.

\bibitem{Bell:1969ts}
J.~S. Bell and R.~Jackiw, {\it {A PCAC puzzle: pi0 --> gamma gamma in the sigma
  model}},  {\em Nuovo Cim.} {\bf A60} (1969) 47--61.

\bibitem{D'Hoker:2009mm}
E.~D'Hoker and P.~Kraus, {\it {Magnetic Brane Solutions in AdS}},  {\em JHEP}
  {\bf 0910} (2009) 088, [\href{http://arxiv.org/abs/0908.3875}{{\tt
  arXiv:0908.3875}}].

\bibitem{Henningson:1998gx}
M.~Henningson and K.~Skenderis, {\it {The Holographic Weyl anomaly}},  {\em
  JHEP} {\bf 07} (1998) 023, [\href{http://arxiv.org/abs/hep-th/9806087}{{\tt
  hep-th/9806087}}].

\bibitem{Balasubramanian:1999re}
V.~Balasubramanian and P.~Kraus, {\it {A Stress tensor for Anti-de Sitter
  gravity}},  {\em Commun.Math.Phys.} {\bf 208} (1999) 413--428,
  [\href{http://arxiv.org/abs/hep-th/9902121}{{\tt hep-th/9902121}}].

\bibitem{Taylor:2000xw}
M.~Taylor, {\it {More on counterterms in the gravitational action and
  anomalies}},  \href{http://arxiv.org/abs/hep-th/0002125}{{\tt
  hep-th/0002125}}.

\bibitem{Witten:1998qj}
E.~Witten, {\it {Anti-de Sitter space and holography}},  {\em
  Adv.Theor.Math.Phys.} {\bf 2} (1998) 253--291,
  [\href{http://arxiv.org/abs/hep-th/9802150}{{\tt hep-th/9802150}}].

\bibitem{Bilal:1999ph}
A.~Bilal and C.-S. Chu, {\it {A Note on the chiral anomaly in the AdS / CFT
  correspondence and 1 / N**2 correction}},  {\em Nucl. Phys.} {\bf B562}
  (1999) 181--190, [\href{http://arxiv.org/abs/hep-th/9907106}{{\tt
  hep-th/9907106}}].

\bibitem{D'Hoker:2009bc}
E.~D'Hoker and P.~Kraus, {\it {Charged Magnetic Brane Solutions in AdS (5) and
  the fate of the third law of thermodynamics}},  {\em JHEP} {\bf 1003} (2010)
  095, [\href{http://arxiv.org/abs/0911.4518}{{\tt arXiv:0911.4518}}].

\bibitem{Jensen:2013kka}
K.~Jensen, R.~Loganayagam, and A.~Yarom, {\it {Anomaly inflow and thermal
  equilibrium}},  {\em JHEP} {\bf 05} (2014) 134,
  [\href{http://arxiv.org/abs/1310.7024}{{\tt arXiv:1310.7024}}].

\bibitem{Jensen:2011xb}
K.~Jensen, M.~Kaminski, P.~Kovtun, R.~Meyer, A.~Ritz, et~al., {\it
  {Parity-Violating Hydrodynamics in 2+1 Dimensions}},  {\em JHEP} {\bf 1205}
  (2012) 102, [\href{http://arxiv.org/abs/1112.4498}{{\tt arXiv:1112.4498}}].

\bibitem{Jensen:2012jh}
K.~Jensen, M.~Kaminski, P.~Kovtun, R.~Meyer, A.~Ritz, et~al., {\it {Towards
  hydrodynamics without an entropy current}},  {\em Phys.Rev.Lett.} {\bf 109}
  (2012) 101601, [\href{http://arxiv.org/abs/1203.3556}{{\tt
  arXiv:1203.3556}}].

\bibitem{Landsteiner:2011cp}
K.~Landsteiner, E.~Megias, and F.~Pena-Benitez, {\it {Gravitational Anomaly and
  Transport}},  {\em Phys.Rev.Lett.} {\bf 107} (2011) 021601,
  [\href{http://arxiv.org/abs/1103.5006}{{\tt arXiv:1103.5006}}].

\bibitem{Gynther:2010ed}
A.~Gynther, K.~Landsteiner, F.~Pena-Benitez, and A.~Rebhan, {\it {Holographic
  Anomalous Conductivities and the Chiral Magnetic Effect}},  {\em JHEP} {\bf
  1102} (2011) 110, [\href{http://arxiv.org/abs/1005.2587}{{\tt
  arXiv:1005.2587}}].

\bibitem{Landsteiner:2011iq}
K.~Landsteiner, E.~Megias, L.~Melgar, and F.~Pena-Benitez, {\it {Holographic
  Gravitational Anomaly and Chiral Vortical Effect}},  {\em JHEP} {\bf 09}
  (2011) 121, [\href{http://arxiv.org/abs/1107.0368}{{\tt arXiv:1107.0368}}].

\bibitem{Amado:2011zx}
I.~Amado, K.~Landsteiner, and F.~Pena-Benitez, {\it {Anomalous transport
  coefficients from Kubo formulas in Holography}},  {\em JHEP} {\bf 1105}
  (2011) 081, [\href{http://arxiv.org/abs/1102.4577}{{\tt arXiv:1102.4577}}].

\bibitem{Ammon:2016fru}
M.~Ammon, S.~Grieninger, A.~Jimenez-Alba, R.~P. Macedo, and L.~Melgar, {\it
  {Holographic quenches and anomalous transport}},  {\em JHEP} {\bf 09} (2016)
  131, [\href{http://arxiv.org/abs/1607.06817}{{\tt arXiv:1607.06817}}].

\bibitem{Edalati:2010hk}
M.~Edalati, J.~I. Jottar, and R.~G. Leigh, {\it {Shear Modes, Criticality and
  Extremal Black Holes}},  {\em JHEP} {\bf 1004} (2010) 075,
  [\href{http://arxiv.org/abs/1001.0779}{{\tt arXiv:1001.0779}}].

\bibitem{Edalati:2010pn}
M.~Edalati, J.~I. Jottar, and R.~G. Leigh, {\it {Holography and the sound of
  criticality}},  {\em JHEP} {\bf 1010} (2010) 058,
  [\href{http://arxiv.org/abs/1005.4075}{{\tt arXiv:1005.4075}}].

\bibitem{Jimenez-Alba:2014pea}
A.~Jimenez-Alba and L.~Melgar, {\it {Anomalous Transport in Holographic Chiral
  Superfluids via Kubo Formulae}},  {\em JHEP} {\bf 10} (2014) 120,
  [\href{http://arxiv.org/abs/1404.2434}{{\tt arXiv:1404.2434}}].

\bibitem{Liu:2016hqb}
Y.~Liu and F.~Pena-Benitez, {\it {Spatially modulated instabilities of
  holographic gauge-gravitational anomaly}},
  \href{http://arxiv.org/abs/1612.00470}{{\tt arXiv:1612.00470}}.

\bibitem{Grozdanov:2016tdf}
S.~Grozdanov, D.~M. Hofman, and N.~Iqbal, {\it {Generalized global symmetries
  and dissipative magnetohydrodynamics}},
  \href{http://arxiv.org/abs/1610.07392}{{\tt arXiv:1610.07392}}.

\bibitem{Rangamani:2009xk}
M.~Rangamani, {\it {Gravity and Hydrodynamics: Lectures on the fluid-gravity
  correspondence}},  {\em Class. Quant. Grav.} {\bf 26} (2009) 224003,
  [\href{http://arxiv.org/abs/0905.4352}{{\tt arXiv:0905.4352}}].

\bibitem{Kovtun:2012rj}
P.~Kovtun, {\it {Lectures on hydrodynamic fluctuations in relativistic
  theories}},  {\em J. Phys.} {\bf A45} (2012) 473001,
  [\href{http://arxiv.org/abs/1205.5040}{{\tt arXiv:1205.5040}}].

\bibitem{Kharzeev:2011ds}
D.~E. Kharzeev and H.-U. Yee, {\it {Anomalies and time reversal invariance in
  relativistic hydrodynamics: the second order and higher dimensional
  formulations}},  {\em Phys. Rev.} {\bf D84} (2011) 045025,
  [\href{http://arxiv.org/abs/1105.6360}{{\tt arXiv:1105.6360}}].

\bibitem{Starinets:2002br}
A.~O. Starinets, {\it {Quasinormal modes of near extremal black branes}},  {\em
  Phys. Rev.} {\bf D66} (2002) 124013,
  [\href{http://arxiv.org/abs/hep-th/0207133}{{\tt hep-th/0207133}}].

\bibitem{Janiszewski:2011sx}
S.~Janiszewski, {\it {Perturbations of Moving Membranes in $AdS_7$}},  {\em
  JHEP} {\bf 09} (2012) 093, [\href{http://arxiv.org/abs/1112.0085}{{\tt
  arXiv:1112.0085}}].

\bibitem{Nakamura:2009tf}
S.~Nakamura, H.~Ooguri, and C.-S. Park, {\it {Gravity Dual of Spatially
  Modulated Phase}},  {\em Phys.Rev.} {\bf D81} (2010) 044018,
  [\href{http://arxiv.org/abs/0911.0679}{{\tt arXiv:0911.0679}}].

\bibitem{Donos:2012wi}
A.~Donos and J.~P. Gauntlett, {\it {Black holes dual to helical current
  phases}},  {\em Phys. Rev.} {\bf D86} (2012) 064010,
  [\href{http://arxiv.org/abs/1204.1734}{{\tt arXiv:1204.1734}}].

\end{thebibliography}\endgroup

\end{document}